\documentclass[prd,twocolumn,showpacs,superscriptaddress]{revtex4}

\usepackage{amsfonts}
\usepackage{amsmath}
\usepackage{amssymb}
\usepackage{upgreek}
\usepackage{bm}
\usepackage{dcolumn}
\usepackage{epsfig}
\usepackage{graphicx}
\usepackage{graphics}
\usepackage[latin1]{inputenc}
\usepackage{latexsym}
\usepackage{rotating}
\usepackage{hyperref}
\usepackage{xspace} 
\usepackage[usenames]{color}

\usepackage{ulem}
\definecolor {darkgreen}{rgb}{0.2,0.7,0.2}

\newcommand\be{\begin{equation}}
\newcommand\ee{\end{equation}}
\def\ba#1\ea{\begin{align}#1\end{align}}

\newcommand{\DK}{{\mbox{\tiny DK}}}
\newcommand{\I}{{\cal{I}}}

\newcommand{\Kerr}{{\mbox{\tiny K}}}

\newcommand{\Kep}{{\mbox{\tiny Kep}}}

\newcommand{\GR}{{\mbox{\tiny GR}}}

\newcommand{\CS}{{\mbox{\tiny CS}}}

\begin{document}
\title{Approximate Waveforms for Extreme-Mass-Ratio Inspirals \\ in Modified Gravity Spacetimes}

\author{Jonathan Gair}
\affiliation{Institute of Astronomy, Madingley Road, Cambridge, CB30HA, United Kingdom.}

\author{Nicol\'as Yunes}
\affiliation{MIT and Kavli Institute, Cambridge, MA 02139, USA.}

\date{\today}

\begin{abstract} 
Extreme mass-ratio inspirals, in which a stellar-mass compact object spirals into a supermassive black hole, are prime candidates for detection with space-borne milliHertz gravitational wave detectors, similar to the Laser Interferometer Space Antenna. The gravitational waves generated during such inspirals encode information about the background in which the small object is moving, providing a tracer of the spacetime geometry and a probe of strong-field physics. In this paper, we construct approximate, ``analytic-kludge'' waveforms for such inspirals with parameterized post-Einsteinian corrections that allow for generic, model-independent deformations of the supermassive black hole background away from the Kerr metric. These approximate waveforms include all of the qualitative features of true waveforms for generic inspirals, including orbital eccentricity and relativistic precession. The deformations of the Kerr metric are modeled using a  recently proposed, modified gravity bumpy metric, which parametrically deforms the Kerr spacetime while ensuring that three approximate constants of the motion remain for geodesic orbits: a conserved energy, azimuthal angular momentum and Carter constant. The deformations represent modified gravity effects and have been analytically mapped to several modified gravity black hole solutions in four dimensions. In the analytic kludge waveforms, the conservative motion is modeled by a post-Newtonian expansion of the geodesic equations in the deformed spacetimes, which in turn induce modifications to the radiation-reaction force. These analytic-kludge waveforms serve as a first step toward complete and model-independent tests of General Relativity with extreme mass-ratio inspirals.
\end{abstract}


\maketitle

\section{Introduction}
``I seem to have been only like a boy playing on the seashore, and diverting myself in now and then finding a smoother pebble or a prettier shell than ordinary, whilst the great ocean of truth lay all undiscovered before me.''~\cite{Newton} Isaac Newton's quote reminds us of the great unknowns that remain 
in gravitational astrophysics~\cite{Schutz:2009tz}. Our understanding of this field should soon be revolutionized by the detection of gravitational waves (GWs) with ground~\cite{Giazotto:1988gw,Abramovici:1992ah} and space-based interferometers~\cite{Danzmann:2003tv,Danzmann:2003ad,Prince:2003aa,Prince:2009uc}. Indeed, the ground-based detectors LIGO~\cite{ligo} and Virgo~\cite{virgo} are currently undergoing upgrades toward a sensitivity at which a first direct detection of gravitational waves is likely. Space-borne detectors are being planned and will hopefully be operational in the next decade. 

Detectors in space will be sensitive to low-frequency [$(10^{-5}$-$10^{-1})$ Hz] GWs. One of the most promising sources in this frequency range are extreme-mass ratio inspirals (EMRIS)~\cite{AmaroSeoane:2007aw}. These systems consist of a stellar-mass compact object [$(10^{0}$-$10^{2}) M_{\odot}$], such as a neutron star or black hole (BH), that spirals into a supermassive BH [$(10^{5}$-$10^{7}) M_{\odot}$],typically  on an inclined and eccentric orbit. Due to their extreme mass-ratio, these inspirals proceed slowly, generating hundreds of thousands of cycles of gravitational radiation while the smaller object is in the strong-field region close to the central supermassive black hole. These GWs encode detailed information about the structure of the spacetime exterior to massive compact objects and the non-linear, ``strong-field'' nature of the gravitational theory that describes the dynamics of the inspiral.

The detection of such GWs requires the construction of accurate waveform templates that allow the extraction of signals from noisy data via matched filtering. EMRI waveforms, however, are particularly difficult to construct, as one requires these to be accurate over hundreds of thousands of cycles. Progress has been made by constructing approximate waveforms that encode all the key features of EMRI waveforms, while being computationally inexpensive. There are two families of such ``kludge'' waveforms that have been used extensively --- the ``analytic kludge'' developed by Barack and Cutler~\cite{AK}, which we will focus on in this paper, and the numerical kludge~\cite{Gair:2005ih,Babak:2006uv}. In the numerical kludge, the orbital trajectories are based on exact Kerr geodesics, the parameters of which are slowly evolving under the influence of radiation-reaction. Once the orbital trajectories have been obtained, the waveforms are constructed using a quadrupole approximation. The AK model is built around the gravitational waveforms generated by particles describing Keplerian ellipses~\cite{1963PhRv..131..435P,1964PhRv..136.1224P}, whose semi-major axis, eccentricity and inclination angle evolve according to certain post-Newtonian (PN) equations~\cite{1963PhRv..131..435P,1964PhRv..136.1224P,Blanchet:2002av}. Relativistic precession of the orbital plane and the perihelion are also included using post-Newtonian approximations.

In the past, attempts have been made to modify EMRI waveforms to study their ability to test General Relativity (GR). Such attempts can be divided into two classes: ``extrinsic'' or ``intrinsic''. The former picks a concrete alternative theory, such as dynamical Chern-Simons (CS) gravity~\cite{jackiw:2003:cmo,Alexander:2009tp}, constructs waveforms for that particular theory~\cite{Alexander:2007kv,Yunes:2007ss,Yunes:2009hc,Sopuerta:2009iy} and compares these to the predictions of GR. The latter are null-tests of GR, where one assumes the validity of GR a priori and concentrates on internal consistency tests, such as measuring the multipolar structure of the metric~\cite{Ryan:1995wh,Collins:2004ex,2010PhRvD..81b4030V,2010PhRvD..82j4041V}, or multi-modal spectroscopy of inspiral and ringdown waveforms~\cite{Barack:2006pq,Berti:2005ys,Berti:2007zu}. Both classes of tests are intrinsically valuable, but they are not ideal to search for  departures from GR in a systematic and model-independent way.

Recently, there has been a focused effort to develop such a systematic and model-independent approach. Yunes and Pretorius~\cite{2009PhRvD..80l2003Y,Yunes:2010qb,Cornish:2011ys} proposed the parameterized post-Einsteinian (ppE) framework, a model-independent template family for the quasi-circular inspiral of comparable mass, non-spinning, compact binaries.
In this approach, deformations of the conservative Hamiltonian and of the radiation-reaction force are mapped onto the waveform observable, i.e.~the frequency-domain GW response function. The meta-template family allows for generic deformations away from GR, that have been shown to reproduce waveform predictions from all known alternative theories proposed to date. However, this approach is ill-suited to EMRI waveforms, as the orbits are likely to be inclined and eccentric and the mass ratios are extreme.

To address this shortcoming, Vigeland, Yunes and Stein~\cite{Vigeland:2011ji} took the first steps toward extending the ppE framework to EMRIs. They concentrated on conservative corrections to the orbit, realizing that these could be parameterized by deformations of the metric tensor. Previous attempts to construct such ``bumpy spacetimes''~\cite{Collins:2004ex,2010PhRvD..81b4030V,2010PhRvD..82j4041V} had focused on intrinsic or null tests, and had therefore used metrics that satisfied the Einstein equations, but with potentially unphysical matter distributions, such as naked singularities, or unphysical spacetime regions, e.g., closed timelike curves.The approach of~\cite{Vigeland:2011ji} was to only allow for the subclass of metric deviations that would ensure the existence of an approximately conserved energy, (z-component of) angular momentum, a second-order Killing tensor and a Carter constant, without requiring that the Einstein equations be satisfied. Such an approach led to a parametrically deformed metric that was shown to map to metrics in known alternative theories in four dimensions~\cite{Yunes:2009hc,Yunes:2011we}.  

In this paper, we construct corrections to the AK waveforms by considering geodesics in the modified gravity, bumpy metric of~\cite{Vigeland:2011ji}. We begin by providing explicit expressions for all components of the metric deformation as an expansion in $r \gg M$, where $r$ is the field-point distance to the supermassive BH with mass $M$; we require that the metric remain asymptotically flat, with the same scaling at spatial infinity as that predicted by the GR peeling theorems, and disallowing pure angular deformations. We show that even with these restrictions, all known metrics in alternative theories are still recovered by an appropriate choice of the parameters characterising the metric functions~\cite{Yunes:2009hc,Yunes:2011we}. Arbitrary choices of these parameters lead to parametric, metric deformations at ${\cal{O}}(1/r^{2})$, ${\cal{O}}(1/r^{3})$, ${\cal{O}}(1/r^{4})$ and ${\cal{O}}(1/r^{5})$ relative to the Kerr metric.

We then calculate the geodesics equations associated with this background. 
We parameterise the orbits via the location of their turning points and will relate various quantities to the Kerr geodesic orbit with the same turning points. The existence of three approximately conserved quantities allows us to explicitly separate the geodesic equations into first order form. From these, we calculate the three orbital frequencies associated with the orbital motion. Although we do not introduce explicit corrections to the radiation-reaction force, modifications to the fluxes of energy, angular momentum and Carter constant will be introduced due to modifications to the orbit itself. We calculate these implicit modifications with a quadrupolar approximation for the fluxes of energy and angular momentum. The Carter constant flux is calculated by assuming that the inclination angle remains approximately constant and by requiring that exactly circular orbits remain circular under radiation reaction. Finally, we collect all pieces of the calculation, providing an explicit prescription to build AK waveforms in these families of parametrically deformed spacetimes.

The study performed here is similar, yet more generic than others already carried out in the literature. For example, Barack and Cutler~\cite{Barack:2006pq} considered modifications to AK waveforms induced by a perturbation to the quadrupole moment of a Kerr BH. Using results in~\cite{1995ApJ...452..819L}, they introduced modifications to the precession frequencies and rate of change of orbital frequency that would be induced by a quadrupole moment deformation; they then searched for the accuracy with which LISA could measure the size of such a deformation. This study, however, neglected modifications to the eccentricity evolution. Glampedakis and Babak~\cite{Glampedakis:2005cf} also considered a class of bumpy spacetimes which differed from Kerr in the quadrupole moment only. Unfortunately these metrics are of Petrov Type I, and thus do not allow for the existence of a Carter constant or for the separability of the geodesic equations~\cite{2006gr.qc.....1094W}. Moreover, both studies neglected the possibility that the metric could be modified at multipole orders higher than quadrupole. This is a critical disadvantage, as strong-field modifications of GR are likely to introduce corrections at higher than quadrupole order, e.g.,~dynamical CS gravity modifies the metric at hexadecapole order, leaving the quadrupole and octopole unchanged. 

Although the waveforms presented here will be useful for studies that determine the accuracy to which instruments like LISA could constrain model-independent deviations from GR, they are not unique or complete. To demand uniqueness is futile in a ppE scheme, as it is also in the parameterized post-Newtonian or parameterized post-Keplerian schemes that are used to search for deviations from GR in the Solar System and in binary pulsar observations respectively. An infinite number of theories predict an infinite number of possible deviations in the observables of interest, while the schemes above can only parameterize a subset of them, i.e., the subset that can reproduce all the predictions of alternative theories that are known to date~\cite{Yunes:2009hc,Yunes:2011we}. 

These waveforms are also not complete because here we will modify only one of the three main ingredients that go into waveform construction, i.e.,~the metric tensor, and will neglect modifications to the first-order equations of motion (that prescribe how GWs are sourced by matter distributions) and the second-order equations of motion (that describe the self-force and radiation-reaction). However, a large sub-class of quadratic gravity theories exist in which the corrections to the metric are dominant over modifications to the wave generation and radiation-reaction~\cite{2011PhRvD..83f4038S}. Nonetheless, a more complete analysis should also investigate the excitation of scalar and vectorial modes in the metric perturbation, which could arise from modifications to the wave generation. We leave an investigation of these other effects to future work, but the results found here will still hold when these other modifications are introduced.

The remainder of this paper is organized as follows: 
Section~\ref{sec:bumpy-description} introduces the new bumpy framework to model modified gravity theories; 
Section~\ref{sec:geodesic-eqs} calculates the geodesic equations that preserve the Kerr turning points;
Section~\ref{sec:fund-freq} computes the orbital frequencies associated with the modified orbital motion;
Section~\ref{sec:rr} calculates the implicit deformations introduced to the radiation-reaction force;
Section~\ref{sec:MG-AK} builds AK waveforms from the modified orbital frequencies and fluxes;
Section~\ref{sec:Conclusions} concludes and points to future research.

Throughout this paper we use the conventions of Misner, Thorne and Wheeler~\cite{Misner:1973cw}. We use Greek letters to denote spacetime indices, while Latin ones in the middle of the alphabet $i,j,\ldots$ stand for spatial indices only. We also use geometric units with $G=c=1$. Background quantities are denoted with an overhead bar, while quantities associated with geodesics in the Kerr metric are denoted with a subscript $K$. 
 
\section{Bumpy Spacetimes for Modified Gravity}
\label{sec:bumpy-description}

In this section we discuss the parametrically deformed BH metric 
that we will later use to study geodesics. We begin by recapitulating
the most important results of~\cite{Vigeland:2011ji} for this paper. 
We then simplify this metric prescription by considering expansions
in $M/r \ll 1$, thus allowing us to compute explicit expressions for all
components of the metric deformation. Finally, we describe some properties
of the new metric.  

\subsection{Spacetime Construction}
We decompose the metric tensor that is to describe the background spacetime of a supermassive BH as 
\be
g_{\mu \nu} = \bar{g}_{\mu \nu} + \epsilon \; h_{\mu \nu}\,,
\ee
where $\epsilon \ll 1$ is a ``deformation'' book-keeping parameter that reminds us that $|h_{\mu \nu}|/|\bar{g}_{\mu \nu}| \ll 1$. The background metric is assumed to be the Kerr metric $\bar{g}_{\mu \nu} = g^{\Kerr}_{\mu \nu}$, which in Boyer-Lindquist coordinates has components:
\ba
\label{sol:metric_elements1}
{g}^{\Kerr}_{tt} &= - \left(1 - \frac{2 M r}{\rho^{2}} \right) \,, 
\qquad
{g}^{\Kerr}_{t\phi} = - \frac{2 M^{2} a r}{\rho^{2}} \sin^{2}{\theta}\,, 
\\
{g}^{\Kerr}_{rr} &= \frac{\rho^{2}}{\Delta}\,,
\qquad
{g}^{\Kerr}_{\theta \theta} = \rho^{2}\,,
\qquad
{g}^{\Kerr}_{\phi \phi} = \frac{\Sigma}{\rho^{2}} \sin^{2}{\theta}\,,
\label{sol:metric_elements}
\ea
for a BH with mass $M$ and spin angular momentum directed along the symmetry axis 
 of magnitude $S = M^{2} a$, where $a$ is the {\emph{dimensionless}} Kerr spin parameter. Equations~\eqref{sol:metric_elements1}-\eqref{sol:metric_elements} depend on the functions 
\ba
\rho^{2} &\equiv r^{2} + a^{2} M^{2} \cos^{2}{\theta}\,, 
\\
\Delta &\equiv r^{2} f + M^{2} a^{2}\,, 
\qquad  
f \equiv 1 - \frac{2 M}{r}\,,
\\ 
\Sigma &\equiv (r^{2} + M^{2} a^{2})^{2} - M^{2} a^{2} \Delta \sin^{2}{\theta}\,. 
\ea

We restrict attention to a certain class of metrics. We begin by requiring that the full metric be stationary and axisymmetric, although it need not solve the Einstein equations, i.e.~it is a solution to a more general set of modified gravity field equations that have a smooth GR limit: $g_{\mu \nu} \to \bar{g}_{\mu \nu}$ as $h_{\mu \nu} \to 0$. In addition to the existence of a temporal and an azimuthal Killing vector, we also assume that a certain integrability condition holds (see Eq.~$(49)$ in~\cite{Vigeland:2011ji}) such that the metric can be written in Lewis-Papapetrou form. The deformation of this metric $g_{\mu \nu} = \bar{g}_{\mu \nu} + \epsilon \; h_{\mu \nu}$, can be transformed to Boyer-Lindquist-like coordinates in which all components of the metric perturbation vanish except $(h_{tt},h_{t\phi},h_{rr},h_{r\theta},h_{\theta \theta},h_{\phi \phi})$. These are the only components that are allowed to be non-zero. 

With this at hand, we then force the full metric $g_{\mu \nu}$ to possess three constants of the motion: a conserved energy, azimuthal component of angular momentum and Carter constant. The first two are generated directly from the Killing vectors and are exact, while the last one is built from an approximate, second-order Killing tensor valid at least to ${\cal{O}}(\epsilon^{2})$. This Killing tensor is parameterized as
\be
\xi_{\mu \nu} = \Delta l_{(\mu} k_{\nu)} + r^{2} g_{\mu \nu}\,.
\label{KT-param}
\ee
The parameterization of the Killing tensor as in Eq.~\eqref{KT-param} identifies the coordinate $r$ with a Boyer-Lindquist-like radial coordinate. The vectors $l_{\mu}$ and $k_{\mu}$ are required to be null and $\xi_{\mu \nu}$ is required to satisfy the Killing tensor equation, which determines the components of the null vectors up to some arbitrary functions of the radial coordinate. This in turn determines the final form for the metric perturbation.

The non-vanishing components of the metric deformation can be written as~\cite{Vigeland:2011ji}\footnote{The expressions here differ slightly from~\cite{Vigeland:2011ji}, because we use a dimensionless Kerr parameter.} 
\begin{widetext}
\allowdisplaybreaks[1]
\ba
	h^{}_{tt} &= -a \frac{P^{\DK}_2}{P^{\DK}_1} h^{}_{t\phi} - \frac{a}{2} \frac{\rho^4 \Delta}{P^{\DK}_1} \frac{\partial h^{}_{t\phi}}{\partial r} - \frac{2M^{2}a^2r (r^2+M^{2}a^2) \Delta \sin\theta \cos\theta}{\rho^2 P^{\DK}_1} h^{}_{r\theta} + \frac{(r^2+M^{2}a^2)\hat{\rho}^{2} \Delta}{\rho^2 P^{\DK}_1} \I \nonumber \\
			&+  \frac{2M^{2}a^2r^2 \Delta \sin^2\theta}{P^{\DK}_1} \gamma_1 + \frac{\hat{\rho}^{2} (r^2+M^{2}a^2) \Delta}{\rho^2 P^{\DK}_1} \Theta_3 - a \frac{\Delta\sin^2\theta}{\rho^2} \frac{P^{\DK}_3}{P^{\DK}_1} \gamma_3 + \frac{2\Delta}{\rho^2} \frac{P^{\DK}_4}{P^{\DK}_1} \gamma_4 \nonumber \\
			&-  \frac{a^2 M}{2} \frac{\rho^2 \Delta^2 \sin^2\theta}{P^{\DK}_1} \frac{d\gamma_1}{dr} - \frac{a}{2} \frac{\Delta^2 (\Sigma+2a^2M^{3}r\sin^2\theta) \sin^2\theta}{P^{\DK}_1} \frac{d \gamma_3}{dr} - \frac{a^2 M}{2} \frac{\Delta^2 (\rho^2-4Mr) \sin^2\theta}{P^{\DK}_1} \frac{d\gamma_4}{dr} \,,
\label{Carter-conds-DK1}
\ea
\allowdisplaybreaks[1]
\ba
h^{}_{rr} &= - \frac{1}{\Delta} \I -\frac{1}{\Delta} \Theta_3 \,,
\ea
\allowdisplaybreaks[1]
\ba
h^{}_{\phi\phi} &= -\frac{(r^2+M^{2}a^2)^2}{M^{2}a^2} h^{}_{tt} + \frac{\Delta}{M^{2}a^2} \I - \frac{2(r^2+M^{2}a^2)}{a} h^{}_{t\phi} + \frac{\Delta}{M^{2}a^2} \Theta_3 - \frac{2\Delta^2 \sin^2\theta}{Ma} \gamma_3 + \frac{2\Delta^2}{M^{2}a^2} \gamma_4 \,,
\ea
\allowdisplaybreaks[1]
\ba
\frac{\partial h^{}_{\theta\theta}}{\partial r} &= \frac{2r}{\rho^2} h^{}_{\theta\theta} + \frac{2 M^{2}a^2 \sin\theta \cos\theta}{\rho^2} h^{}_{r\theta} + 2 \frac{\partial h^{}_{r\theta}}{\partial \theta} + \frac{2r}{\rho^2} \I - 2r \, \gamma_1 + \frac{2r}{\rho^2} \Theta_3 \,, 
\ea
\allowdisplaybreaks[1]
\ba
\frac{\partial^2 h^{}_{t\phi}}{\partial r^2} &= \frac{8aM^{2} \sin\theta \cos\theta}{\rho^8} \frac{P^{\DK}_5}{P^{\DK}_1} h^{}_{r\theta} - \frac{4aM^{2}r(r^2+M^{2}a^2) \sin\theta \cos\theta}{\rho^6} \frac{\partial h^{}_{r\theta}}{\partial r} + \frac{2M^{2}a^2\sin^2\theta}{\rho^4} \frac{P^{\DK}_6}{P^{\DK}_1} h^{}_{t\phi} - \frac{2r}{\rho^2}\frac{P^{\DK}_7}{P^{\DK}_1} h^{}_{t\phi} \nonumber \\
		&+  \frac{4aM^{2}r\sin^2\theta}{\rho^4}\frac{P^{\DK}_{15}}{P^{\DK}_{16}} \I - \frac{4aM^{2}r\sin^2\theta}{\rho^4}\frac{P^{\DK}_8}{P^{\DK}_1} \gamma_1 + \frac{4aM^{2}r}{\rho^4}\frac{P^{\DK}_9}{P^{\DK}_1} \Theta_3 + \frac{2\sin^2\theta}{\rho^4}\frac{P^{\DK}_{10}}{P^{\DK}_1} \gamma_3 \nonumber \\
		&-  \frac{16aM^{2} \sin^2\theta}{\rho^4}\frac{P^{\DK}_{11}}{P^{\DK}_1} \gamma_4 - \frac{2aM}{\rho^4}\frac{P^{\DK}_{12}}{P^{\DK}_1} \frac{d\gamma_1}{dr} - \frac{2\sin^2\theta}{\rho^4}\frac{P^{\DK}_{13}}{P^{\DK}_1} \frac{d\gamma_3}{dr} - \frac{2aM\sin^2\theta}{\rho^4}\frac{P^{\DK}_{14}}{P^{\DK}_1} \frac{d\gamma_4}{dr} - \frac{a M \Delta \sin^2\theta}{\rho^2} \frac{d^{2} \gamma_1}{dr^{2}} \nonumber \\
		&-  \frac{\Delta \sin^2\theta}{\rho^4}(\Sigma+2a^2M^{3}r\sin^2\theta) \frac{d^{2} \gamma_3}{dr^{2}} - \frac{a M\Delta (\rho^2-4Mr) \sin^2\theta}{\rho^4} \frac{d^{2} \gamma_4}{dr^{2}} \,,
		\label{Carter-conds-DK}
\ea
\end{widetext}
where $\hat{\rho}^{2} \equiv r^{2} - M^{2} a^{2} \cos^{2}{\theta}$ and $P^{\DK}_i$ are polynomials in $r$ and $\cos\theta$, given explicitly in the Appendix of~\cite{Vigeland:2011ji} (we have here adopted the deformed Kerr parameterization of~\cite{Vigeland:2011ji}). The quantities $\gamma_{i} = \gamma_{i}(r)$ are arbitrary functions of radius, while $\Theta_{3}=\Theta_{3}(\theta)$ is an arbitrary function of polar angle. The quantity $\I$ is defined as
\ba
\I &= \int dr \; \left[ \frac{2 M^{2} a^2 \sin\theta \cos\theta}{\rho^2} h^{}_{r\theta} + 2r \, \gamma_1 + \rho^2 \, \frac{d \gamma_1}{dr} \right] \,.
\ea
The metric perturbation component $h^{}_{r\theta}$ is free.

The parametrically deformed metric represents a family, some members of which are well-known BH solutions in modified gravity theories. For example, a certain choice of deformation parameters $\gamma_{A}$ leads to the slowly-rotating BH solution in dynamical CS gravity~\cite{Yunes:2009hc}, while another choice leads to modified Schwarzschild BHs in quadratic gravity theories~\cite{Yunes:2011we}, as was shown in~\cite{Vigeland:2011ji}. In both cases, these solutions derive from field equations that arise from a diffeomorphism invariant theory, with a well-defined Lagrangian density. We refer the interesting reader to~\cite{Vigeland:2011ji} for more details on the metric construction.

\subsection{Simplification of the Parameterization}
\label{subsec:Simplifications}

Let us simplify the metric perturbation of the previous section with the following criteria:
\begin{enumerate}
\item {\bf{Asymptotic flatness}}: Require that $h_{\mu \nu} \to 0$ at spatial infinity.
\item {\bf{Peeling}}: Require that $|h_{\mu \nu}| \sim r^{-2}$ or faster for $r/M \gg 1$. 
\item {\bf{Occam's Razor}}: Set the largest number of metric components to zero that
are not needed to reproduced known modified gravity predictions for the metric tensor in four
dimensions~\cite{Yunes:2009hc,Yunes:2011we}. 
\end{enumerate}
Requirement (1) ensures that there is no constant piece to the metric deformation, such that the total metric $g_{\mu \nu} \to \bar{g}_{\mu \nu}$ at $\iota^{0}$. Since the Kerr metric is asymptotically flat, so would the total metric. 
Requirement (2) ensures that the BH mass is not renormalized by $1/r$ corrections to the $(t,t)$ or $(r,r)$ metric components. The norm in this requirement is to be taken with the flat metric in spherical coordinates, such that the $(\theta,\phi)$ sub-sector is simply the metric on the $2$-sphere.
These two requirements ensure that the metric perturbation satisfies all Solar System constraints, as it introduces modifications at higher than leading, Newtonian-order in a weak-field expansion. Furthermore, any $1/r$ correction to $(h_{tt},h_{t\phi},h_{rr})$ would renormalize the BH mass or spin angular momentum, which would not be measurable. Requirement (3) automatically implies that (i) we can set $h_{r\theta} = 0$; and (ii) we can disallow pure angular deformations by setting $\Theta_{3} = 0$. 

Since the purpose of this paper is to construct ppE AK waveforms, which make extensive use of weak-field expansions in $M/r \ll 1$, we choose to parameterize the remaining free functions as Taylor series:
\ba
\gamma_{A} &= \sum_{n=0}^{\infty} \gamma_{A,n} \left(\frac{M}{r}\right)^{n}\,,
\quad
\gamma_{3} = \frac{1}{r} \sum_{n=0}^{\infty} \gamma_{3,n} \left(\frac{M}{r}\right)^{n}\,, \quad
\ea
where $A=1$ or $4$ and $\gamma_{i,n}$ are dimensionless constants. We have pulled out a factor of $1/r$ in the expansion of $\gamma_{3}$ because this quantity has dimensions of $[M]^{-1}$, as one can see from $h_{\phi \phi}$ in Eqs.~\eqref{Carter-conds-DK1}-\eqref{Carter-conds-DK}. Notice that with these definitions the $\gamma_{m,n}$ constants are dimensionless.

With this at hand, let us simplify the metric components, starting with $h_{rr}$. The integral $\I$ can now be solved exactly: $\I = \rho^{2} \gamma_{1}$. Via requirement (3)
\be
h_{rr} = -\gamma_{1} \frac{\rho^{2}}{\Delta}.
\label{hrreq}
\ee
Requirements (1) and (2) force us to choose $\gamma_{1,0} = 0 = \gamma_{1,1}$, since $\rho^{2}/\Delta \to 1$ for $r \gg M$. 

The next simplest component to analyze is $h_{\theta\theta}$, whose behavior is governed by Eqs.~\eqref{Carter-conds-DK1}-\eqref{Carter-conds-DK}, which using the previous results simplifies to $\partial h^{}_{\theta\theta}/{\partial r} = {2r}/({\rho^2}) h^{}_{\theta\theta}$. The solution to this equation is $h_{\theta \theta} = \Theta_{4}(\theta) \rho^{2}$. Since $\bar{g}_{\theta \theta} = \rho^{2}$, this correction would be leading order in the angular sector. By requirement (3), we disallow it and set $\Theta_{4} = 0$. This simplification, and the restrictions made when deriving the metric perturbation, fix the coordinate system in such a way that $h_{\theta \theta} = 0$ and therefore $g_{\theta \theta} = \rho^2 + O(\epsilon^2)$. The radial coordinate thus preserves, at leading order, its physical interpretation in the Kerr metric --- it may be interpreted as a circumferential radius in the equatorial plane and becomes the  oblate-spheroidal radial coordinate in flat-space at infinity. The fact that the radial coordinate has been fixed in this way will be important for interpretation of the waveforms we derive in this paper. This will be discussed further in Section~\ref{sec:MG-AK}.

To determine the remaining metric components, we must first solve the differential equation for $h_{t\phi}$, as the $h_{tt}$ and $h_{\phi \phi}$ components depend explicitly on the former. This is an elliptic equation that could be solved numerically. Since we seek analytic solutions only, however, we will solve it in the $M/r \ll 1$ limit. To do so, we write 
\ba
h_{t \phi} &= M \sum_{n=2}^{N} h_{t \phi,n}(\theta) \left(\frac{M}{\rho}\right)^{n} + {\cal{O}}\left(\frac{1}{\rho^{N+1}}\right)\,, 
\label{t-phi-exp}
\ea
where $h_{t\phi,n}$ are functions of $\theta$ that we will determine by solving Eqs.~\eqref{Carter-conds-DK1}-\eqref{Carter-conds-DK}. We could have chosen to expand $h_{t \phi}$ in a $1/r$ basis, but we empirically found that a $1/\rho$ basis yields simpler results. Solving Eqs.~\eqref{Carter-conds-DK1}-\eqref{Carter-conds-DK} order by order in $1/\rho$, we find that the first few non-zero terms are
\begin{widetext}
\allowdisplaybreaks[1]
\begin{align}
h_{t\phi,2} &= \sin^{2}{\theta} \left[
-  \gamma_{3,3} - a \left( \gamma_{1,2} + \gamma_{4,2} \right) 
- a^{2} \gamma_{3,1}
\right]\,,
\\
h_{t\phi,3} &= \sin^{2}{\theta} \left[ 
\left(2  \gamma_{3,3} -  \gamma_{3,4} \right) 
+ a \left(6 \gamma_{4,2} - \gamma_{4,3}   
+ 2 \gamma_{1,2} -\gamma_{1,3} \right)
+ 2 a^2 \gamma_{3,1} \left(4  \cos^{2}{\theta} - 3 \right)
\right]
\\
h_{t \phi,4} &= \sin^{2}{\theta} \left[
-a^4  \gamma_{3,1} -
\left(\gamma_{4,2} + \gamma_{1,2}\right) a^3
+ 2 \left( -4 \gamma_{3,1} \cos^{2}{\theta} - \gamma_{3,3} + 4 \gamma_{3,1} \right) a^2 
\right.
\nonumber \\
&+ \left. 
\left(-\gamma_{1,4} - 8 \gamma_{4,2} + 6 \gamma_{4,3} - \gamma_{4,4} +
 2 \gamma_{1,3}\right) a 
+  2 \gamma_{3,4} - \gamma_{3,5} \right]\,,
\\
h_{t \phi,5} &= \sin^{2}{\theta} \left\{
2 \gamma_{3,5} 
+ a \left( - \gamma_{4,5} - 8 \gamma_{4,3} + 2 \gamma_{1,4} 
+ 6 \gamma_{4,4} - \gamma_{1,5} \right)
+ a^{2}\left[ - 2 \gamma_{3,4} + \gamma_{3,3} \left(-2 
+ 5 \cos^{2}{\theta} \right) - (1/2) \gamma_{3,4} \cos^{2}{\theta} \right]
\right. 
\nonumber \\
&+ \left. 
a^{3} \left[ (1/2) \cos^{2}{\theta} \left(2 \gamma_{1,2} 
- \gamma_{4,3} - \gamma_{1,3} - 2 \gamma_{4,2}  \right) - \gamma_{4,3} - \gamma_{1,3} + 4 \gamma_{4,2} \right]
+ a^{4} \left( -4 \gamma_{3,1} \cos^{4}{\theta} + 9 \gamma_{3,1} \cos^{2}{\theta} - 4 \gamma_{3,1} \right)
\right\}\,,
\\
h_{t \phi,6} &= \sin^{4}{\theta} \left[
a^{2} \left(\gamma_{3,5} - 6 \gamma_{3,4} + 8 \gamma_{3,3} \right)
+ a^{3} \left(\gamma_{1,4} - 2 \gamma_{4,3} + \gamma_{4,4} - 2 \gamma_{1,3} \right)
+ 2 \gamma_{3,3} a^{4} 
\right.
\nonumber \\
&+ \left.
a^{5} \left( \gamma_{4,2} + \gamma_{1,2} \right)
+ a^{6}  \gamma_{3,1}  \right]
+ \sin^{2}{\theta} \left[
2 \gamma_{3,6} - \gamma_{3,7} + a \left(6 \gamma_{4,5} + 2 \gamma_{1,5} - 8 \gamma_{4,4} - \gamma_{1,6} - \gamma_{4,6} \right)
\right. 
\nonumber \\
&+ \left.
a^{2} \left(4 \gamma_{3,4} - 3 \gamma_{3,5} \right) 
+ a^{3} \left(-2 \gamma_{1,4} + 2 \gamma_{1,3} + 6 \gamma_{4,3} - 2 \gamma_{4,4} \right)
- 3 \gamma_{3,3} a^{4}
- a^{5} \left(\gamma_{1,2} + \gamma_{4,2}\right)
- \gamma_{3,1} a^{6}  \right]\,,
\label{ang-func}
\end{align}
\end{widetext}
where we have here simplified the solution by setting  
$\gamma_{4,0} = 0$, $\gamma_{4,1} = 0$ and $\gamma_{3,2} = 0$. 
We will find that these conditions are necessary to ensure the metric is 
asymptotically flat. We have also set $\gamma_{3,0} = 0$, 
which is required for the differential equation to be satisfied.

Let us now return to the $h_{tt}$ and $h_{\phi \phi}$ components, which have been completely specified by the above solutions. At spatial infinity we find that 
\ba
h_{tt} &\sim 2 \gamma_{4,0} - \frac{2 \gamma_{4,1} - 8 M  \gamma_{4,0}}{r}\,, 
\\
h_{\phi \phi} &\sim - \frac{r^{2}}{2 a M^{2}} \gamma_{3,2} \sin^{2}{\theta} + {\cal{O}}(1)\,,
\ea
when $M/r \ll 1$. By requirements (1) and (2), this then implies that $(\gamma_{4,0},\gamma_{4,1},\gamma_{3,2})$ must all be set to zero. 

In summary, the requirements of asymptotic flatness and the non-renormalization of the mass, have forced us to the following conditions
\ba
\Theta_{3}(\theta)&=0\,,
\quad
\gamma_{1,0}=0\,, 
\quad
\gamma_{1,1}=0\,, 
\quad
\gamma_{3,0} = 0\,, 
 \\ 
\gamma_{3,1} &= 0\,, 
\quad
\gamma_{4,0} = 0\,, 
\quad
\gamma_{4,1} = 0\,, 
\quad
\gamma_{3,2} = 0\,. 
\ea
With these choices, the metric perturbations $(h_{\theta \theta},h_{r \theta})$ vanish, 
$h_{rr}$ is given by Eq.~\eqref{hrreq} and $h_{t \phi}$ is given in the far field by 
Eq.~\eqref{t-phi-exp}, with the angular functions given in Eq.~\eqref{ang-func}. 
The remaining components $(h_{tt},h_{\phi\phi})$ are given explicitly by 
Eqs.~\eqref{Carter-conds-DK1}-\eqref{Carter-conds-DK}.  

Let us now take the far-field expansion of all the metric perturbations: 
\be
h_{\mu \nu} = \sum_{n} h_{\mu \nu,n} \left(\frac{M}{r}\right)^{n}\,.
\ee
The first few non-zero terms are 
\begin{widetext}
\allowdisplaybreaks[1]
\ba
h_{tt,2} &=  \gamma_{1,2} + 2 \gamma_{4,2} - 2 a \gamma_{3,1} \sin^{2}{\theta}\,,
\nonumber \\
h_{tt,3} &= \gamma_{1,3} - 8 \gamma_{4,2} - 2 \gamma_{1,2} + 2 \gamma_{4,3} + 8 a \gamma_{3,1} \sin^{2}{\theta}\,,
\nonumber \\
h_{tt,4} &= -8 \gamma_{4,3} - 2 \gamma_{1,3} + 2 \gamma_{4,4} + 8 \gamma_{4,2} + \gamma_{1,4} - 8 a  \gamma_{3,1} \sin^{2}{\theta} + a^{2} \left( \gamma_{1,2} + 2 \gamma_{4,2} \right) \sin^{2}{\theta} + 2 a^{3} \gamma_{3,1} \cos^{2}{\theta} \sin^{2}{\theta}\,,
\nonumber \\
h_{tt,5} &= 16 a^{3} \gamma_{3,1} \sin^{4}{\theta}  + \sin^{2}{\theta} \left[ 4 a \gamma_{3,3} + a^{2} \left(\gamma_{1,3} - 2 \gamma_{1,2} - 12 \gamma_{4,2} + 2 \gamma_{4,3} \right) - 12 a^{3} \gamma_{3,1}\right] 
\nonumber \\
&+ a^{2} \left(8 \gamma_{4,2} + 2 \gamma_{1,2} \right) + \gamma_{1,5} + 2 \gamma_{4,5}-2 \gamma_{1,4} + 8 \gamma_{4,3} - 8 \gamma_{4,4} \,,
\label{met-pert-eq1}
\ea
\ba
h_{rr,2} &= - \gamma_{1,2}\,,
\qquad
h_{rr,3} = - \gamma_{1,3} - 2 \gamma_{1,2}\,,
\qquad
h_{rr,4} = - \gamma_{1,4} - 2 \gamma_{1,3} - 4 \gamma_{1,2} + (1/2) \gamma_{1,2} a^{2} \left(1 - \cos{2 \theta} \right)\,,
\nonumber \\
h_{rr,5} &=  a^{2} \sin^{2}{\theta} \left(\gamma_{1,3} + 2 \gamma_{1,2} \right) -\gamma_{1,5} - 2 \gamma_{1,4} - 4 \gamma_{1,3} - 8 \gamma_{1,2} + 2 a^{2} \gamma_{1,2}\,,
\ea
\ba
h_{t \phi,2} &= - M \sin^{2}{\theta} \left[ \gamma_{3,3} + a \left( \gamma_{1,2} + \gamma_{4,2} \right) + a^{2} \gamma_{3,1} \right]\,,
\nonumber \\
h_{t \phi,3} &= -8 M a^{2} \gamma_{3,1} \sin^{4}{\theta} + M \sin^{2}{\theta} \left[  \left(2 \gamma_{3,3} - \gamma_{3,4} \right) + a \left( 6 \gamma_{4,2} - \gamma_{4,3} + 2 \gamma_{1,2} - \gamma_{1,3} \right) + 2 \gamma_{3,1} a^{2} \right]\,,
\nonumber \\
h_{t \phi,4} &= M \sin^{4}{\theta} \left[ a^{2} \left( 8 \gamma_{3,1} - \gamma_{3,3} \right) + a^{3}  \left(-\gamma_{1,3} - \gamma_{4,2} \right) - a^{4} \gamma_{3,1} \right] + \sin^{2}{\theta} \left[ \left(2 \gamma_{3,4} - \gamma_{3,5}\right) 
\right. 
\nonumber \\
&+ \left.
a \left( - \gamma_{4,4} - 8 \gamma_{4,2} + 6 \gamma_{4,3} - \gamma_{1,4} + 2 \gamma_{1,3} \right) - a^{2} \gamma_{3,3} \right]\,
\nonumber \\
h_{t \phi,5} &=-16 M a^{4} \gamma_{3,1} \sin^{6}{\theta} + M \sin^{4}{\theta} \left[ a^{2} \left(-2 \gamma_{3,3}- \gamma_{3,4} \right) + a^{3} \left(-\gamma_{4,3} + 10 \gamma_{4,2} + 2 \gamma_{1,2} - \gamma_{1,3} \right) + 14 a^{4} \gamma_{3,1} \right] 
\nonumber \\
&+ \sin^{2}{\theta} \left[ \left(2 \gamma_{3,5} - \gamma_{3,6}\right) + a \left(-\gamma_{1,5} - 8 \gamma_{4,3} - \gamma_{4,5} + 2 \gamma_{1,4} + 6 \gamma_{4,4}\right) - \gamma_{3,4} a^{2} 
\right. 
\nonumber \\
&+ \left.
a^{3} \left(-2 \gamma_{1,2} - 6 \gamma_{4,2} \right) - 2 a^{4} \gamma_{3,1} \right]\,,
\ea
\ba
h_{\phi \phi,-2} &= 0\,,
\qquad
h_{\phi \phi,-1} = 0\,,
\qquad
h_{\phi \phi,0} = 2 M^{2} a \gamma_{3,1} \sin^{4}{\theta} \,,
\qquad
h_{\phi \phi,1} = 0\,,
\nonumber \\
h_{\phi \phi,2} &= M^{2} \sin^{4}{\theta} \left[ 2 a \gamma_{3,3} + a^{2} \gamma_{1,2} + a^{3} \gamma_{3,1} \left(4- 2 \cos^{2}{\theta} \right)\right]\,,
\nonumber \\
h_{\phi \phi,3} &=  8 M^{2} a^{3} \gamma_{3,1} \sin^{6}{\theta} + M^{2} \sin^{4}{\theta} \left[ a  \left( - 4 \gamma_{3,3} + 2 \gamma_{3,4} \right) + a^{2} \left(-2 \gamma_{1,2} - 4 \gamma_{4,2} + \gamma_{1,3} \right) - 4 a^{3} \gamma_{3,1} \right]\,,
\label{met-pert-eq}
\ea
\end{widetext}
Notice that although $(h_{t t},h_{r r})$ are indeed dimensionless, $h_{t\phi}$ has units of length and $h_{\phi \phi}$ has units of length squared, as expected since these are also the dimensions of the corresponding components of the Kerr metric.

The perturbation is parameterized by a number of constants, depending on how many terms in $M/r$ are kept relative to the leading-order Kerr metric: up to ${\cal{O}}(M^{2}/r^{2})$, the metric deformation is given by the 4 constants ${\cal{B}}_{2} \equiv (\gamma_{1,2},\gamma_{3,1},\gamma_{3,3},\gamma_{4,2})$; up to ${\cal{O}}(M^{3}/r^{3})$ it is given by the 7 constants ${\cal{B}}_{2} \cup {\cal{B}}_{3}$, where ${\cal{B}}_{3} \equiv (\gamma_{1,3},\gamma_{3,4},\gamma_{4,3})$; up to ${\cal{O}}(M^{4}/r^{4})$ it is given by the 10 constants ${\cal{B}}_{2} \cup {\cal{B}}_{3} \cup {\cal{B}}_{4}$, where ${\cal{B}}_{4} \equiv (\gamma_{1,4},\gamma_{3,5},\gamma_{4,4})$; up to ${\cal{O}}(M^{5}/r^{5})$ it is given by the 13 constants ${\cal{B}}_{2} \cup {\cal{B}}_{3} \cup {\cal{B}}_{4} \cup {\cal{B}}_{5}$, where ${\cal{B}}_{5} \equiv (\gamma_{1,5},\gamma_{4,5},\gamma_{3,6})$. Later on in the paper, we will take certain ${\cal{B}}_{N}$ limits, where we mean we will only let the ${\cal{B}}_{N}$ coefficients be non-zero and set all others to zero.   

Lastly, note that known BH solutions in alternative theory of gravity can be reproduced with this parameterization, as shown in~\cite{Vigeland:2011ji}. In particular, in this paper we will frequently compare our results to that of dynamical CS gravity, where the metric of a slowly rotating BH is identical to Kerr, except in its $(t,\phi)$ component, which is given by~\cite{Yunes:2009hc}
\ba
g_{t \phi}^{\CS} &= - \frac{2 M^{2} a r}{\rho^{2}} \sin^{2}{\theta}  
\nonumber \\
&+ 
\frac{5}{8} \zeta \frac{a M^{5}}{r^{4}} \left(1 + \frac{12 M}{7 r} + \frac{27 M^{2}}{10 r^{2}} \right) \sin^{2}{\theta}\,,
\ea
where $\zeta$ is a dimensionless coupling constant of the theory. This metric can be completely reproduced by the above parameterization by choosing all $\gamma_{i,j} = 0$, except for $\gamma_{3}(r)$, whose first non-vanishing terms are
\be
\gamma_{3,5} = -  \frac{5}{8} a \zeta\,,
\quad
\gamma_{3,6} = - \frac{65}{28} a \zeta\,.
\quad
\gamma_{3,7} = - \frac{709}{112} a \zeta\,.
\ee
%

\subsection{Physical Properties of the Deformed Metric}

At this junction, one might wonder how the metric of Eqs.~\eqref{Carter-conds-DK1}-\eqref{Carter-conds-DK} may change physical properties of the spacetime. For a stationary and asymptotically flat spacetime, the event horizon coincides with the Killing horizon, and it is given by the hypersurface where:
\be
\frac{g_{t\phi}^{2}}{g_{\phi\phi}^{2}}-\frac{g_{tt}}{g_{\phi\phi}} = 0\,.
\label{EH-cond}
\ee
Let us now use the fact that at $r=r_{+}\equiv M + M (1 - a^{2})^{1/2}$, $\Delta = 0$ and then 
\be
h_{\phi \phi} = - \frac{(r^{2} + M^{2} a^{2})^{2}}{M^{2} a^{2}} h_{tt} -  \frac{2 (r^{2} + M^{2} a^{2}}{a} h_{t \phi}\,.
\ee
With this at hand, linearizing Eq.~\eqref{EH-cond} in $\epsilon$ leads to
\be
0=- \Delta \left[ \frac{2 h_{t\phi}}{Ma} + h_{tt} \left(2 + \frac{\rho^{2}}{M^{2} a^{2}}\right)\right]\,.
\ee
This quantity clearly vanishes at $r = r_{+}$, which then implies that the event horizon of the full metric remains at its Kerr value. Moreover, one can also show that the horizon coincides with the location of a coordinate singularity, by evaluating the components of the inverse metric. 

Other quantities, however, will be different in the deformed spacetime relative to their Kerr values. For example, it is likely that the geometry of the ergosphere, defined by the hypersurface where $g_{tt} = 0$, will be modified. These and other properties of the spacetime metric will not be needed in this paper, and so we leave them to be explored elsewhere. 

\section{Geodesic Equations}
\label{sec:geodesic-eqs}

In this section, we study the geodesic equations associated with the new metric we computed
in the previous section. We begin by summarizing the most important results of~\cite{Vigeland:2011ji},
regarding the geodesic equations. We then fix the location of the turning points of the orbit and compute the modifications to the conserved quantities relative to the Kerr geodesic with the same turning points. 

\subsection{First-Order Equations}
\label{first-order-equations}

Let us define the {\emph{dimensionless}} constants of the motion associated with this new metric as follows:
\be
E \equiv - t^{\alpha} u_{\alpha}\,,
\quad
L \equiv M \phi^{\alpha} u_{\alpha}\,
\quad
C \equiv M^{2} \xi_{\alpha \beta} u^{\alpha} u^{\beta}\,,
\label{dimlessconst}
\ee
where $t^{\alpha} = [1,0,0,0]$ is a timelike Killing vector, $\phi^{\alpha} = [0,0,0,1]$ is an azimuthal Killing vector
and $u^{\alpha} = [\dot{t},\dot{r},\dot{\theta},\dot{\phi}]$ is the $4$-velocity, with overhead dots standing for differentiation with respect to proper time. The Carter constant is also often defined as $Q \equiv C - (L - a E)^{2}$. Here the approximate, second-order Killing tensor $\xi_{\alpha \beta} = \bar{\xi}_{\alpha \beta} + \delta \xi_{\alpha \beta}$. The background Killing tensor is simply the Kerr one 
\be
\bar{\xi}_{\alpha \beta} =  \Delta\, \bar{k}_{(\alpha} \bar{l}_{\beta)} + r^{2}\, \bar{g}^{}_{\alpha \beta}\,,
\label{KT-def}
\ee
with $\bar{k}^{\alpha}$ and $\bar{l}^{\alpha}$ the principal null directions of Kerr:  
\ba
\label{KPV1}
\bar{k}^{\alpha} &= \left[\frac{r^{2}+M^{2}a^{2}}{\Delta},1,0,\frac{Ma}{\Delta}\right]\,,
\\
\bar{l}^{\alpha} &= \left[\frac{r^{2}+M^{2}a^{2}}{\Delta},-1,0,\frac{Ma}{\Delta}\right]\,.
\label{KPV}
\ea
The Killing tensor deformation is 
\be
\delta \xi_{\alpha \beta} \equiv \Delta \left[ \delta k_{(\alpha} l_{\beta)} + \delta l_{(\alpha} k_{\beta)} + 2 h^{}_{\delta (\alpha} \bar{k}_{\beta)} \bar{l}^{\delta} \right]
+ 3 r^{2} h^{}_{\alpha \beta} \,.
\ee
where the deformed null vectors are
\ba
\delta k_{}^{\alpha} &= \left[ \frac{r^{2} + M^{2} a^{2}}{\Delta} \gamma_{1} + \gamma_{4},
\gamma_{1}, 0, \frac{a M}{\Delta} \gamma_{1} + \gamma_{3}\right]\,,
\\
\delta l_{}^{\alpha} &= \left[  \gamma_{4}, 0, 0, \gamma_{3} \right]\,,
\ea
and we have used the requirements of Sec.~\ref{subsec:Simplifications} to simplify these expressions.

The dimensionless constants of the motion, when evaluated on the perturbed metric, become
\ba
\label{EqE}
E &= \bar{E} + \epsilon \left( h_{\mu \nu} t^{\mu} \bar{u}^{\nu} + \bar{g}_{\mu \nu} t^{\mu} \delta u^{\nu} \right)\,,
\\
L &= \bar{L} + M \epsilon \left( h_{\mu \nu} \phi^{\mu} \bar{u}^{\nu} + \bar{g}_{\mu \nu} \phi^{\mu} \delta u^{\nu} \right)\,,
\\
C &= \bar{C} + M^{2} \epsilon \left( \delta \xi_{\mu \nu} \bar{u}^{\mu} \bar{u}^{\nu} + 2 \bar{\xi}_{\mu \nu} \bar{u}^{(\mu} \delta u^{\nu)} \right)\,,
\label{EqC}
\ea
and the normalization condition for the four-velocity is
\be
\label{norm-cond}
0 = h_{\mu \nu} \bar{u}^{\mu} \bar{u}^{\nu}
+ 2 \bar{g}_{\mu \nu} \bar{u}^{\mu} \delta u^{\nu}\,,
\ee
since by definition $-1 = \bar{g}_{\mu \nu} \bar{u}^{\mu} \bar{u}^{\nu}$. In all of these equations, $\bar{u}^{\mu} = [\dot{\bar{t}},\dot{\bar{r}},\dot{\bar\theta},\dot{\bar\phi}]$ is the unperturbed, Kerr four-velocity, while $\delta u^{\mu}$ is a perturbation of ${\cal{O}}(\epsilon)$. 

The second-order geodesic equations can be rewritten as a first-order set through Eqs.~\eqref{EqE}-\eqref{norm-cond}. To achieve this, one must make a gauge choice for the constants of the motion associated with the full spacetime $(E,L,C)$. In~\cite{Vigeland:2011ji}, the authors chose to keep $E = \bar{E}$, $L = \bar{L}$ and $C = \bar{C}$, which then implies that $\delta u^{\mu}$ must be such that all terms in parentheses in Eqs.~\eqref{EqE}-\eqref{EqC} vanish. This choice, however, forces the turning points of the orbit to be different from the GR orbit with constants $(\bar{E}$, $\bar{L}$, $\bar{Q}$). 

Using this condition and Eq.~\eqref{norm-cond}, the geodesic equations can be rewritten in first-order form:  
\ba
\label{geod-eqs1}
\rho^2\dot{t} &= T_{\Kerr}(r,\theta) + \epsilon \; \delta T(r,\theta) \,, \\
\qquad \rho^4\dot{r}^2 &= R_{\Kerr}(r) + \epsilon \; \delta R(r,\theta) \,, \\
\rho^2\dot{\phi} &= \Phi_{\Kerr}(r,\theta) + \epsilon \; \delta \Phi(r,\theta) \,,  \\
\qquad \rho^4\dot{\theta}^2 &= \Theta_{\Kerr}(\theta) + \epsilon \; \delta \Theta(r,\theta) \,,
\label{geod-eqs}
\ea
where the Kerr potentials $(T_{\Kerr},R_{\Kerr},\Theta_{\Kerr},\Phi_{\Kerr})$ are given by
\allowdisplaybreaks[4]
\ba
\label{tdot-GR}
T_{\Kerr} &= -a M \left(a M \bar{E} \sin^{2}{\theta} - M \bar{L} \right) + \left(r^{2} 
+ M^{2} a^{2} \right) \frac{P}{\Delta}\,,  
\\
\label{rdot-GR}
R_{\Kerr} &=  P^{2} -  \Delta\; \left[M^{2} \bar{Q} + M^{2}\left(a \bar{E} - \bar{L} \right)^{2} + r^{2} \right]\,, 
\\
\label{thetadot-GR}
\Theta_{\Kerr} &= M^{2} \bar{Q} -  M^{2} \bar{L}^{2} \cot^{2}{\theta} 
- M^{2} a^{2} \cos^{2}{\theta} \left(1 - \bar{E}^{2} \right)\,,
\\
\label{phidot-GR}
\Phi_{\Kerr} &= - \left(a M \bar{E} - \frac{M \bar{L}}{\sin^{2}{\theta}} \right) 
+ \frac{a M P}{\Delta}\,,
\ea
where $P \equiv \bar{E} (r^2+ M^{2} a^2)- M^{2} a\bar{L}$. 
The perturbation potentials $(\delta T, \delta R, \delta\Theta, \delta\Phi)$ are given by
\ba
\delta T(r,\theta) &= \left[ \frac{(r^2+ M^{2} a^2)^2}{\Delta} - M^{2} a^2\sin^2\theta \right] h_{t\alpha}\bar{u}^\alpha 
\nonumber \\
&+ \frac{2aM^{2}r}{\Delta} h_{\phi\alpha} \bar{u}^\alpha \,, \label{dT_potential}  \\
\delta R(r,\theta) &= \Delta \left[A(r,\theta) \, r^2+B(r,\theta)\right] \,, \\
\delta \Theta(r,\theta) &= A(r,\theta) \, M^{2} a^2\cos^2\theta - B(r,\theta)\,, \\
\delta \Phi(r,\theta) &= \frac{2aM^{2}r}{\Delta} h_{t\alpha} \bar{u}^\alpha - \frac{\rho^2-2Mr}{\Delta\sin^2\theta} h_{\phi\alpha} \bar{u}^\alpha \,, 
\label{dPhi_potential}
\ea
and the functions $A(r,\theta)$ and $B(r,\theta)$ are given by
\ba
A(r,\theta) &= 2 \left[ h_{\alpha t} \bar{\dot{t}} + h_{\alpha \phi} \bar{\dot{\phi}} \right]  \bar{u}^\alpha - h_{\alpha\beta} \bar{u}^\alpha \bar{u}^\beta \,, \\
B(r,\theta) &= 2 \left[ \left( \bar{\xi}_{tt} \bar{\dot{t}} + \bar{\xi}_{t\phi} \bar{\dot{\phi}} \right) \delta u^{t} + \left( \bar{\xi}_{t\phi} \bar{\dot{t}} + \bar{\xi}_{\phi\phi} \bar{\dot{\phi}} \right) \delta u^{\phi}  \right] \nonumber \\
	&+  \delta\xi_{\alpha\beta} \bar{u}^\alpha \bar{u}^\beta \,,
\ea
in which the four-velocity associated with the background trajectory $\bar{u}^{\mu}$ is
\ba
\bar{u}^{t} &\equiv \dot{\bar{t}} = \rho^{-2} T_{\Kerr}(r,\theta)\,
\quad
\bar{u}^{r} \equiv \dot{\bar{r}} = \rho^{-2} \sqrt{R_{\Kerr}(r)}\,,
\\
\bar{u}^{\theta} &\equiv  \dot{\bar{\theta}} = \rho^{-2} \sqrt{\Theta_{\Kerr}(\theta)}\,,
\qquad
\bar{u}^{\phi} \equiv \dot{\bar{\phi}} = \rho^{-2} \Phi_{\Kerr}(r,\theta),
\ea
and the four-velocity associated with the perturbation is
\ba
\delta{u}^{t} &\equiv \delta \dot{t} = \rho^{-2} \delta T\,,
\qquad
\delta{u}^{\phi} \equiv \delta \dot{\phi} = \rho^{-2} \delta\Phi\,.
\ea

Let us expand the perturbation to the geodesic equations in the far-field limit, using the metric in Eqs.~\eqref{met-pert-eq1}-\eqref{met-pert-eq}. Doing so, we find
%
\ba
\delta T &= M^{2} \sum_{n=0}^{\infty} \delta T_{n} \left(\frac{M}{r}\right)^{n}\,,
\quad
\delta \Phi = M \sum_{n=2}^{\infty} \delta \Phi_{n} \left(\frac{M}{r}\right)^{n}\,,
\\
\delta R &= M^{4} \sum_{n=-2}^{\infty} \delta R_{n} \left(\frac{M}{r}\right)^{n}\,,
\quad
\delta \Theta = M^{2} \sum_{n=0}^{\infty} \delta \Theta_{n} \left(\frac{M}{r}\right)^{n}\,,
\ea
where we note that $(\delta T_{n},\delta R_{n}, \delta \Theta_{n}, \delta \Phi_{n})$ are all dimensionless, unlike $(T_{\Kerr},\delta T,\Theta_{\Kerr},\delta \Theta)$ which have units of $M^{2}$, $(R_{\Kerr},\delta R)$ which have units of $M^{4}$ and $(\Phi_{\Kerr},\delta \Phi)$ which have units of $M$. With this, the first non-vanishing perturbations are
%
\begin{widetext}
\ba
\delta T_{0} &= \left(2 \gamma_{4,2} + \gamma_{1,2} - 2 a \gamma_{3,1} \sin^{2}{\theta} \right) \bar{E}\,,
\qquad
\delta T_{1}  = \left(\gamma_{1,3} + 2 \gamma_{4,3} + 2 \gamma_{1,2}\right) \bar{E}\,,
\nonumber \\
\delta T_{2} &= \bar{E} \left[ \left(\gamma_{1,4} + 2 \gamma_{1,3} + 2 \gamma_{4,4} + 4 \gamma_{1,2}\right) + a^{2} \left(\gamma_{1,2} + 2 \gamma_{4,2} \right) \right] + \bar{L} \left[ - \gamma_{3,3} - a \left(\gamma_{4,2} + \gamma_{1,2} \right) - a^{2} \gamma_{3,1} \right]\,, 
\nonumber \\
\delta T_{3} &= \bar{E} \left[ \left(\gamma_{1,5} + 2 \gamma_{4,5} + 4 \gamma_{1,3} + 2 \gamma_{1,4} + 8 \gamma_{1,2} \right) + \left(\gamma_{1,3} + 2 \gamma_{4,3} \right) a^{2} \right] - \bar{L} \left[  \gamma_{3,4} + \left(\gamma_{1,3} + 2 \gamma_{1,2} + \gamma_{4,3}\right) a \right]\,,
\nonumber \\
\delta T_{4} &= \bar{E} \left[ \left(-4 \gamma_{1,2} + 2 \gamma_{4,4} + \gamma_{1,4} \right) a^{2} + \left( 2 \gamma_{4,6} + 8 \gamma_{1,3} + 2 \gamma_{1,5} + 16 \gamma_{1,2} + \gamma_{1,6} + 4 \gamma_{1,4} \right) \right] 
\nonumber \\
&+ \bar{L} \left[ - a^{2} \gamma_{3,3} - \gamma_{3,5} - \left(\gamma_{4,4} + \gamma_{1,4} + 4 \gamma_{1,2} + 2 \gamma_{1,3} \right) a \right]\,,
\nonumber \\
\delta T_{5} &= \bar{E} \left[ \left(-4 \gamma_{1,3} + 2 \gamma_{4,5} - 16 \gamma_{1,2} + \gamma_{1,5} \right) a^{2} + \left(4 \gamma_{1,5} + 32 \gamma_{1,2} + 8 \gamma_{1,4} + 2 \gamma_{4,7} + \gamma_{1,7} + 2 \gamma_{1,6} + 16 \gamma_{1,3} \right) \right] 
\nonumber \\
&+ \bar{L} \left[ - \gamma_{3,4} a^{2} - \gamma_{3,6} + \left( -4 \gamma_{1,3}- \gamma_{4,5} - \gamma_{1,5} - 2 \gamma_{1,4} - 8 \gamma_{1,2} \right) a + 2 a^{3} \gamma_{1,2} \right]\,,
\ea
\ba
\delta R_{-2} &= \left(2 \gamma_{4,2} + 2\gamma_{1,2} \right) \bar{E}^2  
- 2 \bar{E} \bar{L} \gamma_{3,1}   - \gamma_{1,2}  \,,
\nonumber \\
\delta R_{-1} &= \bar{E}^{2} \left(-4 \gamma_{4,2} + 2 \gamma_{4,3} + 2 \gamma_{1,3} \right) + 4 \bar{E} \bar{L} \gamma_{3,1} + 2 \gamma_{1,2} - \gamma_{1,3} \,,
\nonumber \\
\delta R_{0} &= \bar{E}^{2} \left[ \left(2 \gamma_{4,4} + 2 \gamma_{1,4} - 4 \gamma_{4,3} \right) + \left(4 \gamma_{4,2} + 3 \gamma_{1,2} \right) a^{2} \right] - \bar{L} \bar{E} \left[ 2 \left(\gamma_{4,2} + \gamma_{1,2}\right) a  + 2 \gamma_{3,3} + 4 a^{2} \gamma_{3,1} \right] 
\nonumber \\
&+ \bar{L}^{2} \left(2 a \gamma_{3,1} - \gamma_{1,2} \right) + 2 \gamma_{1,3} - \gamma_{1,4}- \left[a^{2} + \bar{Q} \right] \gamma_{1,2}\,, 
\nonumber \\
\delta R_{1} &= \bar{E}^{2} \left[ \left(2 \gamma_{4,5} + 2 \gamma_{1,5} - 4 \gamma_{4,4} \right) + \left(2 \gamma_{1,2} + 4 \gamma_{4,3} - 4 \gamma_{4,2} + 3 \gamma_{1,3} \right) a^{2} \right] 
+ \bar{E} \bar{L} \left[ \left(-4 \gamma_{1,2} - 2 \gamma_{1,3} - 2 \gamma_{4,3} + 4 \gamma_{4,2} \right) 
\right. 
\nonumber \\
&\times \left.
a + \left(4 \gamma_{3,3}- 2 \gamma_{3,4} \right) + 4 \gamma_{3,1} a^{2} \right] 
+ \bar{L}^{2} \left[ \left(-4 a \gamma_{3,1} \right) + \left(2 \gamma_{1,2} - \gamma_{1,3} \right) \right]
+ \bar{Q} \left(2 \gamma_{1,2} - \gamma_{1,3} \right) 
+ 2 \gamma_{1,4} - \gamma_{1,5} - a^{2} \gamma_{1,3}\,, 
\nonumber \\
\delta R_{2} &= \bar{E}^{2} \left[ \left(-4 \gamma_{4,5} + 2 \gamma_{4,6} + 2 \gamma_{1,6}\right) + \left(4 \gamma_{4,4} - 4 \gamma_{4,3} + 3 \gamma_{1,4} + 2 \gamma_{1,3} \right) a^{2} + a^{4} \left(\gamma_{1,2} + 2 \gamma_{4,2} \right) \right] 
\nonumber \\
&+ \bar{E} \bar{L} \left[ \left(-2 \gamma_{4,2} - 2 \gamma_{1,2} \right) a^{3} + \left(4 \gamma_{3,4} - 2 \gamma_{3,5} \right) + \left(-2 \gamma_{4,4} + 4 \gamma_{4,3} - 4 \gamma_{1,3} - 2 \gamma_{1,4} \right) a - 4 a^{2} \gamma_{3,3} - 2 a^{4} \gamma_{3,1} \right] 
\nonumber \\
&+
\bar{L}^{2} \left[ 2 a \gamma_{3,3} + \left(-\gamma_{1,4} + 2 \gamma_{1,3} \right) + a^{2} \gamma_{1,2} + 2 a^{3} \gamma_{3,1} \right] + \bar{Q} \left[ \left(-\gamma_{1,4} + 2 \gamma_{1,3} \right) - a^{2} \gamma_{1,2}\right] + 2 \gamma_{1,5} - \gamma_{1,6} - a^{2} \gamma_{1,4} \,,
\ea
\ba
\delta \Theta_{0} &= -2 \gamma_{3,1} a \bar{E}^2 \sin^{2}{\theta} + 2 \bar{E} \bar{L} \gamma_{3,1}\,,
\qquad 
\delta \Theta_{n>0} = 0\,,
\ea
\ba
\delta \Phi_{2} &= \bar{E}
\left( \gamma_{3,1} a^2 + a \gamma_{1,2} + a \gamma_{4,2} + \gamma_{3,3} \right)
- 2 a \bar{L} \gamma_{3,1}\,,
\nonumber \\
\delta \Phi_{3} &= \bar{E} \left[ \gamma_{3,4} + \left( \gamma_{4,3} + \gamma_{1,3} + 2 \gamma_{1,2} \right) a  \right]\,,
\nonumber \\
\delta \Phi_{4} &=\bar{E} \left[ \left( 2 \gamma_{1,3} + 4 \gamma_{1,2} + \gamma_{1,4} + \gamma_{4,4} \right) a  + M^{3} a^{2} \gamma_{3,3} + \gamma_{3,5} \right] - \bar{L} \left( a^{2} \gamma_{1,2} + 2 a \gamma_{3,3} \right)\,,
\nonumber \\
\delta \Phi_{5} &= \bar{E} \left[ \left( \gamma_{1,5} + 8 \gamma_{1,2} + 2 \gamma_{1,4} + \gamma_{4,5} + 4 \gamma_{1,3} \right) a + \gamma_{3,4} a^{2} + \gamma_{3,6} - 2 a^{3} \gamma_{1,2} \right] 
- \bar{L} \left[ \left(\gamma_{1,3} + 2 \gamma_{1,2} \right) a^{2} + 2 a \gamma_{3,4} \right]\,,
\nonumber \\
\delta \Phi_{6} &=\bar{E} \left[ \left(\gamma_{4,6} + 16 \gamma_{1,2} + 8 \gamma_{1,3} + 4 \gamma_{1,4} + \gamma_{1,6} + 2 \gamma_{1,5} \right) a + \gamma_{3,5} a^{2} + \gamma_{3,7} - \left(2 \gamma_{1,3} + 8 \gamma_{1,2} \right) a^{3} \right] 
\nonumber \\
&+ \bar{L} \left[ \left(-\gamma_{1,4} - 2 \gamma_{1,3} - 4\gamma_{1,2} \right) M^{4} - 2 a \gamma_{3,5} + a^{4} \gamma_{1,2} \right]\,.
\ea
\end{widetext}

In the CS limit, these separated equations become
\ba
\delta T_{\CS} &= \frac{5}{8} M^{2} \left(\frac{M}{r}\right)^{4} \bar{L} \, a \zeta\,,
\\
\delta R_{\CS} &= \frac{5}{4} M^{4} \left(\frac{M}{r}\right)^{2} \bar{E} \bar{L} \, a \zeta\,,
\\
\delta \Phi_{\CS} &= -\frac{5}{8} M \left(\frac{M}{r}\right)^{4} \bar{E} \, a \zeta\,,
\ea
to leading order and $\delta \Theta_{\CS} = 0$, all of which agrees with the results found in~\cite{Sopuerta:2009iy}. 

We finish this section with a comment on this decomposed set of first order equations. The existence of an approximate, second-order Killing tensor allowed us to construct a Carter constant, which in turn allowed us to rewrite the second-order geodesic equations in first-order form. This does not necessarily imply that the resulting first-order equations will be decoupled (i.e.,~that the $\dot{r}$ source term depends on $r$ only and the $\dot{\theta}$ source term depends on $\theta$ only). Nonetheless, in the far-field limit, we have just found empirically in the above equations that the resulting first-order equations do separate. 

\subsection{Weak-Field Expansion}
The equations presented above are not strictly weak-field expansions, as the GR constants of the motion $(\bar{E},\bar{L},\bar{Q})$ also depend on the radius of the orbit. We choose to parameterize the orbit in terms of a semi-latus rectum $p$, an eccentricity, $e$ and an inclination angle, $\theta_{\rm tp}$. The weak-field limit corresponds to taking $p \gg 1$. We define these constants of the motion from the radial and azimuthal turning points of the motion, which are given by $\dot{r}(r_\pm)=0$ and $\dot{\theta}(\theta_{\rm tp})=0$, with 
\ba
\label{rpm}
r_{\pm} = M p/(1 \mp e)\,.
\ea
For Kerr, setting $R_{\Kerr}(r_\pm) = 0=\Theta_{\Kerr} (\theta_{\rm tp})$, we find that
\ba
E_{\Kerr} &\sim 1 + \frac{1}{2 p} \left(e^{2}-1 \right) + \frac{3}{8 p^{2}} \left(1 - e^{2} \right)^{2} + {\cal{O}}(p^{-3})\,,
\\
L_{\Kerr} &\sim \sqrt{p} \sin{\theta_{\rm tp}} + \frac{1}{2} p^{-1/2} \left(e^{2} + 3\right) \sin{\theta_{\rm tp}} 
\nonumber \\
&- \frac{a}{p} \left( e^{2} + 3 \right) \sin^{2}{\theta_{\rm tp}}
- \frac{a^{2}}{2 p^{3/2}} \left( 3 + e^{2}\right) \sin{\theta_{\rm tp}} \cos^2\theta_{\rm tp}
\nonumber \\
&+ \sin{\theta_{\rm tp}} \left[ \frac{3}{8 p^{3/2}} \left(e^{2} + 3 \right)^{2} + \frac{a^{2}}{p^{3/2}} \left(1 + e^{2} \right) \right]
\nonumber \\
&- \frac{a}{2 p^{2}} \left(5 + 3 e^{2}\right) \left(e^{2} + 3\right) \sin^{2}{\theta_{\rm tp}} + {\cal{O}}(p^{-3})\,,
\\
Q_{\Kerr} &\sim \cos^2\theta_{\rm tp}\left( p + \left(e^{2} + 3 \right) - \frac{2 a}{p^{1/2}} \left(e^{2} + 3 \right)
 \sin{\theta_{\rm tp}} \right.
\nonumber \\
&+
\frac{1}{p} \left[  \left(e^{2} + 3 \right)^{2} + a^2 \left(3 + e^{2}\right) \sin^2\theta_{\rm tp} \right]
\nonumber \\
&+  \left.4\frac{a^{2}}{p^{3/2}} \sin\theta_{\rm tp} (2+e^2)(3+e^2)\right) + {\cal{O}}(p^{-2})\,.
\ea
We distinguish here between $(\bar{E},\bar{L},\bar{Q})$ and $(E_{\Kerr},L_{\Kerr},Q_{\Kerr})$, since we are defining $E_\Kerr$ etc. to be the constants of the motion for the Kerr orbit that has the same turning points as the geodesic in the deformed spacetime.

We can now insert these relations into our expressions for $(\delta T, \delta R, \delta \Theta, \delta \Phi)$, but the resulting equations are quite horrendous. We will thus present results only for special cases in which certain $\gamma_{i,j}$ are non-vanishing. If only ${\cal{B}}_{2} = (\gamma_{1,2},\gamma_{3,1},\gamma_{3,3},\gamma_{4,2})$ is non-vanishing (which corresponds to keeping only the ${\cal{O}}(1/r^{2})$ terms in $h_{\mu \nu}$), then
\ba
\delta T_{{\cal{B}}_{2}} &\sim \left(\gamma_{1,2} + 2 \gamma_{4,2} \right) M^{2} - 2 M^{2} a \gamma_{3,1} \sin^{2}{\theta}\,,
\\
\delta R_{{\cal{B}}_{2}} &\sim -2 M^{2} r^{2} p^{1/2} \gamma_{3,1} \sin{\theta_{\rm tp}} 
\nonumber \\
&+ M^{2} r^{2} \left(\gamma_{1,2} + 2 \gamma_{4,2} \right)\,,
\\
\delta \Theta_{{\cal{B}}_{2}} &\sim 2 M^{2} p^{1/2} \gamma_{3,1} \sin{\theta_{\rm tp}} - 2 M^{2} a \gamma_{3,1} \sin^{2}{\theta}\,,
\\
\delta \Phi_{{\cal{B}}_{2}} &\sim - \frac{2 M^{3} a \, p^{1/2}}{r^{2}} \gamma_{3,1}  \sin{\theta_{\rm tp}}  
 \nonumber  \\
&+ \frac{M^{3}}{r^{2}} \left[ a \left(\gamma_{1,2} + \gamma_{4,2} \right) + \gamma_{3,3} + \gamma_{3,1} a^{2} \right]\,.
\ea
In the higher order cases, we can obtain a more general formula: if ${\cal{B}}_{N>2} = (\gamma_{1,N},\gamma_{3,N+1},\gamma_{4,N})$ is non-vanishing (which corresponds to keeping only the ${\cal{O}}(1/r^{N})$ terms in $h_{\mu \nu}$), then (for $N>2$)
\ba
\delta T_{{\cal{B}}_{N}} &\sim r^{2} \left(\frac{M}{r}\right)^{N} \left(\gamma_{1,N} + 2 \gamma_{4,N} \right)\,,
\\
\delta R_{{\cal{B}}_{N}} &\sim r M^{3} \left(\frac{M}{r}\right)^{N-3} \left(\gamma_{1,N} + 2 \gamma_{4,N} \right)\,,
\\
\delta \Theta_{{\cal{B}}_{N}} &\sim 0\,,
\\
\delta \Phi_{{\cal{B}}_{N}} &\sim M \left(\frac{M}{r}\right)^{N} \left[ a \left(\gamma_{1,N} + \gamma_{4,N} \right) + \gamma_{3,N+1} \right]\,.
\ea
In the CS limit and to leading order in the weak-field, we find
\ba
\delta T_{\CS} &\sim \frac{5}{8} \frac{M^{6} p^{1/2}}{r^{4}} a \zeta \sin{\theta_{\rm tp}}\,,
\\
\delta R_{\CS} &\sim \frac{5}{4} \frac{M^{6} p^{1/2}}{r^{2}} a \zeta \sin{\theta_{\rm tp}}\,,
\\
\delta \Theta_{\CS} &\sim 0\,,
\\
\delta \Phi_{\CS} &\sim - \frac{5}{8} \frac{M^{5}}{r^{4}} a \zeta\,.
\ea
%

\subsection{Orbital Parameterization}
If we define the orbit through the semi-latus rectum, eccentricity and inclination parameter introduced above, then in the parametrically deformed spacetimes studied in this paper, the turning points are not in the same spatial location as the GR geodesic with the same $(\bar{E},\bar{L},\bar{Q})$. We can use the same parametrisation of the orbit if we force the turning points to be in these locations. The corresponding constants of the motion are shifted from those in the Kerr spacetime with the given turning points by an amount of ${\cal{O}}(\epsilon)$: 
\ba
\bar{E} &= E_{\Kerr} + \epsilon \delta E\,, 
\\
\bar{L} &= L_{\Kerr} + \epsilon \delta L\,, 
\\
\bar{Q} &= Q_{\Kerr} + \epsilon \delta Q\,.\
\ea
This only changes the $(T_{\Kerr},R_{\Kerr},\Theta_{\Kerr},\Phi_{\Kerr})$ potentials in the geodesic equations [Eqs.~\eqref{geod-eqs1}-\eqref{geod-eqs}], as $(\delta E,\delta L, \delta Q)$ corrections to $(\delta T,\delta R,\delta\Theta,\delta\Phi)$ will be of ${\cal{O}}(\epsilon^{2})$. 

The shifts in the energy, angular momentum and Carter constant can be obtained by requiring that the turning points of the perturbed spacetime be the same as those of the GR Kerr spacetime. Expanding to $O(\epsilon)$, the radial turning points lead to the conditions
\ba
0 &= \left.\frac{\partial R_{\Kerr}}{\partial E_{\Kerr}}\right| _{r_{\pm}} \delta E + \left.\frac{\partial R_{\Kerr}}{\partial L_{\Kerr}}\right| _{r_{\pm}} \delta L + \left.\frac{\partial R_{\Kerr}}{\partial Q_{\Kerr}}\right| _{r_{\pm}} \delta Q 
\nonumber \\
&+ \delta R\left(r_{\pm}; E_{\Kerr}, L_{\Kerr}, Q_{\Kerr}\right)\,,
\ea
while the location of the polar turning points provides the condition
\ba
0 &= \left.\frac{\partial \Theta_{\Kerr}}{\partial E_{\Kerr}}\right| _{\theta_{\rm tp}} \delta E + \left.\frac{\partial \Theta_{\Kerr}}{\partial L_{\Kerr}}\right| _{\theta_{\rm tp}} \delta L + \left.\frac{\partial \Theta_{\Kerr}}{\partial Q_{\Kerr}}\right| _{\theta_{\rm tp}} \delta Q 
\nonumber\\ 
&+ \delta\Theta(\theta_{\rm tp}; E_{\Kerr}, L_{\Kerr} ,Q_{\Kerr}) \,.
\ea
These equations can be solved for $(\delta E, \delta L, \delta Q)$ to obtain
\begin{widetext}
\ba
\label{dEdLdQeq1}
\delta E&=  \left[ \frac{R_Q^+ \delta \Theta(\theta_{\rm tp})}{\sin^2\theta_{\rm tp}} - \delta R\left(r_{+}\right)\right] \left( R_L^- + \cot^2\theta_{\rm tp}R_Q^-\right) - \left[ \frac{R_Q^-\delta\Theta(\theta_{\rm tp})}{\sin^2\theta_{\rm tp}}-\delta R\left(r_{-}\right)\right]\left( R_L^+ + \cot^2\theta_{\rm tp} R_Q^+\right) S^{-1}\,,
\\
\delta L&= \left[ \frac{R_Q^- \delta \Theta(\theta_{\rm tp})}{\sin^2\theta_{\rm tp}} - \delta R\left(r_{-}\right)\right] \left( R_E^+ -2a^2E \cos^2\theta_{\rm tp} R_Q^+\right) - \left[ \frac{R_Q^+\delta\Theta(\theta_{\rm tp})}{\sin^2\theta_{\rm tp}}-\delta R\left(r_{+}\right)\right]\left( R_E^- -2a^2E  \cos^2\theta_{\rm tp}  R_Q^-\right) 
S^{-1}\,,
\\
\delta Q&=\cot^2\theta_{\rm tp} 2L \delta L - 2a^2 E  \cos^2\theta_{\rm tp}  \delta E -\delta\Theta(\theta_{\rm tp} )\,,
\label{dEdLdQeq}
\ea
where
\ba
R_E^\pm &= 2\left\{\left[\frac{p^2}{(1\pm e)^2}+a^2\right]\frac{p^2}{(1\pm e)^2}+2a^2 \frac{p}{1\pm e} \right\} E - 4a\frac{p}{1\pm e} L  \\
R_L^\pm &= -4a\frac{p}{1\pm e} E - 2 L \frac{p}{1\pm e} \left(\frac{p}{1\pm e} - 2\right), 
\qquad 
R_Q^\pm = -\left[\frac{p^2}{(1\pm e)^2}- 2\frac{p}{1\pm e} + a^2 \right]\,, 
\\
S &\equiv\left(R_E^+ -2a^2E \cos^2\theta_{\rm tp} R^+_Q \right) \left(R_L^-+\cot^2\theta_{\rm tp}R_Q^- \right) - \left(R_E^- - 2a^2 E  \cos^2\theta_{\rm tp} R_Q^- \right) \left(R_L^++\cot^2\theta_{\rm tp}R^+_Q \right)\,, 
\ea
\end{widetext}
and all quantities are to be evaluated at $(E_{\Kerr},L_{\Kerr},Q_{\Kerr})$.

Taking the weak-field limit of the above equations, when only ${\cal B}_2$ is non-vanishing, we find
\ba
\delta E_{{\cal B}_2} &\approx \frac{1}{2p^3} (1-e^2)^2 \left( \gamma_{1,2} + 2 a \gamma_{3,1} \sin^{2}{\theta_{\rm tp}} \right)\,,
 \\
\delta L_{{\cal B}_2} &\approx \frac{1}{\sqrt{p}} \left(-a \sin^3\theta_{\rm tp} \gamma_{3,1} +\sin\theta_{\rm tp} (\gamma_{4,2}+\gamma_{1,2}/2)\right)\,,
\\
\delta Q_{{\cal B}_2} &\approx -2 \sqrt{p} \sin\theta_{\rm tp} \gamma_{3,1}\,,
\ea
while if only ${\cal B}_N$ is non-vanishing for $N>2$ we have
\ba
\delta E_{{\cal B}_N} &\approx \frac{1}{2} p^{-N} f_{N-2}(e) (1-e^2)^2 (\gamma_{1,N}+2\gamma_{4,N})\,,
\\
\delta L_{{\cal B}_N} &\approx \frac{1}{2} p^{3/2-N} f_{N}(e) \sin\theta_{\rm tp}(\gamma_{1,N}+2\gamma_{4,N})\,,
\\
\delta Q_{{\cal B}_N} &\approx p^{2-N} f_{N}(e) \cos^2\theta_{\rm tp}(\gamma_{1,N}+2\gamma_{4,N})\,,
\ea
where we have define the eccentricity function
\be
f_{N}(e) = \left[ \frac{(1+e)^{N} - (1-e)^{N}}{(1+e)^2 - (1-e)^2} \right]\,. \label{eccfnc}
\ee
If we take the CS limit, the leading order pieces are
\ba
\delta E_{\rm CS} &= \frac{5}{4} a \zeta p^{-11/2} \sin\theta_{\rm tp} \left(1 - e^{2}\right)^{2} \left(1 + e^{2} \right)\,,
\\
\delta L_{\rm CS} &= \frac{5}{8} a \zeta p^{-4} \sin^2\theta_{\rm tp} \left(e^{2} + 3 \right) \left(3 e^{2} + 1 \right)\,,
\\
\delta Q_{\rm CS} &= \frac{5}{4} a \zeta p^{-7/2} \sin\theta_{\rm tp} \cos^2\theta_{\rm tp} \left(e^{2} + 3 \right) \left(3 e^{2} + 1 \right)\,,
\ea
and we see that $2 \bar{Q} \delta L_{\rm CS}  = \bar{L} \delta Q_{\rm CS} $.

\section{Fundamental Frequencies}
\label{sec:fund-freq}

In this section, we calculate the fundamental frequencies of orbital motion for the parametrically deformed spacetime. We first compute generic expressions for these frequencies, and then present results valid in the weak-field limit.

\subsection{General Results}

\subsubsection{Radial Frequencies}
The condition that the turning points of the radial motion are at $r_{\pm}$ allows us to write the change to the right- hand side of the radial geodesic equation as
\ba
&\frac{\partial R_{\Kerr}}{\partial E_{\Kerr}} \delta E + \frac{\partial R_{\Kerr}}{\partial L_{\Kerr}} \delta L + \frac{\partial R_{\Kerr}}{\partial Q_{\Kerr}} \delta Q + M^{4} 
\sum_{n=-2}^{\infty} \delta R_n \left(\frac{M}{r}\right)^{n} \nonumber \\
&= \left(\frac{M p}{1-e} - r\right) \left(r-\frac{M p}{1+e}\right) M^{2} \sum_{n=-2}^\infty v_n \left(\frac{M}{r}\right)^n\,, 
\label{radfact}
\ea
where we have here factored out the two turning points $r_{\pm}$. Equating coefficients of $r^n$ allows us to derive a recursion relation for the constants $v_n$:
\ba
v_{-2}&=-2 E_\Kerr \, \delta E\,,
\\
- v_{-1} + \frac{2p}{(1-e^2)} v_{-2} &=0\,, 
\\
- v_0 + \frac{2p}{(1-e^2)} v_{-1} - \frac{p^2}{(1-e^2)} v_{-2} &=  
\nonumber \\
2a^2E_\Kerr \delta E - 2L_\Kerr\delta L-\delta Q&+\delta R_{-2}\,,
 \\
-v_{1} + \frac{2p}{(1-e^2)} v_0 - \frac{p^2}{(1-e^2)} v_{-1} &= 
\nonumber \\
4a(aE_\Kerr-L_\Kerr)\delta E +  4(L_\Kerr-aE_\Kerr) \delta L \nonumber \\
+2\delta Q &+\delta R_{-1}\,,
\\
-v_{2} + \frac{2p}{(1-e^2)} v_{1} - \frac{p^2}{(1-e^2)} v_0 &= -a^2 \delta Q  +\delta R_{0}\,, 
\\
-v_{n+2} + \frac{2p}{(1-e^2)} v_{n+1} - \frac{p^2}{(1-e^2)} v_n &=  \delta R_{n} \,\, \forall n \geq 1\,.
\ea

These equations can be solved to derive successive coefficients. In practice, we will be interested in perturbations for which the series terminates at a finite $N$. If $\delta R_{n \geq N} = 0$, then $v_{n\geq N} = 0$ and the above equations provide $N+2$ equations for $N$ unknowns. The two extra equations are actually redundant, as they give the $\delta E$ and $\delta L$ needed to keep the turning points fixed at $r_{\pm}$. These $(\delta E,\delta L)$ were already computed in the previous section.

The solution to these equations can be best studied by separating out a few special cases. Let us consider the situation where the series contains only one term, $\delta R_N$. Then, the above equations indicate that $v_N \sim p^{-2}$ and $v_{n \leq N} \sim p^{-2+(n-N)}$. Hence, all terms proportional to $v_n/r^n$ have the same leading-order behavior in $p$. Now, let us consider the following special cases:
\begin{itemize}
\item If $N \leq 2$, then the five coefficients $(v_{-2}, \ldots, v_2)$ are non-zero, but $v_{n>2} = 0$. 
\item If $N < 1$, the coefficients are given by
\ba
\label{vm2to21}
v_{-2}&=-2E_\Kerr\delta E\,,  \\
v_{-1}&= -\frac{4pE_\Kerr\delta E}{1-e^2}\,, \\
v_0 &=-2\left[\frac{p^2 (3+e^2)}{(1-e^2)^2} + a^2\right] E_\Kerr\delta E + 2L\delta L+ \delta Q - \delta R_{-2}\,,  \\
v_1 &= -\left[\frac{8 p^3 (1+e^2)}{(1-e^2)^3} +\frac{4p a^2}{1-e^2}\right] E_\Kerr \delta E  \nonumber \\
&-4a(aE_\Kerr-L_\Kerr)\delta E+4aE_\Kerr\delta L \nonumber \\
&+  2(2L_\Kerr \delta L +\delta Q) \left(\frac{p}{1-e^2} - 1\right)  \nonumber \\
&-  \left( \frac{2p}{1-e^2} \delta R_{-2} + \delta R_{-1} \right)\,,   \\
v_2 &= -2\left[\frac{(5+10e^2+e^4) p^4}{(1-e^2)^4} +\frac{(3+e^2)p^2 a^2}{(1-e^2)^2} \right]E_\Kerr\delta E  \nonumber \\
&+ \frac{2p}{1-e^2} \left[4aE_\Kerr\delta L-4a(aE_\Kerr-L_\Kerr)\delta E\right] \nonumber \\
&+  \frac{p[(3+e^2)p-4(1-e^2)]}{(1-e^2)^2} (2L_\Kerr\delta L + \delta Q) + a^2 \delta Q  \nonumber \\
&- \left[ \frac{(3+e^2) p^2}{(1-e^2)^2} \delta R_{-2} + \frac{2p}{1-e^2} \delta R_{-1} +\delta R_0\right]\,, 
\label{vm2to2}
\ea
where we have assumed all but one of the $\delta R_i$'s are zero. 
\item If $N \geq 1$, the previous five equations still hold, with the $\delta R_i$'s set to zero. 
\item If $N=1,2$, we have the additional equations
\ba
\frac{2p}{(1-e^2)} v_2 - \frac{p^2}{(1-e^2)} v_1 = \delta R_1\,, 
 \\
 - \frac{p^2}{(1-e^2)} v_2 = \delta R_2\,,
\ea
but these will be automatically satisfied for the $\delta E$, $\delta L$ and $\delta Q$ derived earlier. 
\item If $N > 2$, we have $v_n = 0$ for $n > N$, $v_{-2}, \cdots, v_2$ are given by Eqs.~\eqref{vm2to21}-\eqref{vm2to2} and
\ba
v_{N-k} &= -(1-e^2) \frac{\delta R_N}{2e p^{2+k}} \left[(1+e)^{k+1} - (1-e)^{k+1}\right] \nonumber \\
& \hspace{1.0in} \mbox{for }0 \leq k < N-2\,.\nonumber
\ea
We note that $v_k$ is proportional to $p^{k-(2+N)}$. 
\end{itemize}

The set of equations for the $v_n$'s, $\delta E$ and $\delta L$ are linear, so to find the general solution we can proceed as follows. We denote by $v_n^{-2}$, $\delta E_{-2}$ and $\delta L_{-2}$ the solution to the above equations with only $\delta R_{-2}\neq 0$ and define
\be
\delta Q_{-2}=  \cot^2\theta_{\rm tp} 2L_\Kerr \delta L_{-2} - 2a^2 E_\Kerr  \cos^2\theta_{\rm tp}  \delta E_{-2} -\delta\Theta( \theta_{\rm tp} ) .
\ee
Then, for $N > -2$, we denote by $v_n^{N}$, $\delta E_{N}$ and $\delta L_{N}$ the solution to the above equations with only $\delta R_{N}\neq 0$ and with
\be
\delta Q_{N}= \cot^2\theta_{\rm tp}  2L_\Kerr \delta L_{N} - 2a^2 E_\Kerr  \cos^2\theta_{\rm tp}  \delta E_{N} .
\ee
The general solution is then given by $v_n = \sum_k v_n^k$, $\delta E = \sum_k \delta E_k$, $\delta L = \sum_k \delta L_k$ and $\delta Q = \sum_k \delta Q_k$.

The radial geodesic equation is most conveniently integrated by changing variables. First, we define a new dimensionless time parameter, $\lambda$:
\be
\frac{d}{d\lambda} \equiv \frac{\rho^{2}}{M} \frac{d}{d\tau}\,,
\ee
where $\tau$ is proper time (differentiation with respect to which was denoted by an overhead dot previously).  Next, we parameterize the orbit in terms of the (dimensionless) semi-latus rectum $p$, the eccentricity $e$ and a phase angle $\psi$ via 
\be
r=\frac{M p}{1+e\cos\psi}\,.
\ee
Equation~(\ref{radfact}) then becomes a differential equation for $\psi$, namely
${{\rm d}\psi}/{{\rm d}\lambda} = \sqrt{V_{\psi}}$, where $V_{\psi} = V_{K \psi} + \epsilon \delta V_{\psi}$ and 
\ba
V_{K\psi} (\psi,p,e,\iota) &= \frac{1-E_{\Kerr}^2}{(1-e^2)^3}
\left[p(1-e^2)-p_3(1+e\cos\psi)\right] 
\nonumber \\ 
&\times 
\left[p(1-e^2)-p_4(1+e\cos\psi)\right]\,,
\\
\delta V_{\psi}(\psi,p,e,\iota) &= \frac{1}{(1-e^2)} \sum_{n=-2}^{\infty} \frac{v_n}{p^n} (1+e\cos\psi)^{n+2}\,.
\ea
The Kerr potential depends on $Mp_3 = (1-e^2)r_3$ and $M p_4 = (1-e^2)r_4$, where $r_{3,4}$ are the other two turning points of the radial motion. These are given in terms of $E_{\Kerr}$, $Q_{\Kerr}$, $p$ and $e$ by the expressions
\be
p_3 = Y + \sqrt{Y^2 - X}, \qquad p_4 = Y - \sqrt{Y^2 - X} 
\ee
where
\be
Y \equiv \frac{1-e^2}{1-E_{\Kerr}^2} - p, \qquad X \equiv \frac{(1-e^2)^3}{1-E_{\Kerr}^2} \frac{a^2 Q_{\Kerr}}{p^{2}}
\ee
These relations imply that $p_3 \sim 1$ and $p_4 \sim a^2$ in the weak field, so $V_{K\,\psi} \approx (1-E_{\Kerr}^2)p^2/(1-e^2)$ when $p \gg 1$.

We can integrate the evolution equation for $\psi$ over a complete orbital cycle to determine $\Lambda_r$, i.e.,~the $\lambda$ time elapsed per radial cycle. Writing this in terms of the difference to the Kerr value, $\Lambda_r = \Lambda_{K \,r}+\epsilon\delta \Lambda_r$, we obtain the correction to the radial period
\ba
\delta \Lambda_r &= \int_{0}^{2\pi} \frac{d \psi}{\sqrt{V_{\psi}}} -\int_{0}^{2\pi} \frac{d \psi}{\sqrt{V_{K \psi}}}  \nonumber \\
&= - \frac{(1-e^2)^{7/2}}{(1-E_\Kerr^2)^{3/2}} \sum_{n=-2}^\infty \frac{v_n}{p^n}\int_0^\pi 
(1+e\cos\psi)^{n+2}
\nonumber \\
&\times
\left[p(1-e^2)-p_3(1+e\cos\psi)\right]^{-3/2}
\nonumber \\
&\times
\left[p(1-e^2)-p_4(1+e\cos\psi)\right]^{-3/2} {\rm d}\psi \,,
\ea
which is valid for arbitrary radius at leading order in $\epsilon$.

\subsubsection{Polar Frequencies}
If we write $z=\cos^2\theta$, the condition that the turning points of the polar motion are at $z= \cos^2\theta_{\rm tp}$ allows us to write the change to the right- hand side of the polar geodesic equation as
\ba
& \frac{M^2}{(1-z)} \left[\delta Q - z(\delta Q + 2L_\Kerr\delta L) + 2a^2 E_\Kerr \delta E z(1-z)\right] \nonumber 
 \\ \nonumber
& \hspace{1cm}+ 2 M^2\gamma_{3,1}\left[E_\Kerr L_\Kerr-a E_\Kerr^2 (1-z) \right] \nonumber \\
& =M^2 \frac{ \cos^2\theta_{\rm tp}  -z}{1-z} \left( t_0 + t_1 z\right)\,,
\ea
and by matching powers of $z$ we find
\ba
t_0&= \frac{1}{ \cos^2\theta_{\rm tp}} \left[\delta Q + 2\gamma_{3,1} (E_\Kerr L_\Kerr-a E_\Kerr^2)\right]\,, 
\\
t_1&= 2a(a E_\Kerr\delta E+\gamma_{3,1} E_\Kerr^2)\,.
\ea
The coefficient of $z$ then gives the relationship between $\delta Q$, $\delta E$ and $\delta L$ we found in Eqs.~\eqref{dEdLdQeq1}-\eqref{dEdLdQeq}.

Let us now rewrite the polar geodesic equation in a simpler way, by parameterizing the polar motion using
\be
\cos^2\theta =  \cos^2\theta_{\rm tp} \cos^2\chi\,.
\ee
The polar equation then becomes
\ba
\left( \frac{{\rm d}\chi}{{\rm d}\lambda} \right)^2 &= V_{K\chi}(\chi,p,e,\iota) +\epsilon (t_0+t_1  \cos^2\theta_{\rm tp} \cos^2\chi)\,,   \\
V_{K\chi}(\chi,p,e,\iota) &= a^2(1-E_\Kerr^2) (z_+- \cos^2\theta_{\rm tp}  \cos^2\chi)\,,  \\
\mbox{where} \qquad & z_+ =\frac{Q_\Kerr}{a^2(1-E_\Kerr^2)  \cos^2\theta_{\rm tp} }\,.
\ea
As in the radial case, we can compute the perturbation to the $\theta$ period in $\lambda$ by writing $\Lambda_\theta=\Lambda_{K\,\theta}+ \epsilon\delta \Lambda_\theta$ and find
\be
\delta \Lambda_\theta = -\frac{1}{a^3(1-E_\Kerr^2)^{3/2}} \int_0^\pi \frac{t_0+t_1 \cos^2\theta_{\rm tp} \cos^2\chi}{\left(z_+- \cos^2\theta_{\rm tp} \cos^2\chi\right)^{3/2}} {\rm d}\chi\,.
\ee

\subsubsection{Azimuthal Frequencies}
We can write the azimuthal geodesic equation as
\be
\frac{{\rm d}\phi}{{\rm d}\lambda} = \frac{\rho^2}{M} \frac{{\rm d}\phi}{{\rm d}\tau} = \Phi_r(r) + \Phi_\theta(\theta),
\ee
which allows us to define a radial and a polar contribution to the average advance of $\phi$ over a radial /polar period
\ba
\Delta\phi_{\Lambda,r} &= \int_0^{2\pi} \frac{\Phi_r[p/(1+e\cos\psi)]}{{\rm d}\psi/{\rm d}\lambda} {\rm d} \psi\,, 
 \\
\Delta\phi_{\Lambda,\theta} &=\int_0^{2\pi} \frac{\Phi_\theta[\cos^{-1}( \cos\theta_{\rm tp} \cos\chi)]}{{\rm d}\chi/{\rm d}\lambda} {\rm d} \chi\,.
\ea
The average rate of advance of $\phi$, $\omega_{\phi \, \Lambda}$, can then be computed as
\ba
\omega_{\phi \, \Lambda} = \left\langle\frac{{\rm d}\phi}{{\rm d}\lambda}\right\rangle_\lambda = \frac{ \Delta\phi_{\Lambda,r}}{\Lambda_r} + \frac{\Delta\phi_{\Lambda,\theta}}{\Lambda_\theta}.
\ea
We can write the individual contributions in the form $\Delta \phi_{\Lambda,r} = \Delta\phi_{K\,\Lambda,r} + \epsilon \delta \Delta\phi_{\Lambda,r}$ and $\Delta\phi_{\Lambda, \theta} = \Delta\phi_{K\,\Lambda,\theta} + \epsilon \delta \Delta\phi_{\Lambda,\theta}$ in which we take the Kerr pieces to be
\ba
 \Delta\phi_{K\,\Lambda,r} &= aM \left( \frac{P}{\Delta} - E_{\Kerr} \right) = aM \left( \frac{2 r E_{\Kerr} - a L_{\Kerr}}{\Delta} \right)\,,
 \\
  \Delta\phi_{K\,\Lambda,\theta} &= \frac{M L_{\Kerr}}{\sin^2 \theta}\,.
\ea
Each of the $\epsilon$ corrections has two contributions --- one from the change in the numerator, $\Phi_r$ etc., and one from the change in the denominator, ${\rm d}\psi/{\rm d}\lambda$. We obtain
\begin{widetext}
\ba
\label{deltaphilambdar}
\delta\Delta\phi_{\Lambda,r} &= a\frac{(1-e^2)^{3/2}}{(1-E_\Kerr^2)^{1/2}} \int_0^{2\pi}  \frac{2 p (1+e\cos\psi)\delta E-a(1+e\cos\psi)^2\delta L}{D_{r}(\psi) [p^2-2p(1+e\cos\psi)+a^2(1+e\cos\psi)^2]} {\rm d}\psi
\nonumber \\
&+\frac{(1-e^2)^{3/2}}{(1-E_\Kerr^2)^{1/2}} \sum_{n=2}^\infty \frac{\delta\Phi_n}{p^n} \int_0^{2\pi} \frac{(1+e\cos\psi)^n}{D_r(\psi)} {\rm d}\psi 
\nonumber \\
&- a \frac{(1-e^2)^{7/2}}{(1-E_\Kerr^2)^{3/2}} \sum_{n=-2}^\infty  \frac{v_n}{p^n}\int_0^\pi \frac{(1+e\cos\psi)^{n+2}\left\{ 2 p(1+e\cos\psi) E_\Kerr-a(1+e\cos\psi)^2 L_\Kerr\right\}}{D_{r}(\psi)^3 \left[p^2-2p(1+e\cos\psi)+a^2(1+e\cos\psi)^2\right]}{\rm d}\psi\,, 
\\
D_{r}(\psi)&=\left\{\left[p(1-e^2)-p_3(1+e\cos\psi)\right]\left[p(1-e^2)-p_4(1+e\cos\psi)\right]\right\}^{1/2}\,,
\ea
and
\ba
\delta\Delta\phi_{\Lambda,\theta} &= \frac{1}{a\sqrt{1-E_\Kerr^2}}\int_0^{2\pi}  \frac{\delta L}{(1- \cos^2\theta_{\rm tp} \cos^2\chi)\sqrt{z_+ -  \cos^2\theta_{\rm tp} \cos^2\chi}}{\rm d} \chi 
\nonumber\\
&- \frac{1}{a^3(1-E_\Kerr^2)^{3/2}}  \int_0^\pi \frac{(t_0+t_1 \cos^2\theta_{\rm tp} \cos^2\chi)L_\Kerr}{(1- \cos^2\theta_{\rm tp} \cos^2\chi) \left(z_+- \cos^2\theta_{\rm tp} \cos^2\chi\right)^{3/2}} {\rm d}\chi\,. 
\label{phifreqpert}
\ea
\end{widetext}
In both the radial and polar corrections, the first term is the contribution from the change in the orbital constants $E$ and $L$ in the Kerr part of the numerator. The second term is the contribution from the change in the numerator due to the addition of the perturbation to the numerator. The third term is the contribution from the change in the denominator. There is no second term in the $\theta$ contribution since the perturbation to $\dot{\Phi}$ is purely radial.

\subsubsection{Temporal Frequencies}
The motion in $t$ can be treated in the same way as the motion in $\phi$. 
With the analogous definitions, the average rate of advance of time with respect to $\lambda$, $\omega_{T \, \Lambda}$, is
\be
\omega_{T\, \Lambda} = \left\langle\frac{{\rm d} t}{{\rm d}\lambda}\right\rangle = \frac{ \Delta T_{\Lambda,r}}{\Lambda_r} + \frac{\Delta T_{\Lambda,\theta}}{\Lambda_\theta},
\ee
where $\Delta T_{\Lambda,r} = \Delta T_{K\,\Lambda,r} + \epsilon \delta \Delta T_{\Lambda,r}$ and $\Delta T_{\Lambda, \theta} = \Delta T_{K\,\Lambda,\theta} + \epsilon \delta \Delta T_{\Lambda,\theta}$. The Kerr pieces of this rate of change are
\ba
\Delta T_{K \Lambda,r} &= (r^2+ M^2a^2) \frac{P}{\Delta}\,,
\nonumber \\
\Delta T_{K \Lambda,\theta} &= -aM (aM E_{\Kerr} \sin^2\theta - ML_{\Kerr})\,, 
\ea
while the ${\cal{O}}(\epsilon)$ corrections are
\begin{widetext}
\ba
\frac{\delta\Delta T_{\Lambda,r}}{M} &= \frac{(1-e^2)^{3/2}}{(1-E_\Kerr^2)^{1/2}} \int_0^{2\pi} \frac{\left[p^2+a^2(1+e\cos\psi)^2\right]\left\{ \left[ p^2+a^2(1+e\cos\psi)^2 \right]\delta E-a(1+e\cos\psi)^2\delta L\right\}}{(1+e\cos\psi)^2 D_{r}(\psi) \left[p^2-2p(1+e\cos\psi)+a^2(1+e\cos\psi)^2 \right]} {\rm d}\psi 
\nonumber \\
&+ \frac{(1-e^2)^{3/2}}{(1-E_\Kerr^2)^{1/2}} \sum_{n=0}^\infty \frac{\delta T^r_n}{p^n} \int_0^{2\pi}  \frac{(1+e\cos\psi)^n}{D_r(\psi)}  {\rm d}\psi 
\nonumber \\
&- \frac{(1-e^2)^{7/2}}{(1-E_\Kerr^2)^{3/2}} \sum_{n=-2}^\infty   \frac{v_n}{p^n}\int_0^\pi \frac{(1+e\cos\psi)^{n} \left[p^2+a^2(1+e\cos\psi)^2\right]}{D_{r}(\psi)^3 \left[p^2-2p(1+e\cos\psi)+a^2(1+e\cos\psi)^2\right]} 
\nonumber\\
&\times \left\{ E_\Kerr \left[p^2+a^2(1+e\cos\psi)^2 \right] - aL_\Kerr(1+e\cos\psi)^2 \right\} {\rm d}\psi\,,
\ea
and
\ba
\frac{\delta \Delta T_{\Lambda,\theta}}{M}&= \frac{1}{a\sqrt{1-E_\Kerr^2}}\int_0^{2\pi}  \frac{a\delta L - a^2 (1- \cos^2\theta_{\rm tp} \cos^2\chi)\delta E}{\sqrt{z_+ -  \cos^2\theta_{\rm tp} \cos^2\chi}} {\rm d} \chi
-\frac{2\gamma_{3,1} E_\Kerr}{\sqrt{1-E_\Kerr^2}}\int_0^{2\pi}  \frac{(1- \cos^2\theta_{\rm tp} \cos^2\chi)}{\sqrt{z_+ -  \cos^2\theta_{\rm tp} \cos^2\chi}} {\rm d} \chi
 \nonumber\\
&- \frac{1}{a^3(1-E_\Kerr^2)^{3/2}}  \int_0^\pi \frac{(t_0+t_1 \cos^2\theta_{\rm tp} \cos^2\chi)\left[aL_\Kerr - a^2E_\Kerr(1- \cos^2\theta_{\rm tp} \cos^2\chi)\right]}{\left(z_+- \cos^2\theta_{\rm tp} \cos^2\chi\right)^{3/2}} {\rm d}\chi\,,
\ea
\end{widetext}
in which $D_r(\psi)$ was defined in Eq.~(\ref{phifreqpert}). The origin of each term is as described in the azimuthal case. We note that $\delta T_n^r$ denotes the radial part of the perturbation to the temporal potential. This is equal to $\delta T_n$, except $\delta T_0^r = (\gamma_{1,2}+2\gamma_{4,2})M^2$.

\subsubsection{Combining Results}
The orbital frequencies quoted so far were written in terms of the $\lambda$ parameter. We compute the perturbations to the physical frequencies, i.e., those expressed in Boyer-Lindquist time, by 
first defining
\ba
\Delta T_\Lambda &= \Lambda_\theta \Delta T_{\Lambda, r} + \Lambda_r \Delta T_{\Lambda, \theta}\,,  \\
\Delta \phi_\Lambda&=\Lambda_\theta \Delta \phi_{\Lambda, r} + \Lambda_r \Delta \phi_{\Lambda, \theta}\,,
\ea
and then writing 
\be
\Omega_\phi = \frac{\left\langle\frac{{\rm d}\phi}{{\rm d}\lambda}\right\rangle}{\left\langle\frac{{\rm d}t}{{\rm d}\lambda}\right\rangle} =\frac{\Delta\phi_\Lambda}{\Delta T_\Lambda}\,,
\quad
\Omega_r = \frac{2\pi\Lambda_\theta}{\Delta T_{\Lambda}}\,,
\quad
\Omega_\theta = \frac{2\pi\Lambda_r}{\Delta T_{\Lambda}}\,.
\ee
The factor of $2\pi$ is included in the definitions of $\Omega_r$ and $\Omega_\theta$ so that all three frequencies are angular frequencies, expressed as radians per second. From the preceding equations, we may find the perturbations to the physical frequencies to be
\ba
& \frac{\delta \Omega_\phi}{\Omega_\phi} = \frac{\delta\Delta\phi_\Lambda}{\Delta\phi_\Lambda} - \frac{\delta \Delta T_\Lambda}{\Delta T_\Lambda}\,,
\quad
\frac{\delta \Omega_r}{\Omega_r} = \frac{\delta\Lambda_\theta}{\Lambda_\theta} - \frac{\delta \Delta T_\Lambda}{\Delta T_\Lambda}\,, \\
& \qquad \qquad \qquad 
\frac{\delta \Omega_\theta}{\Omega_\theta} =\frac{\delta\Lambda_r}{\Lambda_r} - \frac{\delta \Delta T_\Lambda}{\Delta T_\Lambda}\,.
\ea

The frequencies that we need for the construction of AK waveforms, using the notation of Barack and Cutler~\cite{AK}, are the orbital frequency, $2 \pi \nu$, the perihelion precession frequency, $\dot{\tilde{\gamma}}$, and the orbital plane precession frequency, $\dot{\alpha}$ (see Section~\ref{sec:MG-AK}). These are given by
\be
2\pi\nu = \Omega_r\,,
\quad
\dot{\tilde{\gamma}} = \Omega_\theta - \Omega_r\,, 
\quad
\dot{\alpha} =  \Omega_\phi - \Omega_\theta\,,
\ee
and therefore the changes in the frequencies can be computed from
\ba
\label{precpert1}
\frac{\delta \nu}{\nu} &= \frac{\delta\Lambda_\theta}{\Lambda_\theta} - \frac{\delta \Delta T_\Lambda}{\Delta T_\Lambda}\,,  \\
\frac{\delta\dot{\tilde{\gamma}}}{\dot{\tilde{\gamma}}} &= \frac{\delta(\Lambda_r - \Lambda_\theta)}{\Lambda_r - \Lambda_\theta} - \frac{\delta \Delta T_\Lambda}{\Delta T_\Lambda}\,, \\
\frac{\dot{\alpha}}{\dot{\alpha}} &= \frac{\delta(\Delta\phi_\Lambda-2\pi\Lambda_r)}{\Delta\phi_\Lambda-2\pi\Lambda_r} -  \frac{\delta \Delta T_\Lambda}{\Delta T_\Lambda}\,.
\label{precpert}
\ea
\subsection{Weak-Field Expansion}
\subsubsection{Radial Frequencies}
Denoting by $v_{{\cal B}_{N,i}}$ the leading order in $p$ piece of the parameter $v_i$ under assumptions ${\cal B}_N$, we find for ${\cal B}_2$
\ba
v_{{\cal B}_{2,-2}} &= -\frac{(1-e^2)^2}{p^3} \left[\gamma_{1,2}+2a \sin^2\theta_{\rm tp}  \gamma_{3,1}\right]\,,
\\
v_{{\cal B}_{2,-1}} &= -2\frac{(1-e^2)}{p^2} \left[\gamma_{1,2}+2a \sin^2\theta_{\rm tp}  \gamma_{3,1}\right]\,,
\\
v_{{\cal B}_{2,0}} &= \frac{(1-e^2)}{p} \left[\gamma_{1,2}-2a \sin^2\theta_{\rm tp}  \gamma_{3,1}\right]\,, 
\ea
and the other $v_{{\cal B}_{2,i}}$'s are subdominant in the sense that $v_{{\cal B}_{2,i}}/p^i$ is higher order in $1/p$. We see that for these dominant terms $v_{{\cal B}_{2,i}}/p^i \sim O(1/p)$, which motivates us to define $p$-independent terms $\tilde{v}_{{\cal B}_{2,i}}= v_{{\cal B}_{2,i}}/p^{i-1}$. For ${\cal B}_N$ with $N > 2$, we find $v_{{\cal B}_{N,i}}/p^i \sim O(1/p^{N-2})$ so we can define the $p$-independent quantities $\tilde{v}_{{\cal B}_{N,i}}= v_{{\cal B}_{N,i}}/p^{i+2-N}$. We find $v_{{\cal B}_{N,i}} \sim 0$ for $i > N-4$,
\be
\tilde{v}_{{\cal B}_{N,-2}} = -(1-e^2)^2 f_{N-2}(e) (\gamma_{1,N}+2\gamma_{4,N})\,, 
\ee
and for $-1 \leq i \leq N-4$
\be
\tilde{v}_{{\cal B}_{N,i}} = -2(1-e^2) f_{N-i-3}(e) (\gamma_{1,N}+2\gamma_{4,N})\,,
\ee
in which $f_N(e)$ is as defined in Eq.~(\ref{eccfnc}).

The corresponding expressions in the CS limit are
\ba
\tilde{v}_{{\rm CS},-2} &= -\frac{5}{2}(1-e^2)^2 (1+e^2) \zeta a \sin\theta_{\rm tp}\,, \\
\tilde{v}_{{\rm CS},-1} &=-5 (1-e^2) (1+e^2) \zeta a \sin\theta_{\rm tp}\,,  \\
\tilde{v}_{{\rm CS},0} &= -\frac{5}{4}(1-e^2) (3+e^2) \zeta a \sin\theta_{\rm tp}\,, \\
\tilde{v}_{{\rm CS},1} &= -\frac{5}{2}(1-e^2) \zeta a \sin\theta_{\rm tp}\,, \\
\tilde{v}_{{\rm CS},2} &=-\frac{5}{4}(1-e^2) \zeta a \sin\theta_{\rm tp}\,,
\ea
where in this case $\tilde{v}_{{\rm CS}_{N,i}}= v_{{\rm CS}_{N,i}}/p^{i-7/2}$.

With this identification of the leading-order parts of the $v_i$'s, the integral for the change to the radial frequency reduces to
\ba
\delta \Lambda_r &= - p^{-K} (1-e^2)^{-1}\sum_{n=-2}^\infty \tilde{v}_n \, I_{n+2}\,,  \\
\mbox{in which } \qquad & I_n =  \int_0^\pi (1+e\cos\psi)^{n}{\rm d}\psi\,, 
\ea
the $\tilde{v}_n$'s are as defined above and $K$ represents the scaling of the frequency correction with $p$, which is $K=5/2$ for ${\cal B}_2$, $K=N-1/2$ for ${\cal B}_N$ with $N \geq 3$ and $K=5$ in the CS limit. We note that although we have written $\infty$ as the upper limit of the sum for convenience, in general we will be able to terminate the summation as described above, e.g., at $n=0$ for ${\cal B}_2$, at $n=N-4$ for ${\cal B}_N$ with $N\geq 3$ and at $n=2$ for CS. The integral $I_n$ is evaluated in Appendix~\ref{intapp} and shown to be equal to
\be
I_{n} = \pi\sum_{k=0}^{\lfloor \frac{n}{2} \rfloor} \left(\frac{e}{2}\right)^{2k} \frac{n!}{(n-2k)!\,(k!)^2}\,,
\ee
in which  $\lfloor x\rfloor$ denotes the largest integer smaller than $x$. For the various special cases we have been considering elsewhere, the leading-order shifts in the radial frequency can be simplified to
\ba
\delta \Lambda_{r}^{{\cal{B}}_{2}} &= \frac{\pi}{2 p^{5/2}}
\left[\left(4-3 e^{2}\right) \gamma_{1,2} 
\right. 
\nonumber \\
&- \left.
2 a \left(8-e^{2} \right) \sin^{2}{\theta_{\rm tp}} \gamma_{3,1} \right]\,,
\\
\delta \Lambda_{r}^{{\cal{B}}_{3}} &= \frac{\pi (3-e^{2})}{2 p^{5/2}}
\left(\gamma_{1,3} + 2 \gamma_{4,3} \right)\,,
 \\
\delta \Lambda_{r}^{{\cal{B}}_{4}} &= \frac{\pi (8-e^{2})}{2 p^{7/2}} 
\left(\gamma_{1,4} + 2 \gamma_{4,4} \right)\,,
 \\
\delta \Lambda_{r}^{{\cal{B}}_{5}} &= \frac{\pi (15 + 5 e^{2} - e^{4})}{2 p^{9/2}}
\left(\gamma_{1,5} + 2 \gamma_{4,5} \right)\,,
 \\
\delta \Lambda_{r}^{\CS} &= \frac{5 \pi}{32 p^{5}} a \zeta \sin{\theta_{\rm tp}} 
\left(12 - e^{2}\right)\left(8 + 9 e^{2}\right)\,.
\ea
%

\subsubsection{Polar Frequencies}

For $p \gg 1$, we find $z_+ \sim (Q+L^2)/a^2(1-E^2)$ and $Q+L^2 \approx p$, so $V_{K\chi}(\chi,p,e,\iota) \approx p$. The weak-field limit is therefore
\be
\delta \Lambda_\theta \approx -\frac{\pi}{p^{3/2}} \left(t_0+ \frac{t_1 \cos^2\theta_{\rm tp} }{2}\right)\,,
\ee
which for assumptions ${\cal B}_2$ gives
\ba
\delta \Lambda_{\theta\,{\cal B}_2} &\approx -\frac{\pi}{p^{3/2}} \left\{\gamma_{1,2}+ 2\gamma_{4,2}
+ \left[3a\cos^2\theta_{\rm tp} - 4a\right] \gamma_{3,1} \right\}\,,
\ea
for assumptions ${\cal B}_N$ with $N \geq 3$ gives
\be
\delta \Lambda_{\theta\,{\cal B}_N} \approx -\frac{\pi}{p^{N-1/2}} f_N(e) \left(\gamma_{1,N}+2\gamma_{4,N}\right)\,,
\ee
and in the CS limit gives
\be
\delta \Lambda_{\theta\,{\rm CS}} \approx -\frac{5\pi}{4p^{5}} a \zeta \sin\theta_{\rm tp} (3+e^2)(1+3e^2) .
\ee

\subsubsection{Azimuthal Frequencies}
In the weak-field, the leading order part of $\delta \Delta\phi_{\Lambda, r}$ comes from the terms
\ba
\delta \Delta\phi_{\Lambda, r} &= \frac{4\pi a \delta E}{p^{3/2}} +2\pi\sqrt{p} \sum_{l=N_\phi}^{\infty} \frac{\delta \Phi_l}{p^{l+1}} \sum_{k=0}^{\lfloor \frac{l}{2}\rfloor} \left(\frac{e}{2}\right)^{2k} 
\nonumber \\ 
&\times 
\frac{l!}{(l-2k)!\,(k!)^2} -\frac{2a}{(1-e^2) p^{K+1}} \sum_{n=-2}^\infty \tilde{v}_n I_{n+3}\,,
\ea
in which we are using $N_\phi$ to denote the first non-zero $\delta\Phi_l$, $I_n$ is as defined above and $K$ has the same meaning as in the radial case: $K=5/2$ for ${\cal B}_2$, $K=N-1/2$ for ${\cal B}_N$ with $N \geq 3$ and $K=5$ in the CS limit. Under assumptions ${\cal B}_2$, the second term can be seen to dominate, $N_\phi = 2$ and $\delta \Phi_2 \sim -2a \sqrt{p} \sin\theta_{\rm tp} \gamma_{3,1}$. We therefore find
\be
\delta\Delta\phi_{\Lambda,r}^{{\cal{B}}_{2}} = - \frac{2 \pi}{p^{2}} a \gamma_{3,1} \left(e^{2}+2\right) \sin{\theta_{\rm tp}}.
\ee
Under assumptions ${\cal B}_N$ with $N \geq 3$, the second and third terms both contribute and we find
\ba
\delta\Delta\phi_{\Lambda,r}^{{\cal{B}}_{N>2}} &= \frac{1}{p^{N+1/2}} \Big\{2[\gamma_{3,N+1}+ a (\gamma_{1,N}+\gamma_{4,N})] I_N  \nonumber \\
&+ 2a\left[\pi(1-e^2) f_{N-2}(e) + 2\sum_{k=2}^{N-1} f_{N-k}(e) I_{k}\right] \nonumber \\
&\times \left(\gamma_{1,N}+2 \gamma_{4,N}\right)\Big\}\,,
\ea
which for the first few $N$'s gives
\ba
\delta\Delta\phi_{\Lambda,r}^{{\cal{B}}_{3}} &= \frac{1}{p^{7/2}} \left( (2+3e^2)\gamma_{3,4} + a(5+3e^2)\gamma_{1,3}\right. \nonumber\\
&+ \left. a (8+3e^2)\gamma_{4,3}\right) \,,
\\
\delta\Delta\phi_{\Lambda,r}^{{\cal{B}}_{4}} &=\frac{1}{4p^{9/2}} \left( (8+24e^2+3e^4)\gamma_{3,5} \right. \nonumber\\
&+ \left. a(40+36e^2+3e^4)\gamma_{1,4}\right. \nonumber\\
&+ \left. 3a (24+16e^2+e^4)\gamma_{4,4}\right)\,,
 \\
\delta\Delta\phi_{\Lambda,r}^{{\cal{B}}_{5}} &=\frac{1}{4p^{11/2}} \left( (8+40e^2+15e^4)\gamma_{3,6} \right. \nonumber\\
&+ \left. 2a(34+50e^2+9e^4)\gamma_{1,5}\right. \nonumber\\
&+ \left. a (128+160e^2+21e^4)\gamma_{4,5}\right).
\ea
In the CS limit we find that the second term again dominates, which gives
\be
\delta\Delta\phi_{\Lambda,r}^{\CS} = - \frac{5\pi}{4p^{9/2}} a \zeta \left(1 + 3 e^{2}+\frac{3}{8} e^{4}\right)\,.
\ee
The dominant contribution to $\delta \Delta\phi_{\Lambda, \theta}$ is given by
\ba
\delta \Delta\phi_{\Lambda, \theta} &= \frac{\pi}{p^{3/2}} \left[ \frac{2 Q_{\Kerr} \delta L -  \cos^2\theta_{\rm tp} t_0 L_\Kerr}{ \cos^2\theta_{\rm tp} \sin\theta_{\rm tp}} \right. 
\nonumber  \\ 
&+ \left. \left(\frac{2 a^2 (1-e^2)}{p} \delta L +t_1 L_\Kerr \right) 
\left( 1-  \frac{1}{\sin\theta_{\rm tp}}\right) \right]\,,  
\ea
With the ${\cal B}_2$ assumptions, we find
\ba
\delta\Delta\phi_{\Lambda,\theta}^{{\cal{B}}_{2}} = \frac{2 \pi a}{p} \sin\theta_{\rm tp}\gamma_{3,1}\,,
\ea
For assumptions ${\cal B}_N$ with $N\geq 3$ and in the CS limit, we have $z_- t_0 = \delta Q$ and using Eqs.~\eqref{dEdLdQeq1}-\eqref{dEdLdQeq}, we see that 
\ba
\frac{2 Q_{\Kerr}\delta L - L_{\Kerr} \delta Q}{ \cos^2\theta_{\rm tp}} &= 2 \left( \frac{Q_{\Kerr}}{ \cos^2\theta_{\rm tp}} - \frac{L_{\Kerr}^2}{ \sin^2\theta_{\rm tp}}\right) \delta L 
\nonumber \\ 
& + 2 a^2 E_\Kerr L_\Kerr \delta E\,, \nonumber \\
&= \frac{2 a^2}{p}(1-e^2)\delta L + 2a^2 E_\Kerr L_\Kerr \delta E\,.  
\ea
Both terms therefore contribute at the same order giving
\ba
\delta\Delta\phi_{\Lambda,\theta}^{{\cal{B}}_{N>2}} &= \frac{a^2 \pi (1-e^2)}{p^{N+1}} \left[f_{N}(e) +(1-e^2)f_{N-2}(e)\right] \nonumber \\
&\times \left(\gamma_{1,N} + 2 \gamma_{4,N} \right) \sin\theta_{\rm tp} \,,
\ea
which for the first few $N$'s gives
\ba
\delta\Delta\phi_{\Lambda,\theta}^{{\cal{B}}_{3}} &= \frac{2 a^2 \pi (1-e^2)}{p^{4}} \left(\gamma_{1,3} + 2 \gamma_{4,3} \right) \sin\theta_{\rm tp}\,, \\
\delta\Delta\phi_{\Lambda,\theta}^{{\cal{B}}_{4}} &= \frac{3 a^2 \pi (1-e^2)}{p^{5}} \left(1+\frac{e^2}{3}\right)\left(\gamma_{1,4} + 2 \gamma_{4,4} \right) \sin\theta_{\rm tp}\,,  \\
\delta\Delta\phi_{\Lambda,\theta}^{{\cal{B}}_{5}} &= \frac{4 a^2 \pi (1-e^2)}{p^{6}} (1+e^2) \left(\gamma_{1,5} + 2 \gamma_{4,5} \right) \sin\theta_{\rm tp}\,. 
\ea
In the CS case, we find, at leading order in $p$,
\be
\delta\Delta\phi_{\Lambda,\theta}^{\CS} = \frac{25 \pi}{4} \frac{a^3 \zeta}{p^{13/2}} \left(1-e^{2}\right) \left(1+2 e^{2}+\frac{e^4}{5} \right) 
\sin^2{\theta_{\rm tp}}\,.
\ee
If we linearize in $a$, for consistency with the order to which the CS limit was derived, this term can be seen to vanish 
$\delta\Delta\phi_{\Lambda,\theta}^{\CS} = 0$.

\subsubsection{Temporal Frequencies}
In the weak-field, the leading order part of $\delta \Delta T_{\Lambda, r}$ comes from the terms
\ba
\frac{\delta \Delta T_{\Lambda, r}}{M} &= \frac{2\pi p^{3/2} \delta E}{(1-e^2)^{3/2}} +2\sqrt{p} \sum_{l=N_T}^\infty \frac{\delta T_l}{p^{l+1}} \sum_{k=0}^{\lfloor \frac{l}{2}\rfloor} \left(\frac{e}{2}\right)^{2k} 
\nonumber \\
&\times
\frac{l!}{(l-2k)!\,(k!)^2} -\frac{1}{(1-e^2)p^{K-2}}\sum_{n=-2}^{N-4}  \tilde{v}_n I_{n}\,, 
\ea
in which $N_T$ denotes the first non-zero $\delta T_n$ and $I_n$, $K$ have the same meaning as in the radial case and as in the preceding section.

We can now evaluate the dominant contribution to $\delta \Delta T_{\Lambda, r}$ in the weak-field for the various cases we have been considering. For assumptions ${\cal B}_2$, the second and third terms dominate and we find
\ba
\delta \Delta T_{\Lambda,r}^{{\cal{B}}_{2}} &= \frac{\pi M}{p^{1/2}} 
\left[ \gamma_{1,2} + 4 \gamma_{4,2} - 2 a \sin^2\theta_{\rm tp} \gamma_{3,1}  \right. \nonumber \\
&+\left. \frac{3}{(1-e^2)^{1/2}} (\gamma_{1,2} + 2 a \sin^2\theta_{\rm tp} \gamma_{3,1}) \right]\,.
\ea
Under assumptions ${\cal B}_N$ for $N \geq 3$, the third term dominates. For the first few $N$'s we obtain
\ba
 \delta \Delta T_{\Lambda,r}^{{\cal{B}}_{3}} &= \frac{3 \pi M}{2 p^{1/2}  (1-e^2)^{1/2}}
\left(2 \gamma_{4,3} + \gamma_{1,3} \right)\,,
 \\
 \delta \Delta T_{\Lambda,r}^{{\cal{B}}_{4}} &= \frac{\pi M}{p^{3/2}}  \left[\frac{3}{(1-e^2)^{1/2}} + 1\right]\,  \left(2 \gamma_{4,4} + \gamma_{1,4} \right)\,, 
\\
 \delta \Delta T_{\Lambda,r}^{{\cal{B}}_{5}} &= \frac{3\pi M}{2 p^{5/2}} \left[\frac{3+e^2}{(1-e^2)^{1/2}} + 2\right]  \left(2 \gamma_{4,5} + \gamma_{1,5} \right)\,.
\ea
In the CS limit, at leading order in $p$, the third term again dominates and we obtain
\ba
\delta \Delta T_{\Lambda,r}^{\CS} &=  \frac{15 \pi M}{8} \frac{\zeta a}{p^{3}} \sin{\theta_{\rm tp}} \nonumber \\
&\times \left[\frac{4(1+e^{2})}{(1-e^2)^{1/2}} +(4+e^2)\right] \,.
\ea

The dominant contribution to $\delta \Delta T_{\Lambda, \theta}$ comes from the terms
\ba
\frac{\delta \Delta T_{\Lambda, \theta}}{M} &=\frac{2\pi a\delta L}{\sqrt{p}} -\frac{2\pi a \gamma_{3,1} E_\Kerr (2- \cos^2\theta_{\rm tp})}{\sqrt{p}} \nonumber \\
&- \frac{\pi aL_\Kerr}{p^{3/2}} \left(t_0 + \frac{ \cos^2\theta_{\rm tp} t_1}{2}\right) \nonumber \\ & +\frac{\pi a^2 E_\Kerr}{p^{3/2}} \left[\left(1-\frac{ \cos^2\theta_{\rm tp}}{2}\right)t_0\right. \nonumber \\
&+ \left. \left(1-\frac{3  \cos^2\theta_{\rm tp}}{4}\right)  \cos^2\theta_{\rm tp}t_1 \right]\,.
\ea
Under assumptions ${\cal B}_2$, the second term dominates and we find
\be
\frac{\delta \Delta T_{\Lambda,\theta}^{{\cal{B}}_{2}}}{M} = - \frac{2\pi a}{\sqrt{p}}\gamma_{3,1}(2-\cos^2\theta_{\rm tp} )\,. 
\ee
Under assumptions ${\cal B}_N$ for $N \geq 3$, the final term dominates giving
\be
\frac{\delta \Delta T_{\Lambda,\theta}^{{\cal{B}}_{N>2}}}{M}= \frac{\pi a^2}{2 p^{N-1/2}} f_{N}(e) (2-\cos^2\theta_{\rm tp} ) \left(2 \gamma_{4,N} + \gamma_{1,N} \right)\,.
\ee
At leading order in $p$ the same term dominates in the CS limit to give
\be
\frac{\delta \Delta T_{\Lambda,\theta}^{\CS}}{M} = \frac{5 \pi}{8} \frac{a^3 \zeta}{p^{5}} (2-\cos^2\theta_{\rm tp}) \sin{\theta_{\rm tp}}\, \left(e^{2} + 3 \right) \left(3 e^{2} + 1 \right) \,. 
\ee
If we linearise in $a$ for consistency, then we find $\delta \Delta T_{\Lambda,\theta}^{\CS} = 0$ in the CS limit.

\subsubsection{Physical frequencies}
In the weak-field Kerr metric, the leading-order parts of the frequencies are given by
\ba
\Lambda_{r K} &= \frac{2\pi}{\sqrt{p}} \qquad
\Lambda_{\theta K} = \frac{2\pi}{\sqrt{p}}\,, \\
\Delta \phi_{\Lambda, r K} &= \frac{2\pi a}{\sqrt{p}}  \qquad
\Delta \phi_{\Lambda, \theta K} = 2\pi\,,    \\
\frac{\Delta T_{\Lambda, r K}}{M} &= 2\pi \left(\frac{p}{1-e^2}\right)^{\frac{3}{2}}\,, \\
\frac{\Delta T_{\Lambda, \theta K}}{M} &= 2\pi a \sin\theta_{\rm tp}\,,  \\
M\Omega_{\phi K} &=M \Omega_{r K}  = M\Omega_{\theta K} = \left(\frac{1-e^2}{p}\right)^{\frac{3}{2}}\,, \\
\Delta T_{\Lambda K}&= 4\pi^2 M p(1-e^2)^{-3/2}\,,
 \\ 
  \Lambda_{r K}-\Lambda_{\theta K} &= 6\pi p^{-3/2}\,,   
  \\
\Delta\phi_{\Lambda K} -2\pi\Lambda_{r K} &= 8\pi^2 a p^{-2}\,, 
\ea
from which we obtain
\be
M\dot{\tilde{\gamma}}_{\Kerr} = \frac{3 (1-e^2)^{3/2}}{p^{5/2}}, \qquad
M\dot{\alpha}_{\Kerr} = 2 a \frac{(1-e^2)^{3/2}}{p^3}.
\ee
We can now put together the pieces from the preceding sections to derive the leading order corrections to the three frequencies, which we require in order to construct the modified AK waveforms.

\paragraph{Orbital frequency.}
The perturbation to the orbital frequency, $2\pi\nu$, is
\ba
M\delta(2\pi\nu)\equiv M\delta\Omega_r &= M\Omega_r\left(\frac{\delta\Lambda_\theta}{\Lambda_\theta} - \frac{\delta{\Delta T_\Lambda}}{\Delta T_\Lambda}\right)
\nonumber \\
&=-\frac{(1-e^2)^3}{2\pi p^{5/2}} \left( \frac{\delta\Delta T_{\Lambda, r}}{M\sqrt{p}} + \frac{\delta\Delta T_{\Lambda, \theta}}{M\sqrt{p}} \right. \nonumber \\
&+ \left. a\sin\theta_{\rm tp} \left[\delta\Lambda_r - \delta\Lambda_\theta\right]\right).
\ea
Under assumptions ${\cal B}_2$, the first two terms dominate and we find
\ba
M\delta\Omega_r^{{\cal B}_2} &= -\frac{(1-e^2)^3}{2p^{7/2}} \left[ \left(1+\frac{3}{\sqrt{1-e^2}}\right)\gamma_{1,2} + 4\gamma_{4,2} \right. \nonumber \\
&-\left. 2a\left(1+\left[4-\frac{3}{\sqrt{1-e^2}}\right]\sin^2\theta_{\rm tp}\right)\gamma_{3,1}\right] .
\nonumber \\
\ea
Under assumptions ${\cal B}_N$ with $N \geq 3$, the first term alone dominates and we find
\ba
M\delta\Omega_r^{{\cal B}_{N>2}} &= \frac{(1-e^2)^2}{2\pi p^{N+1/2}} \sum_{n=-2}^{N-4} \tilde{v}_n I_n .
\ea
For the first few $N$'s we have
\ba
M\delta\Omega_r^{{\cal B}_3} &= -\frac{3 (1-e^2)^{5/2}}{4 p^{7/2}} \left(2 \gamma_{4,3} + \gamma_{1,3} \right)\,,  
\\
M\delta\Omega_r^{{\cal B}_4} &=-\frac{(1-e^2)^{5/2}}{2 p^{9/2}}  \left[3 + (1-e^2)^{1/2}\right]\,  \left(2 \gamma_{4,4} + \gamma_{1,4} \right)\,, 
 \\
M\delta\Omega_r^{{\cal B}_5} &=-\frac{3 (1-e^2)^{5/2}}{4 p^{11/2}} \left[3+e^2 + 2 (1-e^2)^{1/2} \right]  \nonumber \\ 
&\times \left(2 \gamma_{4,5} + \gamma_{1,5} \right)\,.
\ea
In the CS limit, the first term again dominates and, linearizing in $a$ for consistency, we have
\ba
\delta\Omega_r^{CS} &= -\frac{15 M (1-e^2)^{5/2}}{16p^6} \zeta a \sin\theta_{\rm tp} 
\nonumber  \\ 
&\times \left[4(1+e^{2}) +(4+e^2)(1-e^2)^{1/2} \right]\,.
\ea

\paragraph{Perihelion precession frequency.}
The perturbation to the perihelion precession rate, $\dot{\tilde{\gamma}}$, is given by Eqs.~\eqref{precpert1}-\eqref{precpert}, which can be simplified to
\ba
\frac{\delta\dot{\tilde{\gamma}}}{\dot{\tilde{\gamma}}} &= \frac{(\Lambda_\theta \delta\Lambda_r - \Lambda_r\delta\Lambda_\theta)(\Delta T_{\Lambda r} +\Delta T_{\Lambda \theta})}{(\Lambda_r-\Lambda_\theta)\Delta T_\Lambda} \nonumber \\
&- \frac{(\Lambda_r\delta\Delta T_{\Lambda \theta}+\Lambda_\theta\delta\Delta T_{\Lambda r})}{\Delta T_\Lambda}.
\ea
In all the cases we have been considering, the first term makes the dominant contribution in the weak-field and we find
\be
M\delta\dot{\tilde{\gamma}} = \frac{(1-e^2)^{3/2}}{2\pi p} \left(\delta \Lambda_r - \delta \Lambda_\theta\right) .
\ee
Under assumptions ${\cal B}_2$, only the $\delta\Lambda_\theta$ term contributes giving
\ba
M\delta\dot{\tilde{\gamma}}^{{\cal B}_2} &= \frac{(1-e^2)^{3/2}}{2 p^{5/2}} \left[ \gamma_{1,2} + 2\gamma_{4,2}
 \right.\nonumber\\
&+ \left. \left(3a\cos^2\theta_{\rm tp} - 4a\right)\gamma_{3,1}\right]\,.
\ea
Under assumptions ${\cal B}_N$ with $N\geq 3$, both terms contribute and we find
\ba
M\delta\dot{\tilde{\gamma}}^{{\cal B}_{N>2}} &= \frac{(1-e^2)^{3/2}}{2 p^{N+1/2}}\left[f_N(e) + (1-e^2)f_{N-2}(e) \right. \nonumber \\
&+ \left. 2\sum_{k=1}^{N-2} f_{N-k-1}(e) I_k\right]\left(\gamma_{1,N}+2\gamma_{4,N}\right)\,,
\ea
which for the first few $N$'s gives
\ba
M\delta\dot{\tilde{\gamma}}^{{\cal B}_{3}} &= \frac{3(1-e^2)^{3/2}}{2 p^{7/2}} (\gamma_{1,3}+2\gamma_{4,3})\,,  \\
M\delta\dot{\tilde{\gamma}}^{{\cal B}_{4}} &= \frac{3(1-e^2)^{3/2}}{4 p^{9/2}} (4+e^2) (\gamma_{1,4}+2\gamma_{4,4})\,,  \\
M\delta\dot{\tilde{\gamma}}^{{\cal B}_{5}} &= \frac{5(1-e^2)^{3/2}}{4 p^{11/2}} (4+3e^2) (\gamma_{1,5}+2\gamma_{4,5})\,.
\ea
In the CS limit, both terms again contribute and we find
\be
M\delta\dot{\tilde{\gamma}}^{CS} =\frac{75 (1-e^2)^{3/2}}{64 p^6} a\zeta\sin\theta_{\rm tp} (8+12e^2+e^4) .
\ee

\paragraph{Orbital plane precession frequency.}
Equations~\eqref{precpert1}-\eqref{precpert} for the orbital plane precession frequency can be written as
\ba
\frac{\delta\dot{\alpha}}{\dot{\alpha}} &= \frac{\Lambda_r \delta\Delta\phi_\theta + \Lambda_\theta \delta\Delta\phi_r + \delta\Lambda_\theta \Delta\phi_r + (\Delta\phi_\theta-2\pi)\delta\Lambda_r}{\Delta\phi_\Lambda-2\pi\Lambda_r} \nonumber \\
&\qquad\qquad-  \frac{\delta\Delta T_\Lambda}{\Delta T_\Lambda}.
\ea
In the Kerr limit, $\Delta \phi_\theta - 2\pi \sim 1/p^2$. Using the preceding results we can see that in all the special cases we have considered in this paper, the term on the first line dominates over the term in the second line. For ${\cal B}_2$, the dominant contribution comes from the $\delta\Delta\phi_\theta$ term 
\ba
M\delta\dot{\alpha}^{{\cal{B}}_{2}} &= \frac{(1-e^2)^{3/2}}{p^{5/2}} a \sin\theta_{\rm tp} \gamma_{3,1}.
\ea
If $\gamma_{3,1}=0$, the ${\cal B}_2$ limit instead gives
\ba
M\delta\dot{\alpha}^{{\cal{B}}_{2}} &= -\frac{a(1-e^2)^{3/2}}{p^{3}} (\gamma_{1,2}+2\gamma_{4,2})\,.
\ea
For assumptions ${\cal B}_N$ with $N \geq 3$, the dominant contribution comes from the $\delta\Lambda_\theta$ term and we obtain
\ba
M\delta\dot{\alpha}^{{\cal{B}}_{N>2}} &= -\frac{a (1-e^2)^{3/2}}{2 p^{N+1}} f_N(e) (\gamma_{1,N}+2\gamma_{4,N})\,,
\ea
which for the first few $N$s gives
\ba
M\delta\dot{\alpha}^{{\cal{B}}_{3}} &= - \frac{a}{4 p^{4}} \left(1 - e^{2} \right)^{3/2} \left(3 + e^{2} \right) \left(\gamma_{1,3} + 2 \gamma_{4,3} \right)\,,
 \\
M\delta\dot{\alpha}^{{\cal{B}}_{4}} &= - \frac{a}{p^{5}} \left(1 - e^{2} \right)^{3/2} \left(1 + e^{2} \right) \left(\gamma_{1,4} + 2 \gamma_{4,4} \right)\,,
\\  
M\delta\dot{\alpha}^{{\cal{B}}_{5}} &= - \frac{a}{4 p^{6}} \left(1 - e^{2}\right)^{3/2} \left(5 + 10 e^{2} + e^{4}\right) \left(\gamma_{1,5} + 2 \gamma_{4,5}\right)\,.
\ea
In the CS limit, the dominant contribution comes from the $\delta\Delta\phi_r$ term and we find
\ba
M\delta\dot{\alpha}^{\CS} &= -\frac{5}{64} a\zeta \frac{(1-e^2)^{3/2}}{p^6} (8 + 24 e^2 + 3 e^4) .
\ea

\paragraph{Kepler Law in the Circular Limit.}
With all of these frequencies, it is instructive to compute the correction to the Kepler orbital 
frequency when considering orbits in the equatorial plane. This frequency  
is given by $\Omega_{\Kep} = \Omega_\phi = 2 \pi \nu + \dot{\alpha} + \dot{\gamma}$, and for GR this is simply
\be
M \Omega_{\Kep,\GR} = \frac{\left(1 - e^{2}\right)^{3/2}}{p^{3/2}}\,.
\ee
For the Kerr deformed metric, $\Omega_{\Kep} = \Omega_{\Kep,\GR} + \epsilon\delta \Omega_{\Kep}$,
where in the ${\cal{B}}_{N}$ limits 
\ba
\delta \Omega_{\Kep,{\cal{B}}_{2}} &= -\frac{2}{p^{5/2}} 
\left(\gamma_{1,2} + 2 \gamma_{4,2} + a \gamma_{3,1} \right)
\\
\delta \Omega_{\Kep,{\cal{B}}_{3}} &=  \frac{3}{4} \frac{\gamma_{1,3} + 2 \gamma_{4,3}}{p^{7/2}} 
\,,
\\
\delta \Omega_{\Kep,{\cal{B}}_{4}} &=  \frac{\gamma_{1,4} + 2 \gamma_{4,4}}{p^{9/2}},
\\
\delta \Omega_{\Kep,{\cal{B}}_{5}} &=  \frac{5}{4} \frac{\gamma_{1,5} + 2 \gamma_{4,5}}{p^{11/2}}
\,,
\ea
and in the CS limit
\ba
\delta \Omega_{\Kep,\CS} &= \frac{5}{4} \frac{\zeta a}{p^{6}}\,,
\ea
which agrees with~\cite{2010CQGra..27v5006C}. Notice that this correction is of $4.5$PN absolute order, 
which is consistent with the fact that the CS correction to the Kerr solution enters
at $3$PN order relative to the $1.5$PN order spin-orbit coupling in the gravitomagnetic sector.

\section{Implicit Deformation of Radiation-Reaction}
\label{sec:rr}
We have so far concentrated on corrections to the conservative sector of GR, i.e.,~to the metric and the shape of orbits, but have not yet considered dissipative corrections, that is, modifications to the radiation-reaction force and the fluxes of energy, angular momentum and Carter constant. Modifications to the orbital shape will implicitly translate to modifications to the fluxes, as the latter are computed from derivatives of the quadrupole moment, with the particle moving on a geodesic orbit as a source. In what follows, we will compute these implicit modifications to the radiated fluxes. 

We will, however, neglect any direct modifications to the fluxes. Stein and Yunes~\cite{2009PhRvD..80l2003Y} have shown that there is a wide class of quadratic gravity theories for which the effective gravitational energy flux is indeed still described by the Isaacson effective stress-energy tensor. An example of such a theory is dynamical CS gravity~\cite{Alexander:2009tp}. There can be, however, theories in which dynamical scalar fields emit dipolar radiation, which in turn might lead to modifications of similar order as those considered here. We leave the inclusion of such direct dissipative modifications to future work.

\subsection{Equations of Motion}
To estimate the implicit corrections to the radiated fluxes, we must evaluate the leading order in $\epsilon$ and $p$ correction to the quadrupole moment tensor. This can be accomplished by linearizing the geodesic equations in $\epsilon$ and then separately linearising the Kerr geodesic part, i.e., the $O(\epsilon^0)$ part, and the linear correction,  i.e.,~the $O(\epsilon)$ part, in $p$. Using the same orbital parameterization as before, in terms of the three phase angles $\left(\psi,\chi,\phi\right)$, we find that in this approximation
\ba
\frac{{\rm d}\psi}{{\rm d}t} &= \frac{1}{p^{3/2}} (1+e\cos\psi)^2 \nonumber \\
&+  \epsilon \left[ \frac{1}{2(1-e^2)\,p^{K+1}} \sum_{n=-2}^{\infty} \tilde{v}_n (1+e\cos\psi)^{n+4} \right. \nonumber
\\ \nonumber 
&- \frac{\delta E}{p^{3/2}} (1+e\cos\psi)^2  \nonumber \\
& \left.- \frac{\delta T_{N_T}}{p^{N_T+7/2}} (1+e\cos\psi)^{N_T+4} \right]\,,  
\\
\frac{{\rm d}\chi}{{\rm d}t} &= \frac{1}{p^{3/2}} (1+e\cos\psi)^2  
+ \epsilon \left[ \frac{t_0}{2p^{5/2}} (1+e\cos\psi)^2 
\right.\nonumber\\
&+ \frac{t_1  \cos^2\theta_{\rm tp}\cos^2\chi}{2p^{5/2}}(1+e\cos\psi)^2 
\nonumber \\ & - \frac{\delta E}{p^{3/2}} (1+e\cos\psi)^2 \nonumber\\
&-  \left. \frac{\delta T_{N_T}}{p^{N_T+3/2}} (1+e\cos\psi)^{N_T+4}
\right]\,, 
\\
\frac{{\rm d}\phi}{{\rm d}t} &= \frac{\sin\theta_{\rm tp}}{p^{3/2}} \frac{(1+e\cos\psi)^2}{1- \cos^2\theta_{\rm tp}\cos^2\chi} 
\nonumber \\ & + \epsilon \left[ \frac{2a\delta E}{p^3} (1+e\cos\psi)^3  \right.
\nonumber \\
&+ \frac{\delta L}{p^2} \frac{(1+e\cos\psi)^2}{1- \cos^2\theta_{\rm tp}\cos^2\chi} 
\nonumber\\&+\frac{\delta\Phi_{N_\phi}}{p^{N_\phi+2}} (1+e\cos\psi)^{N_\phi+2} 
\nonumber \\
&- \frac{\delta E\,\sin\theta_{\rm tp}}{p^{3/2}} \frac{(1+e\cos\psi)^2}{1- \cos^2\theta_{\rm tp}\cos^2\chi} \nonumber \\
&-\left. \frac{\delta T_{N_T}\,\sin\theta_{\rm tp}}{p^{N_T+7/2}} \frac{(1+e\cos\psi)^{N_T+4}}{1- \cos^2\theta_{\rm tp}\cos^2\chi} \right]\,.
\ea
The first term in each of these equations is the Keplerian solution and $(N_T,N_\phi)$ denote the leading-order terms in the perturbations $\delta T$ and $\delta \Phi$ as before. The terms inside the brackets are of different orders in $p$ in general, but the dominant terms differ under the various sets of assumptions we have been considering.

Under assumptions ${\cal B}_2$, the first and third terms are dominant in ${\rm d}\psi/{\rm d}t$, the first two terms are dominant for ${\rm d}\chi/{\rm d}t$ and the second term is dominant for  ${\rm d}\phi/{\rm d}t$. Our approximation to the orbit then becomes
\ba
\frac{{\rm d}\psi^{{\cal{B}}_{2}}}{{\rm d}t}  &= \frac{1}{p^{3/2}} (1+e\cos\psi)^2 \nonumber \\
&-  \epsilon \frac{1}{p^{7/2}} \left. \bigg[ \left(\frac{1}{2}\gamma_{1,2} + 2\gamma_{4,2} \right) (1+e\cos\psi)^4 \right. \nonumber \\
&+ (1+e\cos\psi)^3 \gamma_{1,2} 
+ \frac{(1-e^2)}{2} (1+e\cos\psi)^2 \gamma_{1,2} \nonumber \\
&+ 2a \left[ \frac{(1-e^2)}{2}(1+e\cos\psi)^4+  (1+e\cos\psi)^3 \right. \nonumber \\
&+\left.\frac{1}{2} (1+e\cos\psi)^2\right] \gamma_{3,1}\sin^2\theta_{\rm tp}  \nonumber \\
&+ 2a(1+e\cos\psi)^4 \gamma_{3,1}(1- \cos^2\theta_{\rm tp}\cos^2\chi) \bigg]\,,
\\
\frac{{\rm d}\chi^{{\cal{B}}_{2}}}{{\rm d}t}  &= \frac{1}{p^{3/2}} (1+e\cos\psi)^2 
+ \epsilon \frac{1}{2p^{5/2}} \big( \gamma_{1,2}+2 \gamma_{4,2} \nonumber 
\\ \nonumber  
&+ 2( \cos^2\theta_{\rm tp}-2)a\gamma_{3,1}
+ 2a\gamma_{3,1}  \cos^2\theta_{\rm tp} \cos^2\chi\big) \nonumber \\ 
&  \times(1+e\cos\psi)^2\,, \\
\frac{{\rm d}\phi^{{\cal{B}}_{2}}}{{\rm d}t}
&= \frac{\sin\theta_{\rm tp}}{p^{3/2}} \frac{(1+e\cos\psi)^2}{1- \cos^2\theta_{\rm tp}\cos^2\chi} 
+ \epsilon \frac{1}{p^{5/2}} \left(\frac{\gamma_{1,2}}{2}+\gamma_{4,2}  \right.
\nonumber \\
&- \left. a\sin^2\theta_{\rm tp}\gamma_{3,1} \right)\sin\theta_{\rm tp} 
\frac{(1+e\cos\psi)^2}{1- \cos^2\theta_{\rm tp}\cos^2\chi}\,.
\ea
Under assumptions ${\cal B}_N$, the first term dominates in ${\rm d}\psi/{\rm d}t$ and in ${\rm d}\chi/{\rm d}t$, while the second term is dominant in  ${\rm d}\phi/{\rm d}t$. Our approximation to the orbit then becomes
\ba
\frac{{\rm d}\psi^{{\cal{B}}_{N>2}}}{{\rm d}t} &= \frac{1}{p^{3/2}} (1+e\cos\psi)^2 
\nonumber \\
&- \epsilon \frac{1}{p^{N+1/2}} \left( \frac{(1-e^2)}{2} f_{N-2}(e) (1+e\cos\psi)^2 \right. \nonumber 
\\ \nonumber 
&+ \left. \sum_{l=-1}^{N-4} f_{N-l-3}(e) (1+e\cos\psi)^{l+4} \right) \nonumber \\
&\times \left(\gamma_{1,N}+2\gamma_{4,N}\right)\,,
\\
\frac{{\rm d}\chi^{{\cal{B}}_{N>2}}}{{\rm d}t} &= \frac{1}{p^{3/2}} (1+e\cos\psi)^2 \nonumber \\
&+  \epsilon \frac{1}{2 p^{N+1/2}} f_N(e)  (1+e\cos\psi)^2 \left(\gamma_{1,N}+2\gamma_{4,N}\right)\,, 
\\
\frac{{\rm d}\phi^{{\cal{B}}_{N>2}}}{{\rm d}t} &= \frac{\sin\theta_{\rm tp}}{p^{3/2}} \frac{(1+e\cos\psi)^2}{1- \cos^2\theta_{\rm tp}\cos^2\chi} 
+ \epsilon \frac{1}{2 p^{N+1/2}}f_N(e) 
\nonumber \\
&\times \left(\gamma_{1,N}+2\gamma_{4,N}\right)  \frac{(1+e\cos\psi)^2}{1- \cos^2\theta_{\rm tp}\cos^2\chi}\,.
\ea
In the CS case, the first term dominates in ${\rm d}\psi/{\rm d}t$ and in ${\rm d}\chi/{\rm d}t$, while the second and third terms are dominant in ${\rm d}\phi/{\rm d}t$. Our approximation to the orbit then becomes
\ba
\frac{{\rm d}\psi^{\CS}}{{\rm d}t} &= \frac{1}{p^{3/2}} (1+e\cos\psi)^2 
-  \epsilon \frac{5}{8 p^6} a\zeta\sin\theta_{\rm tp} \nonumber \\ 
&\times \big[2 (1-e^2)(1+e^2)  (1+e\cos\psi)^2 \nonumber \\ 
&+ 4(1+e^2)  (1+e\cos\psi)^3 \nonumber \\ &+ (3+e^2)  (1+e\cos\psi)^4 
\nonumber \\ 
&+  2 (1+e\cos\psi)^5 + (1+e\cos\psi)^6 \big]\,, 
\\
\frac{{\rm d}\chi^{\CS}}{{\rm d}t} &= \frac{1}{p^{3/2}} (1+e\cos\psi)^2 \nonumber \\
&+  \epsilon \frac{5}{8p^6} a\zeta\sin\theta_{\rm tp} (3+e^2)(1+3e^2) (1+e\cos\psi)^2\,,   
\\
\frac{{\rm d}\phi^{\CS}}{{\rm d}t} &= \frac{\sin\theta_{\rm tp}}{p^{3/2}} \frac{(1+e\cos\psi)^2}{1- \cos^2\theta_{\rm tp}\cos^2\chi} \nonumber \\
&+ \epsilon \frac{5}{8p^6} a\zeta \Bigg( (3+e^2)(1+3e^2) 
\nonumber \\
&\times
\sin^2\theta_{\rm tp} \frac{ (1+e\cos\psi)^2}{1- \cos^2\theta_{\rm tp}\cos^2\chi} 
- (1+e\cos\psi)^6\Bigg)\,.
\ea

\subsection{Fluxes}
Using  these simplified equations of motion, we can now compute the corrected fluxes. For this, we assume that~\cite{1963PhRv..131..435P,1964PhRv..136.1224P} 
\ba
\left<\dot{E}\right> &=- \frac{1}{5\mu} \left< \dddot{I}_{ij} \dddot{I}^{ij} - \frac{1}{3} \dddot{I}_{ii} \dddot{I}_{jj} \right>\,,
\nonumber \\
\left<\dot{L}_{i}\right> &= - \frac{2}{5\mu M} \epsilon_{ijk} \left< \ddot{I}_{jm} \dddot{I}^{km}\right>\,,
\label{Edot-Ldot}
\ea
where $\mu$ is the reduced mass, $E$ and $L_i$ are the dimensionless energy and momentum defined in Eq.~(\ref{dimlessconst}), the angle-brackets stand for an average over several wavelengths and we have defined the quadrupole moment
\be
I^{ij} = \mu \; z^{i}(t) \; z^{j}(t)\,,
\ee
in which the particle trajectories are $z^{i} = [r \cos{\phi} \sin{\theta}, r \sin{\phi} \sin{\theta}, r \cos{\theta}]$. 
In the AK model, it is assumed that the orbital inclination remains constant, as the change in inclination is a higher order PN effect. If we make the same assumption here, $\dot{\theta_{\rm tp}}=0$, we can find the rate of change of Carter constant in terms of the rates of change of energy and angular momentum. This gives
\allowdisplaybreaks[1]
\begin{align}
\dot{\delta Q} &= \frac{2\sqrt{p}  \cos^2\theta_{\rm tp}}{\sin\theta_{\rm tp}} \dot{\delta L} - 2 a^{2}  \cos^2\theta_{\rm tp} \delta \dot{E}
\nonumber \\
&+ 2 a^{2}  \cos^2\theta_{\rm tp}\left( \frac{\partial \delta E}{\partial p} \dot{p}_{\Kerr}  + \frac{\partial  \delta E}{\partial e} \dot{e}_{\Kerr}  \right)
\nonumber \\
&-  \frac{2 \sqrt{p}  \cos^2\theta_{\rm tp}}{\sin{\theta_{\rm tp}}} \left( \frac{\partial  \delta L}{\partial p} \dot{p}_{\Kerr}  + \frac{\partial  \delta L}{\partial e} \dot{e}_{\Kerr}  \right)
\nonumber \\
&+  \left( \frac{\partial  \delta Q}{\partial p} \dot{p}_{\Kerr}  + \frac{\partial  \delta Q}{\partial e} \dot{e}_{\Kerr}  \right)\,.\label{constinceq}
\end{align}
In all cases except ${\cal B}_2$, at leading order this is equivalent to writing
\be 
\left<\dot{Q}\right> = \frac{2 Q}{L} \left<\dot{L}\right>\,,
\ee
where, as before, an $L$ without a suffix is being used to denote the $z$ component of the angular momentum. This is the statement that $L^2/Q =$constant, which is equivalent to constant inclination in the Keplerian limit. In the case of ${\cal B}_2$, the result is different, but only if $\gamma_{3,1}\neq0$. This leads to problems, and so in the following we will make this additional restriction and use the above to compute $\delta{\dot{Q}}$ in all cases.

With this at hand, we can now compute $\dot{E}$ and $\dot{L}$ through Eq.~\eqref{Edot-Ldot}. To do so, every time we take a time derivative of the quadrupole moment, we use the equations of motion to reduce their order. The averaging is performed by taking $\int dt \to \int d\psi/\dot{\psi}$. This, of course, works well for the ${\cal{O}}(\epsilon^{0})$ terms (the GR contributions), which only depend on $\psi$. The non-GR deformations, however, depend both on $\psi$ and $\chi$, and thus, more care is needed. Since these are ${\cal{O}}(\epsilon)$ corrections, it suffices to use the ${\cal{O}}(\epsilon^{0})$ piece of $\dot\psi$ and $\dot\chi$ in the averaging of these terms, but at this order in $\epsilon$ and to leading order in $1/p$, $\dot{\psi} = \dot{\chi}$, which then implies $\psi = \chi + \delta \chi$. After integration, one then obtains corrections that depend on $\cos^{2}{\delta \chi}$, but since these quantities are slowly varying (at next order in $1/p$, the frequencies are no longer commensurate), one can average over them by writing $\cos^{2}{\delta \chi} \sim 1/2$. 

Let us decompose the fluxes into $<\dot{E}> = <\dot{E}>_{\GR} + \epsilon \delta {\dot{E}}$,  $<\dot{L}> = <\dot{L}>_{\GR} + \epsilon \delta {\dot{L}}$, and  $<\dot{Q}> = <\dot{Q}>_{\GR} + \epsilon \delta {\dot{Q}}$, where the GR fluxes are~\cite{1963PhRv..131..435P,1964PhRv..136.1224P}
\ba
M\left<\dot{E} \right>_{\GR} &= -\frac{32}{5} \frac{\eta}{p^{5}} \left(1 - e^{2}\right)^{3/2}
\left(1 + \frac{73}{24} e^{2} + \frac{37}{96} e^{4} \right)\,,\nonumber\\
\\
M\left<\dot{L}\right>_{\GR} &= - \frac{32}{5} \frac{\eta}{p^{7/2}} \left(1 - e^{2}\right)^{3/2}
\left(1 + \frac{7}{8} e^{2} \right)\,,
\\
M\left<\dot{Q}\right>_{\GR} &= - \frac{64}{5} \frac{\eta}{p^{3}} \left(1 - e^{2}\right)^{3/2}
\left(1 + \frac{7}{8} e^{2} \right)  \cos^2\theta_{\rm tp}\,. \nonumber \\ 
\ea
The correction to these fluxes are, in the ${\cal{B}}_{2}$ limit, with $\gamma_{3,1}=0$:
\allowdisplaybreaks[1]
\begin{align}
M\delta \dot{E}_{{\cal{B}}_{2}} &= -\frac{96}{5} \frac{\eta}{p^{6}} \left(1 - e^{2}\right)^{3/2}
\left(\gamma_{12} + 2 \gamma_{42}\right) \nonumber \\ & \times 
\left(1 + \frac{13}{3} e^{2} + \frac{11}{8} e^{4} \right)\,, 
\\
M\delta \dot{L}_{{\cal{B}}_{2}} &= - 16 \frac{\eta\sin\theta_{\rm tp}}{p^{9/2}}\left(1 - e^{2}\right)^{3/2} \left(\gamma_{12} + 2 \gamma_{42}\right) \nonumber \\ &
\times\left(1 + \frac{63}{40} e^{2} + \frac{e^{4}}{20} \right)\,, \\
M\delta \dot{Q}_{{\cal{B}}_{2}} &= -\frac{8}{5} \frac{\eta}{p^{7/2}} \cos^2\theta_{\rm tp} \left(1 - e^{2}\right)^{3/2}
\nonumber \\ & \times \left(24+35e^2 + e^{4} \right)\left(\gamma_{1,2}+2\gamma_{4,2}\right)\,,
\end{align}
in the ${\cal{B}}_{3}$ limit:
\begin{align}
M\delta \dot{E}_{{\cal{B}}_{3}} &= -\frac{144}{5} \frac{\eta}{p^{7}} \left(1 - e^{2}\right)^{3/2}
\left(\gamma_{13} + 2 \gamma_{43} \right) 
\nonumber \\
&\times \left(1 + \frac{461}{72} e^{2} + \frac{1345}{288} e^{4} + \frac{251}{864} e^{6} \right)\,,
\\
M\delta \dot{L}_{{\cal{B}}_{3}} &= - 24 \frac{\eta}{p^{11/2}}  \sin\theta_{\rm tp} \left(1 - e^{2}\right)^{3/2}
\left(\gamma_{13} + 2 \gamma_{43} \right)
\nonumber \\
&\times
\left(1 + \frac{361}{120} e^{2} + \frac{7}{10} e^{4} \right)\,,
\\
M\delta \dot{Q}_{{\cal{B}}_{3}} &= - \frac{288}{5} \frac{\eta}{p^{5}}  \cos^2\theta_{\rm tp} \left(1 - e^{2}\right)^{3/2}
\left(\gamma_{13} + 2 \gamma_{43} \right)
\nonumber \\
&\times
\left(1 + \frac{65}{24} e^{2} + \frac{91}{144} e^{4} \right)\,,
\end{align}
in the ${\cal{B}}_{4}$ limit:
\begin{align}
M\delta \dot{E}_{{\cal{B}}_{4}} &= -\frac{208}{5} \frac{\eta}{p^{8}} \left(1 - e^{2}\right)^{3/2}
\left(\gamma_{14} + 2 \gamma_{44} \right) 
\nonumber \\
&\times \left[1 + \frac{2701}{312} e^{2} + \frac{3155}{312} e^{4} + \frac{3779}{2496} e^{6} 
\right. 
\nonumber \\
&- \left.
\frac{96}{1248} \left(1 - e^{2}\right)^{3/2}  \left(1 + \frac{73}{24} e^{2} + \frac{37}{96} e^{4} \right)\right]\,, \nonumber
\\ & \\
M\delta \dot{L}_{{\cal{B}}_{4}} &= - \frac{176}{5} \frac{\eta}{p^{13/2}}  \sin\theta_{\rm tp} \left(1 - e^{2}\right)^{3/2}
\left(\gamma_{14} + 2 \gamma_{44} \right)
\nonumber \\
&\times
\left[1 + \frac{409}{88} e^{2} + \frac{757}{352} e^{4} + \frac{3}{88} e^{6}
\right.
\nonumber \\
&- \left. 
\frac{1}{11} \left(1 - e^{2}\right)^{3/2} \left(1 + \frac{7}{8} e^{2} \right) \right]\,,
\\
M\delta \dot{Q}_{{\cal{B}}_{4}} &= - \frac{416}{5} \frac{\eta}{p^{6}}  \cos^2\theta_{\rm tp} \left(1 - e^{2}\right)^{3/2}
\left(\gamma_{14} + 2 \gamma_{44} \right)
\nonumber \\
&\times
\left[1 + \frac{439}{104} e^{2} + \frac{3}{104} e^{6} + \frac{813}{416} e^{4} 
\right. 
\nonumber \\
&- \frac{1}{13} \left.
 \left(1 - e^{2}\right)^{3/2}  \left(1 + \frac{7}{8} e^{2} \right) \right]\,,
\end{align}
in the ${\cal{B}}_{5}$ limit:
\begin{align}
M\delta \dot{E}_{{\cal{B}}_{5}} &= -\frac{288}{5} \frac{\eta}{p^{9}} \left(1 - e^{2}\right)^{3/2}
\left(\gamma_{15} + 2 \gamma_{45} \right) 
\nonumber \\
&\times \left[1 + \frac{791}{72} e^{2} + \frac{31805}{1728} e^{4} \right. \nonumber \\
& +\frac{19277}{3456} e^{6} + \frac{2617}{13824} e^{8}   
\nonumber \\
&- \left.
\frac{1}{6} \left(1 - e^{2}\right)^{3/2}  \left(1 + \frac{73}{24} e^{2} + \frac{37}{96} e^{4} \right)\right]\,,
\\
M\delta \dot{L}_{{\cal{B}}_{5}} &= - \frac{248}{5} \frac{\eta}{p^{15/2}}  \sin\theta_{\rm tp} \left(1 - e^{2}\right)^{3/2}
\left(\gamma_{15} + 2 \gamma_{45} \right)
\nonumber \\
&\times
\left[\left(1 + \frac{1585}{248} e^{2} + \frac{2493}{496} e^{4} + \frac{237}{496} e^{6}\right)
\right.
\nonumber \\
&- \left. 
\frac{6}{31} \left(1 - e^{2}\right)^{3/2} \left(1 + \frac{7}{8} e^{2} \right) \right]\,,
\\
M\delta \dot{Q}_{{\cal{B}}_{5}} &= - \frac{576}{5} \frac{\eta}{p^{7}}  \cos^2\theta_{\rm tp} \left(1 - e^{2}\right)^{3/2}
\left(\gamma_{15} + 2 \gamma_{45} \right)
\nonumber \\
&\times
\left[ 1 + \frac{425}{72} e^{2} + \frac{883}{192} e^{4} + \frac{251}{576} e^{6} 
\right. 
\nonumber \\
&- \left. 
\frac{1}{6} \left(1 - e^{2}\right)^{3/2} \left(1 + \frac{7}{8} e^{2} \right) \right]\,,
\end{align}
and in the CS limit:
\begin{widetext}
\begin{align}
M\delta \dot{E}_{\CS} &= -72 \frac{\eta \zeta a}{p^{19/2}} \left(1 - e^{2}\right)^{3/2} \sin\theta_{\rm tp}
\left[1 + \frac{97}{8} e^{2} + \frac{15065}{576} e^{4} + \frac{1865}{144} e^{6} 
+ \frac{31555}{27648} e^{8}  
\right. 
\nonumber \\
&- \left. \frac{1}{3} \left(1 - e^{2}\right)^{3/2} \left(1 + \frac{e^{2}}{4} \right) 
\left(1 + \frac{73}{24} e^{2} + \frac{37}{96} e^{4} \right)\right]\,,
\\
M\delta \dot{L}_{\CS} &= - 64 \frac{\eta\zeta a}{p^{8}} \left(1 - e^{2}\right)^{3/2}
\left[\left(1 + \frac{121}{16} e^{2} + \frac{4585}{512} e^{4} 
+ \frac{1039}{512} e^{6} + \frac{3}{128} e^{8}\right)
- \frac{3}{8} \left(1 - e^{2}\right)^{3/2} \left(1 + \frac{e^{2}}{4} \right)   
\right. 
\nonumber \\
&\times \left.
\sin^2\theta_{\rm tp} \left(1 + \frac{7}{8} e^{2} \right)
-\frac{69}{64}  \cos^2\theta_{\rm tp} \left(1 + \frac{715}{92} e^{2} + \frac{223}{24} e^{4} 
+ \frac{4597}{2208} e^{6} + \frac{9}{368} e^{8} \right) \right]\,,
\\
M\delta \dot{Q}_{\CS} &= \frac{1}{16} \frac{\eta \zeta a}{p^{15/2}} \frac{ \cos^2\theta_{\rm tp}}{ \sin\theta_{\rm tp}} \left(1 - e^{2}\right)^{3/2} \left[ -2432
\left(1 + \frac{1069}{152} e^{2} + \frac{4961}{608} e^{4} 
+ \frac{1123}{608} e^{6} + \frac{3}{152} e^{8} \right)
\right. 
\nonumber \\
&+ \left.
2592  \cos^2\theta_{\rm tp} \left(1 + \frac{2347}{324} e^{2} + \frac{1835}{216} e^{4} 
+ \frac{4933}{2592} e^{6} + \frac{e^{8}}{48} \right)
+ 768 \left(1 - e^{2}\right)^{3/2}  \sin^2\theta_{\rm tp} \left(1 + \frac{e^{2}}{4} \right) 
\left(1 + \frac{7}{8} e^{2} \right) \right]\,, 
\label{CSfluxes}
\end{align}
\end{widetext}
We can now compare these results to those obtained by Ryan~\cite{Ryan:1995wh} [see eg. Eq.~$(55)$ in that paper]. We see that the CS correction to the energy flux corresponds to a $\ell = 4$ (hexadecapole) correction to the current multipoles. Similarly, the ${\cal{B}}_{3}$ modification to the energy flux corresponds to a $\ell = 2$ (quadrupole) correction to the mass multipoles. 

In the following section we will combine the results described in this section, and the precession rates derived in Section~\ref{sec:fund-freq} to obtain the leading order corrections to the AK waveforms in these modified spacetimes.

\section{Modified Gravity AK Waveforms}
\label{sec:MG-AK}
The AK waveform model of Barack and Cutler~\cite{AK} is built around the gravitational waveform from a weak-field, Keplerian orbit, as computed in~\cite{1963PhRv..131..435P,1964PhRv..136.1224P}. The geometry of the orbit is characterized by the eccentricity, $e$, and the radial orbital frequency, $\nu$, while the instantaneous phase of the object in the orbit is specified by the mean anomaly, $\Phi$, which for a geodesic orbit satisfies ${\rm d}\Phi/{\rm d}t = 2\pi \nu$. The orientation of the orbit within the frame of reference of the BH is defined by three angles --- the inclination of the orbital plane to the spin axis of the BH, $\lambda$, an angle, $\alpha$, describing the rotation of the orbital plane about the spin axis and the angle, $\tilde{\gamma}$, between the direction of pericenter and the direction $\hat{L}\times\hat{S}$, where $\hat{L}$ is a unit vector perpendicular to the orbital plane, and $\hat{S}$ is a unit vector parallel to the BH spin axis.

For a Keplerian orbit, all of these quantities are fixed, but the AK model includes relativistic effects by imposing precession of the angles $\tilde{\gamma}$ and $\alpha$ and evolution of the orbital geometry through changes of $e$ and $\nu$, while the inclination angle, $\lambda$, is kept constant. This evolution is computed through the integration of a set of five coupled ordinary differential equations, which at leading order take the form
\ba
\label{AKode1}
\frac{{\rm d}\Phi}{{\rm d}t} &= 2\pi\nu\,, \\
\frac{{\rm d}\nu}{{\rm d}t} &= \frac{96}{10\pi}\frac{\eta}{M^2} (2\pi M\nu)^{11/3} (1-e^2)^{-7/2} \nonumber \\
& \times\left(1+\frac{73}{24}e^2 +\frac{37}{96}e^4\right)\,, \\
\frac{{\rm d}\tilde{\gamma}}{{\rm d}t} &= 6\pi\nu (2\pi M\nu)^{2/3} (1-e^2)^{-1}\,, \\
\frac{{\rm d}e}{{\rm d}t} &=-\frac{32}{5} \frac{\eta}{M} \frac{(1-e^2)^{3/2}}{p^4}\, e \left(\frac{19}{6} +\frac{121}{96}e^2\right)\,,  \\
\frac{{\rm d}\alpha}{{\rm d}t} &= 8\pi^2 M \nu^2 a\cos\lambda(1-e^2)^{-3/2}\,.
\label{AKode}
\ea
To construct AK waveforms in the modified gravity spacetimes we have been considering here, we need only add corrections to these ordinary differential equations. The corrections to $\dot{\tilde{\gamma}}$ and $\dot\alpha$ were computed in Section~\ref{sec:fund-freq}, although we must change variables to rewrite these in terms of $e$ and $\nu$. The corrections to $\dot{e}$ and $\dot{\nu}$ are determined by the corrections to the fluxes of energy, angular momentum and Carter constant computed in Section~\ref{sec:rr}, but we must change variables to put these into the AK form. We will describe this below. The inclination angle, $\lambda$, is equivalent to the inclination angle $\theta_{\rm tp}$ we have used in this paper and so the AK assumption that $\dot{\lambda}=0$ is consistent with keeping this angle constant, as we have done elsewhere in this paper.

\subsection{Rate of change of eccentricity and frequency}
To convert the fluxes of energy, angular momentum and Carter constant into fluxes of $e$, $p$ and $\theta_{\rm tp}$ we write, as usual, $\dot{e} = \dot{e}_{\Kerr} + \epsilon \; \delta\dot{e}$ etc., where the Kerr subscript denotes the corresponding flux in the Kerr metric. The Kerr fluxes are
\ba
M\dot{p}_{\Kerr} &= -\frac{64}{5} \eta \frac{(1-e^2)^{3/2}}{p^3} \left(1+\frac{7}{8}e^2\right)\,,  \\
M\dot{e}_{\Kerr} &= -\frac{32}{5} \eta \frac{(1-e^2)^{3/2}}{p^4}\, e \left(\frac{19}{6} +\frac{121}{96}e^2\right)\,, \\
\dot{\theta_{\rm tp}}_{\Kerr} &= 0 .
\ea
Transforming from $E$, $L$ and $Q$ to $p$, $e$ and $\theta_{\rm tp}$, we find
\ba
\left(\begin{array}{c} \delta\dot{E}\\\delta\dot{L}\\\delta\dot{Q}\end{array}\right) &= \left( \begin{array}{ccc}\frac{\partial \delta E}{\partial p}&\frac{\partial \delta E}{\partial e}&\frac{\partial \delta E}{\partial \theta_{\rm tp}}\\ \frac{\partial \delta L}{\partial p}&\frac{\partial \delta L}{\partial e}& \frac{\partial \delta L}{\partial \theta_{\rm tp}}\\ \frac{\partial \delta Q}{\partial p}&\frac{\partial \delta Q}{\partial e}&\frac{\partial \delta Q}{\partial \theta_{\rm tp}} \end{array}\right) \left( \begin{array}{c}\dot{p}_{\Kerr}\\\dot{e}_{\Kerr}\\\dot{\theta_{\rm tp}}_{\Kerr} \end{array}\right)\nonumber \\
&+ \left( \begin{array}{ccc} \frac{\partial E_{\Kerr}}{\partial p}&\frac{\partial E_{\Kerr}}{\partial e}&\frac{\partial E_{\Kerr}}{\partial \theta_{\rm tp}}\\ \frac{\partial {L}_{\Kerr}}{\partial p}&\frac{\partial {L}_{\Kerr}}{\partial e}&\frac{\partial {L}_{\Kerr}}{\partial \theta_{\rm tp}}\\ \frac{\partial Q_{\Kerr}}{\partial p}&\frac{\partial Q_{\Kerr}}{\partial e}&\frac{\partial Q_{\Kerr}}{\partial \theta_{\rm tp}} \end{array}\right) \left( \begin{array}{c}\delta\dot{p}\\\delta\dot{e}\\\delta\dot{\theta_{\rm tp}} \end{array}\right).
\label{perttrans}
\ea
To leading order in $p$, the second matrix on the right hand side of this equation is
\be
{\cal M}_{\Kerr}=\left(\begin{array}{ccc}\frac{(1-e^2)}{2p^2}&\frac{e}{p}& \frac{(1 - e^{2})^{2}}{p^{5/2}} a \cos{\theta_{\rm tp}}\\\frac{\sin{\theta_{\rm tp}}}{2\sqrt{p}}&\frac{e\sin{\theta_{\rm tp}}}{\sqrt{p}}&\sqrt{p}\cos\theta_{\rm tp}\\ \cos^2\theta_{\rm tp}&2e \cos^2\theta_{\rm tp}&-2p\cos\theta_{\rm tp}\sin\theta_{\rm tp}\end{array}\right)\,.
\ee
Rearranging Eq.~(\ref{perttrans}) we obtain
\ba
\left( \begin{array}{c}\delta\dot{p}\\\delta\dot{e}\\\delta\dot{\theta_{\rm tp}} \end{array}\right) &= {\cal M}_{\Kerr}^{-1} \left[ \left(\begin{array}{c} \delta\dot{E}\\\delta\dot{L}\\\delta\dot{Q}\end{array}\right) \right.\nonumber\\
& \left. \hspace{0.5in}  - \left( \begin{array}{ccc}\frac{\partial \delta E}{\partial p}&\frac{\partial \delta E}{\partial e}&\frac{\partial \delta E}{\partial \theta_{\rm tp}}\\ \frac{\partial \delta L}{\partial p}&\frac{\partial \delta L}{\partial e}& \frac{\partial \delta L}{\partial \theta_{\rm tp}}\\ \frac{\partial \delta Q}{\partial p}&\frac{\partial \delta Q}{\partial e}&\frac{\partial \delta Q}{\partial \theta_{\rm tp}} \end{array}\right) \left( \begin{array}{c}\dot{p}_{\Kerr}\\\dot{e}_{\Kerr}\\\dot{\theta_{\rm tp}}_{\Kerr} \end{array}\right)\right]\,,
\label{deltapdot}
\ea
and, expanding to leading order in $p$, ${\cal M}_{\Kerr}^{-1}$ is given by
\be
\left(\begin{array}{ccc}
-2p & 2 \sqrt{p} \sin{\theta_{\rm tp}} & 1
\\
\frac{p}{e} &  -(1-e^2) \frac{\sin{\theta_{\rm tp}}}{e\sqrt{p}} &-(1-e^2) \frac{1}{2 e p}
\\
-\cos{\theta}_{\rm tp} \sin{\theta}_{\rm tp} \frac{a^{2}}{p} & \frac{\cos{\theta}_{\rm tp}}{\sqrt{p}} & -\frac{\tan{\theta_{\rm tp}}}{2 p}
\end{array}\right) .
\ee
As described above, for consistency with the AK model, we have adopted the assumption that the inclination angle is constant, therefore $\delta\dot{\theta_{\rm tp}} = \dot{\theta_{\rm tp}}_{\Kerr} = 0$. We see from the above matrix that $\partial Q_{\Kerr}/\partial p/ \partial {L}_{\Kerr}/\partial p=\partial Q_{\Kerr}/\partial e/ \partial {L}_{\Kerr}/\partial e= 2z_-\sqrt{p}/\sin\theta_{\rm tp}$. This result and the constancy of the inclination angle allows us to relate $\delta\dot{Q}$ to $\delta\dot{L}$, which gives expression~(\ref{constinceq}), which was used to derive the values of $\delta \dot{Q}$ presented in the previous section.

The constant inclination angle assumption also allows us to simplify Eq.~(\ref{deltapdot}) to
\ba
\delta\dot{p} &= -2p \left(\delta\dot{E} - \frac{\partial\delta E}{\partial p} \dot{p}_{\Kerr} -  \frac{\partial\delta E}{\partial e} \dot{e}_{\Kerr}\right) \nonumber \\
&+ \frac{2\sqrt{p}}{\sin\theta_{\rm tp}}\left(\delta\dot{L} - \frac{\partial\delta L}{\partial p} \dot{p}_{\Kerr} -  \frac{\partial\delta L}{\partial e} \dot{e}_{\Kerr} \right)\,, \\
\delta\dot{e} &= \frac{p}{e} \left(\delta\dot{E} - \frac{\partial\delta E}{\partial p} \dot{p}_{\Kerr} -  \frac{\partial\delta E}{\partial e} \dot{e}_{\Kerr}\right) \nonumber \\
&- \frac{(1-e^2)}{e\sqrt{p}\sin\theta_{\rm tp}}\left(\delta\dot{L} - \frac{\partial\delta L}{\partial p} \dot{p}_{\Kerr} -  \frac{\partial\delta L}{\partial e} \dot{e}_{\Kerr} \right)\,.
\ea
In general, only the second term contributes to $\delta\dot{p}$, while both terms contribute to $\delta\dot{e}$.

Once $\delta\dot{p}$ and $\delta\dot{e}$ have been determined, the correction to the rate of change of frequency can be found from
\ba
2\pi M\delta\dot{\nu} &= -\frac{3(1-e^2)^{3/2}}{2p^{5/2}} \delta\dot{p} - \frac{3e(1-e^2)^{1/2}}{p^{3/2}} \delta\dot{e} \nonumber \\ &+2\pi M\frac{\partial\delta\nu}{\partial p}\dot{p}_{\Kerr}+2\pi M\frac{\partial\delta\nu}{\partial e}\dot{e}_{\Kerr}\,,
\ea
where we have written $\dot{\nu} = \dot{\nu}_{\Kerr}+\epsilon \delta\dot{\nu}$ as usual.

Using these expressions, we find under assumptions ${\cal B}_2$ (and setting $\gamma_{3,1} = 0$)
\ba
M \delta \dot{p}_{{\cal B}_{2}} &= - \frac{16}{5} \frac{\eta}{p^{4}} \left(1 - e^{2}\right)^{3/2} 
g_{p,2} \left(\gamma_{1,2} + 2 \gamma_{4,2}\right) \,,
 \\
M \delta \dot{e}_{{\cal B}_{2}} &= - \frac{16}{5} \frac{\eta}{p^{5}} \left(1 - e^{2}\right)^{3/2}  
g_{e,2} \left(\gamma_{1,2} + 2 \gamma_{4,2}\right)\,,
 \\
2 \pi M^{2} \delta \dot{\nu}_{{\cal B}_{2}} &= \frac{16}{5} \eta\frac{(1 - e^{2})^{2}}{p^{13/2}} 
g_{\nu,2} \left(\gamma_{1,2} + 2 \gamma_{4,2}\right)\,,
\ea
where 
\ba
\label{B2gs1}
g_{p,2} &= 12 + \frac{35}{2} e^{2} + \frac{1}{2} e^{4}\,,
\\
g_{e,2} &= \frac{93}{4} e + \frac{67}{4} e^{3} + \frac{1}{4} e^{5}\,,
 \\
g_{\nu,2} &= 18 + 78 e^{2} + \frac{99}{4} e^{4}\,.
\label{B2gs}
\ea
Under assumptions ${\cal B}_N$ we find
\ba
M\delta\dot{p}_{{\cal B}_{N}} &= -\frac{16}{5}\eta \frac{(1-e^2)^{3/2}}{p^{N+2}} g_{p,N}(e) \left(\gamma_{1,N}+2\gamma_{4,N}\right)\,, \\
M\delta\dot{e}_{{\cal B}_{N}} &= -\frac{16}{5}\eta \frac{(1-e^2)^{3/2}}{p^{N+3}} g_{e,N}(e) \left(\gamma_{1,N}+2\gamma_{4,N}\right)\,, \\
2\pi M^2\delta\dot{\nu}_{{\cal B}_{N}} &=  \frac{16}{5}\eta \frac{(1-e^2)^{2}}{p^{N+9/2}} g_{\nu,N}(e) \left(\gamma_{1,N}+2\gamma_{4,N}\right)\,,
\ea
where the various eccentricity-dependent factors are
\ba
\label{BNgs1}
g_{p,3}(e) &= 24 + \frac{149}{3}e^2 + \frac{509}{48} e^4\,, 
\\
g_{e,3}(e) &=  \frac{191}{4}e + \frac{111}{2}e^3+\frac{385}{48}e^5\,,
 \\
g_{\nu,3}(e) &= \frac{51}{2}+\frac{2717}{16}e^2+\frac{8325}{64}e^4+\frac{539}{64}e^6\,,
\\
g_{p,4}(e) &= 42+\frac{1373}{12}e^2+\frac{2627}{48}e^4+\frac{3}{4}e^6
\nonumber \\
&- 2(1-e^2)^{3/2} \left(1+\frac{7}{8}e^2\right)\,,
 \\
g_{e,4}(e) &= 80 e + \frac{14177}{96}e^3+\frac{9337}{192}e^5 +\frac{3}{8}e^7
 \nonumber \\
&- (1-e^2)^{3/2} e\left( \frac{19}{6}+\frac{121}{96}e^2\right)\,,
 \\
g_{\nu,4}(e) &= 36 + \frac{663}{2} e^{2} + \frac{12875}{32} e^{4} + \frac{3853}{64} e^{6}
\nonumber\\
&- \frac{3}{32} \left(1 - e^{2}\right)^{3/2} \left(128 + 288 e^{2} + 9 e^{4} \right)\,,
 \\ 
g_{p,5}(e) &= 66+\frac{2825}{12}e^2+\frac{2979}{16}e^4+\frac{763}{48}e^6
\nonumber \\
&- 6(1-e^2)^{3/2} \left(1+\frac{7}{8}e^2\right)\,, 
 \\
g_{e,5}(e) &= 117 e + \frac{32545}{96} e^{3} + \frac{35117}{192} e^{5} + \frac{3059}{256} e^{7}
\nonumber \\
&- \frac{1}{32} e \left(1 - e^{2}\right)^{3/2} \left(304 + 121 e^{2} \right)\,,
 \\
g_{\nu,5}(e) &= \frac{99}{2} + \frac{9321}{16} e^{2} + \frac{64179}{64} e^{4} + \frac{19643}{64} e^{6} + \frac{2765}{256} e^{8}
\nonumber \\
&+ \frac{3}{32} \left(1 - e^{2}\right)^{3/2} \left(29 e^{4} - 856 e^{2} - 448 \right)\,.
\label{BNgs}
\ea
In the CS limit, if we carry out this same procedure, we find
\ba
M \delta \dot{p}_{\CS} &= 10 \frac{\eta}{p^{15/2}} \left(1 - e^{2}\right)^{3/2} 
\frac{\zeta a}{\sin{\theta_{\rm tp}}} g_{p,\CS}\,,
 \\
M \delta \dot{e}_{\CS} &= - 5 \frac{\eta}{p^{17/2}} \left(1 - e^{2}\right)^{3/2}  
\frac{\zeta a}{e \sin{\theta_{\rm tp}}} g_{e,\CS} \,,
 \\
2 \pi M^{2} \delta \dot{\nu}_{\CS} &= 192 \eta\frac{(1 - e^{2})^{2}}{p^{10}} 
\zeta a \sin{\theta_{\rm tp}} g_{\nu,\CS}\,,
\ea
where
\ba
\label{CSgs1}
g_{p,\CS} &= 1 + \frac{209}{20} e^{2} + \frac{68}{5} e^{4} + \frac{441}{160} e^{6} + \frac{3}{89} e^{8}
\nonumber \\
&+ \sin^{2}{\theta_{\rm tp}} \left[ - 33 - \frac{8243}{60} e^{2} - \frac{18343}{120} e^{4} - \frac{5349}{160} e^{6}
\right. 
\nonumber \\
&- \left.
 \frac{27}{80} e^{8}
+\frac{6}{5} \left(1 + \frac{7}{8} e^{2}\right) \left(4 + e^{2}\right) \left(1 - e^{2}\right)^{3/2}
\right]\,,
 \\
g_{e,\CS} &= 1 + \frac{189}{20} e^{2} + \frac{63}{20} e^{4} - \frac{347}{32} e^{6} - \frac{87}{32} e^{8}
+ \frac{3}{80} e^{10}\nonumber \\
& -\sin^{2}{\theta_{\rm tp}}
\left[1 - \frac{1563}{20} e^{2} - \frac{7049}{20} e^{4} - \frac{45727}{160} e^{6} \right.\nonumber\\
& - \frac{101431}{1920} e^{8} 
-\frac{27}{80} e^{10} \nonumber \\
&\left.+\frac{1}{80} e^{2} \left(304 + 121 e^{2} \right) \left(4 + e^{2}\right) \left(1 - e^{2}\right)^{3/2}
\right]\,,
 \\
g_{\nu,\CS} &= 1 + \frac{1273}{96} e^{2} + \frac{45313}{1536} e^{4} + \frac{5741}{384} e^{6} + \frac{10913}{8192} e^{8} 
\nonumber \\
&- \frac{15}{8} \left(1 - e^{2}\right)^{3/2} \left(1 + \frac{691}{192} e^{2} + \frac{373}{512}e^{4} + \frac{71}{3072} e^{6} \right)\,,
\nonumber \\
\label{CSgs}
\ea
We notice that there is a problem with the $\dot{e}$ result, as there are terms proportional to $1/e$ that do not cancel. This implies that circular orbits would not remain circular under radiation reaction. It is known that circular orbits do remain circular in GR~\cite{Kennefick:1995za}. This proof carries over to the modified gravity spacetimes we consider here, and this is proven in Appendix~\ref{app:circ}. The unphysical result we have derived above has arisen from the assumption that the inclination of the orbital plane, $\theta_{\rm tp}$, remains constant at leading order. It is clear that this assumption is incompatible with the physical requirement that circular orbits remain circular. We used the constant inclination assumption to derive $\dot{Q}_\CS$ from $\dot{L}_\CS$ for simplicity, and to ensure consistency with the AK model which also makes this assumption. For general relativity, a PN formula similar to Eqs.~(\ref{Edot-Ldot}) exists to compute the rate of change of the Carter constant~\cite{Flanagan:2007tv}. Additionally, for extreme-mass-ratio inspirals, a formula for the evolution of the Carter constant based on an expansion of the Teukolsky equation is also known~\cite{2005PThPh.114..509S,2006PThPh.115..873S}. The constant inclination assumption is known to lead to some pathologies in numerical kludge models of extreme-mass-ratio inspirals~\cite{Babak:2006uv}, but these arise only when this assumption is used in conjunction with exact Kerr geodesic trajectories, and the constant inclination assumption does give the correct leading order expression for $\dot{Q}$. The fact that we have found an inconsistency only in the CS case suggests that the assumption of constant inclination also gives the correct $\dot{Q}$ at leading order for each of the other modified gravity spacetimes we have considered here. A full computation of $\dot{Q}$, using either a PN expression or an expansion of the Teukolsky equation, is beyond the scope of the current paper, but should be explored in the future.

In the CS case, we can derive the leading order correction to $\dot{Q}$ {\it in the circular orbit limit}, $e=0$, from the requirement that circular orbits remain circular under radiation reaction. That is, we can solve for the $\delta \dot{Q}_{\CS}$ such that $\delta \dot{e}_{\CS}$ does not contain a $1/e$ piece. Doing so, we find
\be
M\delta\dot{Q}_\CS(e=0) = -114 \frac{1}{p^{15/2}} \eta a\zeta\sin\theta_{\rm tp}\cos^2\theta_{\rm tp}\,,
\ee
which in fact guarantees that $\delta \dot{e}_{\CS}(e = 0) = 0$. With this at hand, the remaining elements become
\ba
M\delta\dot{p}_\CS(e=0) &= -272  \frac{1}{p^{15/2}} \eta a\zeta\sin\theta_{\rm tp}\,,  \\ 
M\delta\dot{\theta}_{{\rm tp}\CS}(e=0) &= 5 \frac{1}{p^{17/2}} \eta a\zeta \cos\theta_{\rm tp}\,,  \\ 
2\pi M\delta\dot{\nu}_\CS(e=0) &= -168  \frac{1}{p^{10}} \eta a\zeta\sin\theta_{\rm tp}\,. 
\ea
We cannot derive $\dot{Q}$ for eccentric orbits in this way, nor the corresponding results for $\delta\dot{p}$, $\delta\dot{e}$ or $\delta\dot{\theta}_{\rm tp}$, although the eccentricity dependences of $\dot{E}$ and $\dot{L}$ given in expression~(\ref{CSfluxes}) are correct.

\subsection{Corrections to AK waveforms}
We now combine the above results to obtain final expressions for the corrections to the AK waveforms. The final step is to rewrite the results in terms of the orbital frequency, $\nu$, rather than the semi-latus rectum, $p$. This can be achieved by replacing $p$ in the preceding expressions by $p_{\Kerr} + \epsilon \delta p$, in which
\ba
p_{\Kerr} &= \frac{(1-e^2)}{(2\pi M\nu)^{2/3}}\,,
\\
\delta p &= \frac{2}{3} (1-e^2) \frac{2\pi M\delta\nu}{(2\pi M\nu)^{5/3}} .
\ea
In this way, we get a modification to the $O(\epsilon)$ part of $\delta\dot{\tilde{\gamma}}$ etc., from the GR piece of these terms, evaluated at $p_{\Kerr}+\epsilon\delta p$. This additional correction is sub-dominant for $\delta\dot{\alpha}$ and $\delta\dot{\tilde{\gamma}}$, but can contribute at leading order to $\delta\dot{\nu}$ and $\delta\dot{e}$. This gives final expressions for the fluxes as follows.

Under assumptions ${\cal B}_2$ (and setting $\gamma_{3,1} = 0$), these latter corrections are subdominant. The final expressions for the corrections to the AK waveforms are
\ba
M \delta \dot{e}_{{\cal B}_{2}} &= - \frac{16}{5} \eta\frac{(2\pi M\nu)^{10/3}}{\left(1 - e^{2}\right)^{7/2}}  
g_{e,2} \left(\gamma_{1,2} + 2 \gamma_{4,2}\right)\,,
 \\
2 \pi M^{2} \delta \dot{\nu}_{{\cal B}_{2}} &= \frac{16}{5} \eta\frac{(2\pi M\nu)^{13/3}} {(1 - e^{2})^{9/2}}
g_{\nu,2} \left(\gamma_{1,2} + 2 \gamma_{4,2}\right)\,,  \\
M\delta\dot{\tilde{\gamma}}^{{\cal B}_2} &= \frac{(2\pi M\nu)^{5/3}}{2(1-e^2)} \left( \gamma_{1,2} + 2\gamma_{4,2}\right)\,,  \\
M\delta\dot{\alpha}^{{\cal{B}}_{2}} &= -\frac{a(2\pi M\nu)^{2}}{(1-e^2)^{3/2}} (\gamma_{1,2}+2\gamma_{4,2})\,,
\ea
where $g_{e,2}$ and $g_{\nu,2}$ are as given in Eqs.~\eqref{B2gs1}-\eqref{B2gs}.

Under assumptions ${\cal B}_N$, we have
\ba
M\delta\dot{e}_{{\cal B}_{N}} &= -\frac{16}{5}\eta \frac{(2\pi M\nu)^{2N/3+2}}{(1-e^2)^{N+3/2}} g_{e,N}(e) \left(\gamma_{1,N}+2\gamma_{4,N}\right)\,,  
 \\
2\pi M^2\delta\dot{\nu}_{{\cal B}_{N}} &=  \frac{16}{5}\eta \frac{(2\pi M\nu)^{2N/3+3}}{(1-e^2)^{N+5/2}} g_{\nu,N}(e) \left(\gamma_{1,N}+2\gamma_{4,N}\right)\,, 
 \\ 
M\delta\dot{\tilde{\gamma}}^{{\cal B}_N} &= \frac{(2\pi M\nu)^{(2N+1)/3}}{(1-e^2)^{N-1}} g_{\gamma,N}(e) (\gamma_{1,N}+2\gamma_{4,N})\,, 
 \\
M\delta\dot{\alpha}^{{\cal B}_N} &= - a \frac{(2\pi M\nu)^{2(N+1)/3}}{(1-e^2)^{N-1/2}} g_{\alpha,N}(e) (\gamma_{1,N}+2\gamma_{4,N})\,,
\ea
where the eccentricity-dependent factors have now been modified to
\ba
g_{e,3}(e) &=  \frac{725}{12}e + \frac{383}{8}e^3+\frac{143}{48}e^5\,,
  \\
g_{\nu,3}(e) &= 42+\frac{407}{2}e^2+\frac{345}{4}e^4+\frac{33}{16}e^6\,,
\\
g_{\gamma,3}(e) &=  \frac{3}{2}\,,  \\
g_{\alpha,3}(e) &=\frac{3}{4}+\frac{1}{4}e^2\,,  \\
g_{e,4}(e) &= \frac{316}{3} e + \frac{12713}{96}e^3+\frac{2467}{64}e^5 +\frac{3}{8}e^7
 \nonumber \\
&+ (1-e^2)^{1/2} e\left( \frac{95}{18}-\frac{305}{96}e^2-\frac{605}{288}e^4\right)\,,
 \\
g_{\nu,4}(e) &= 69 + \frac{3191}{8} e^{2} + \frac{5035}{16} e^{4} + \frac{3039}{64} e^{6}
\nonumber\\
&- \frac{1}{48} \left(1 - e^{2}\right)^{1/2} \left(48 - 358 e^{2} + 147 e^{4} + 163e^6 \right)\,,
 \\ 
g_{\gamma,4}(e) &= 3+ \frac{3}{4}e^2\,, \\
g_{\alpha,4}(e) &= 1+e^2\,,  \\
g_{e,5}(e) &= 155 e + \frac{31565}{96} e^{3} + \frac{30749}{192} e^{5} + \frac{5305}{768} e^{7}
\nonumber \\
&+ \frac{5}{96} e \left(1 - e^{2}\right)^{1/2} \left(304 -183 e^{2} - 121 e^4 \right)\,,
 \\
g_{\nu,5}(e) &= 99+ \frac{5601}{8} e^{2} +905 e^{4} + \frac{15617}{64} e^{6} + \frac{1137}{256} e^{8}
\nonumber \\
&- \frac{1}{16} \left(1 - e^{2}\right)^{1/2} \left(144-466 e^{2} + 75 e^{4} + 247e^6 \right)\,
 \\
g_{\gamma,5}(e) &=5 + \frac{15}{4} e^2\,,  \\
g_{\alpha,5}(e) &=\frac{5}{4} + \frac{5}{2} e^2 + \frac{1}{4} e^4\,.
\ea
Finally, for circular orbits in the CS limit we have
\ba
M\delta\dot{e}_\CS (e=0) &= 0 \,,
 \\
2 \pi M^{2} \delta \dot{\nu}_{\CS} (e=0) &= 360 (2\pi M\nu)^{20/3} \eta a \zeta \sin\theta_{\rm tp} \,,
 \\
M\delta\dot{\tilde{\gamma}}_\CS (e=0) &= \frac{75}{8} (2\pi M\nu)^4 a\zeta\sin\theta_{\rm tp} \,,
 \\
M\delta\dot{\alpha}_\CS (e=0) &=  -\frac{5}{8} a\zeta (2\pi M\nu)^4 .
\ea
We can compare our results to previous results in the literature. Barack and Cutler~\cite{Barack:2006pq} considered a quadrupolar deformation of the Kerr metric within GR and constructed the associated AK waveforms. These waveforms are similar to those found here in the ${\cal{B}}_{3}$ limit. In both cases, $\delta \dot{\nu}$ and $\delta \dot{\tilde{\gamma}}$ scale as $\nu^{5}$ and $\nu^{7/3}$ [see Eqs.~$(6)$ and $(7)$ in~\cite{Barack:2006pq}]. Differences arise in the evolution equation for $\delta \dot{e}$ and $\delta{\dot{\alpha}}$, which are probably due to the fact that the metrics used here and in~\cite{Barack:2006pq} cannot be mapped into each other. This is because our metric is constructed such that an approximate Carter constant remains, while that used in~\cite{Barack:2006pq} is of Petrov type I and does not have a Carter constant. Even then, however, our modified AK waveforms serve as a superset of the Barack and Cutler study~\cite{Barack:2006pq} and also the Glampedakis and Babak study~\cite{Glampedakis:2005cf}.

We note that the expressions given above depend on the eccentricity and inclination parameters, $e$ and $\theta_{\rm tp}$, which are gauge dependent quantities. If we had constructed the waveforms using a different definition of eccentricity or inclination, or by parameterising the orbits using the energy, angular momentum and Carter constant, we would have obtained somewhat different corrections. Some care must therefore be taken when interpreting the meaning of the gauge dependent waveform parameters $p$, $e$ and $\theta_{\rm tp}$. Nonetheless, we have constructed the waveform corrections to ensure that quantities which are measurable by an observer at infinity, specifically the three frequencies characterising the motion and their time derivatives, are consistent with the perturbed metric (modulo the omission of the ``direct'' contribution to the radiation field). Therefore, if the preceding waveform corrections were now rewritten as functions of these three frequencies, the same, gauge-independent, waveform model should result, independently of the parameterisation used to derive it.

The corrections depend on the $\gamma_{i,j}$ parameters, and on $N$, which are both related to the dependence of the metric coefficients on the radial coordinate $r$. One might therefore expect these quantities to have a dependence on the choice of radial coordinate. If we were to make an infinitesimal coordinate transformation of the form $r \rightarrow r + \epsilon \delta r(r)$, we could change the $r$ dependence of the various metric corrections, $h_{\mu\nu}$. However, any such transformation would introduce $h_{\theta\theta} \neq 0$, which puts the metric into a different form than the one we have analysed here. This apparent discrepancy can be understood from how the metric was constructed --- certain choices were made in the form assumed for the Killing tensor of the perturbed spacetime, as described in Section~\ref{sec:bumpy-description}, that result in the identification of the coordinate $r$ with a Boyer-Lindquist-like radius. It can be interpreted as the circumferential radius in the equatorial plane, and becomes the oblate-spheroidal radial coordinate at infinity. It is the fact that this coordinate has been fixed in this way that allows quantities that relate to it to be asymptotically measurable and to appear in the waveforms.

\subsection{Example Waveform}
\begin{figure}
\begin{tabular}{c}
\includegraphics[angle=270,width=0.5\textwidth]{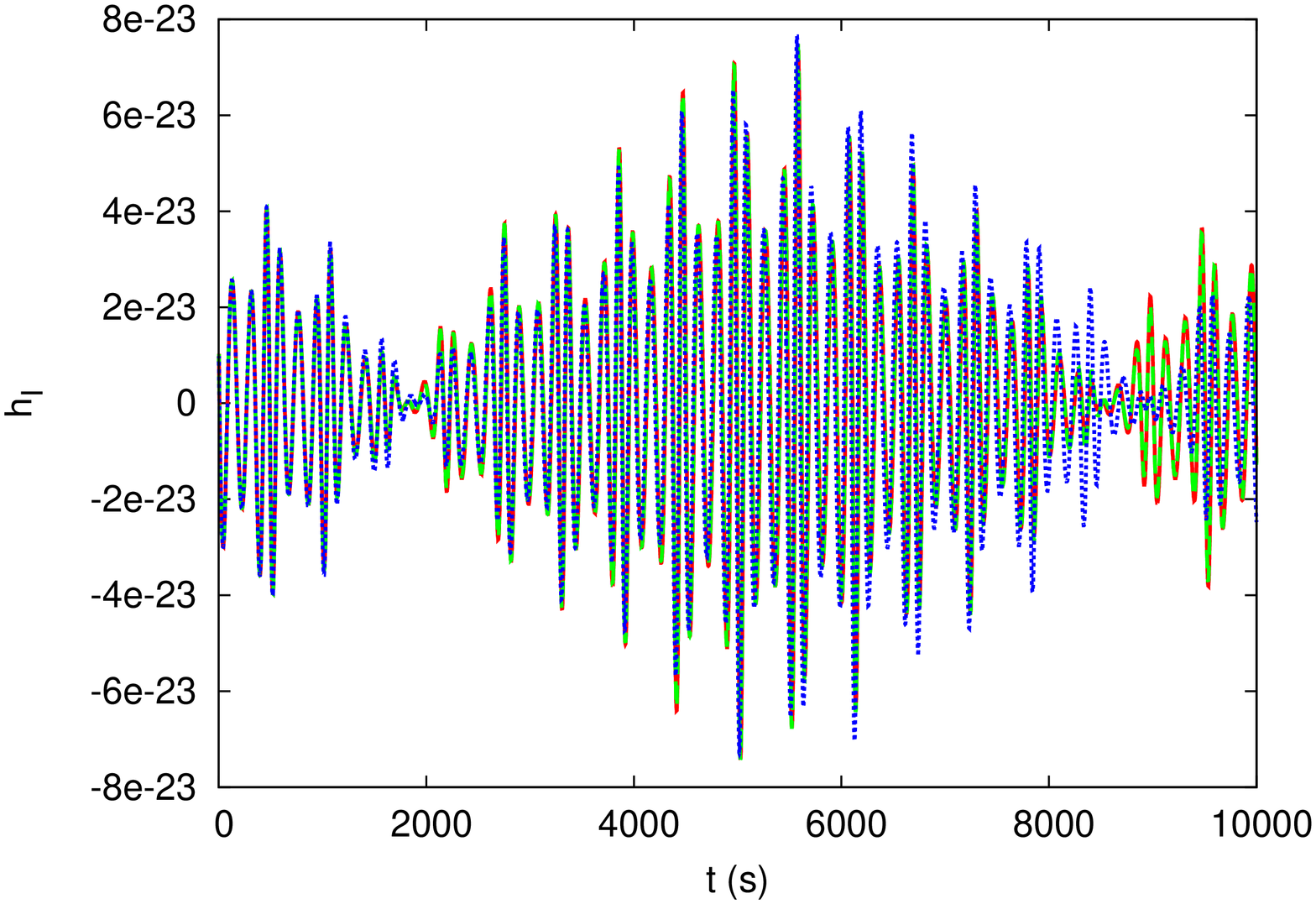}\\
\includegraphics[angle=270,width=0.5\textwidth]{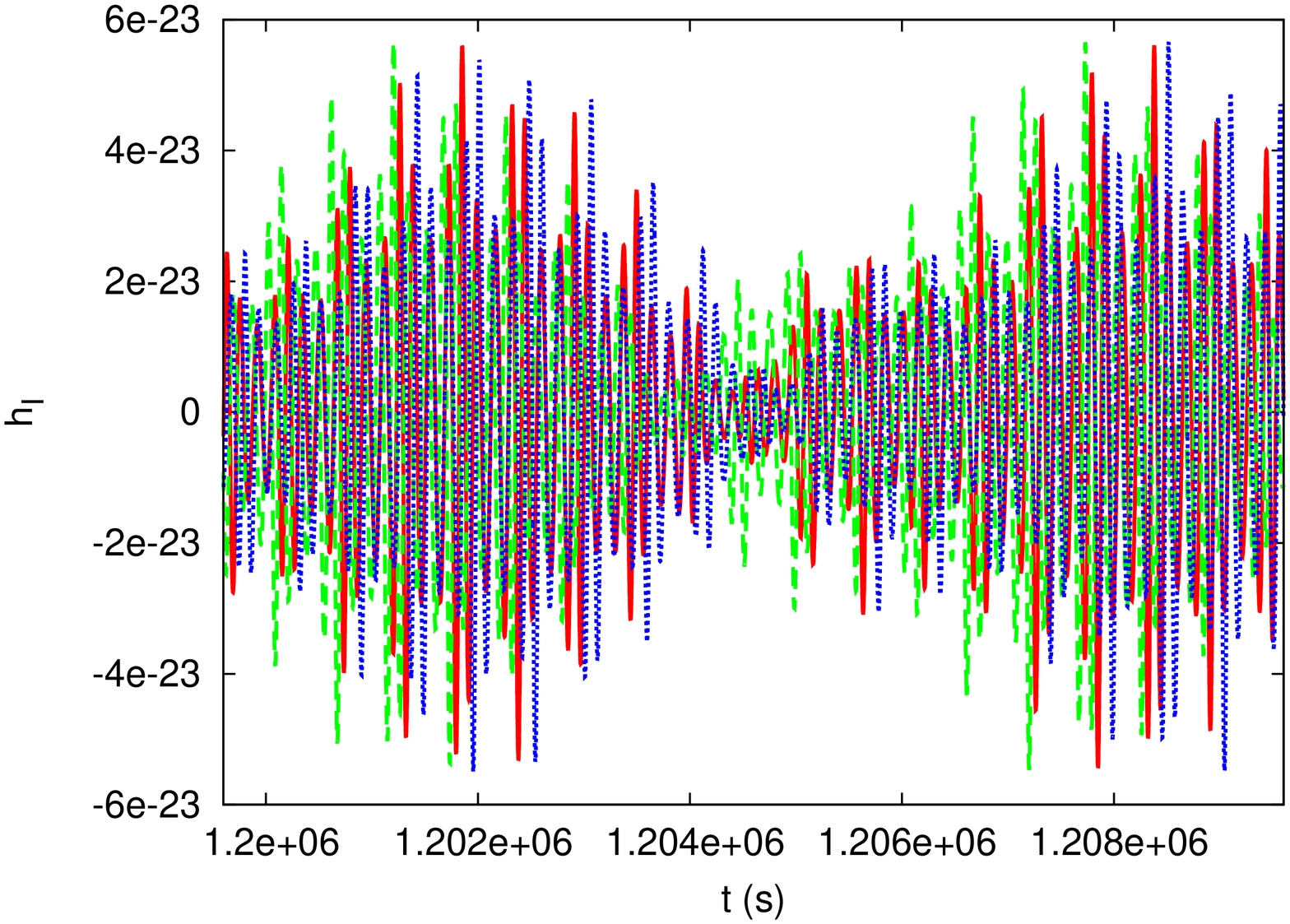}
\end{tabular}
\caption{\label{wavesamp}Example of a modified gravity waveform. In each panel, we show the $h_I$ Michelson response for EMRIs in the Kerr metric (red solid line) and in a ${\cal B}_2$-type modified gravity deformed metric with $\epsilon (\gamma_{1,2}+2\gamma_{4,2}) = 0.01$ (green dashed line) and  $\epsilon (\gamma_{1,2}+2\gamma_{4,2}) = 0.1$ (blue dotted line). The two panels show the first (upper) and last (lower) $10^4$s of a $2$ week long waveform.
}
\end{figure}

To illustrate the modified gravity waveforms we show an example in Figure~\ref{wavesamp}. This plot shows one of the two low-frequency Michelson response channels of the LISA detector, $h_I$, as described in~\cite{AK}, for an EMRI occurring in the Kerr metric, and for EMRIs occurring in a deformed metric with a ${\cal B}_2$-type deviation. We use two different values for the size of the deformation of the metric, $\epsilon (\gamma_{1,2}+2\gamma_{4,2}) = 0.01$ and $\epsilon (\gamma_{1,2}+2\gamma_{4,2}) = 0.1$. The other waveform parameters were taken to be $M=10^6M_{\odot}$, $\mu = 10M_{\odot}$, $S/M^2 = 0.7$, $e_0=0.25$, $\lambda=\pi/4$, $\nu_0=0.00177$, $\theta_S=\pi/4$, $\phi_S=\pi/2.34$, $\theta_K=\pi/8$, $D=1$Gpc and $\phi_K=\alpha_0=\tilde{\gamma}_0=\Phi_0=0$, using the notation of the analytic kludge model~\cite{AK}. We show two sections of the waveform, each of duration $10^4$s, taken at the beginning and the end of a $2$ week long observation. 

We see that the ${\cal B}_2$ waveform for the smaller perturbation is almost indistinguishable from the Kerr waveform for the first $10^4$s, while for the larger perturbation the waveform can be seen to be beginning to drift out of phase already by the end of the first segment. All three waveforms are clearly out of phase in the second segment and would therefore be clearly distinguishable in a LISA observation of two weeks duration. We note, however, that we have fixed the parameters for all three waveforms to the same values, and have not made any attempt to improve the overlap between the modified gravity waveforms and the Kerr waveform templates by making small adjustments to the parameters. Such parameter adjustment would better approximate the process of parameter estimation through matched filtering that will be used to analyse LISA data in practice and it is likely that this would significantly improve the phase coherence between the Kerr and modified gravity waveforms. Nonetheless, this example illustrates that LISA should be able to place some kind of constraint on the size of any deviations from GR of this form that would be consistent with EMRIs observed  by LISA. We leave a full study of such constraints, accounting for parameter correlations, to the future.

\section{Discussion}
\label{sec:Conclusions}
We have here taken the first steps toward the construction of ppE waveforms for EMRIs. We have taken the parametrically deformed, bumpy metric for modified gravity theories described in~\cite{Vigeland:2011ji} and calculated the geodesic equations in second and first-order form, the orbital frequencies and the implicit deformations to the fluxes of energy, angular momentum and Carter constant. With all these ingredients, we then explicitly described how to construct ppE AK waveforms. 

A natural follow-up to this work would be to study the accuracy with which space-borne detectors, such as LISA, could constrain such ppE deformations. It was in the context of such a study for EMRIs in a Kerr background that the original AK model was developed by Barack and Cutler~\cite{AK}. Now that we have obtained the leading order corrections to the AK waveforms, we can follow a similar analysis to ask with what precision LISA or a LISA-like mission might be able to measure the bump parameters $\gamma_{N,M}$ that characterize the deviations from GR. The modifications described here could also be incorporated into the numerical kludge waveform model~\cite{Babak:2006uv} for a similar study. Higher-order PN corrections have already been included in that framework for Kerr EMRIs. 

Another possible application of these results would be to study how these GR deformations could be constrained with observations in the electromagnetic spectrum, such as from the emission spectra of AGN or iron line profiles~\cite{2008LRR....11....9P}. The emission properties of discs are determined in part by the geodesic structure of the BH spacetime, which is characterized by the orbital frequencies and precession rates computed in this paper. We note, however, that such an analysis would have to account for the gauge dependence of these results, and work with gauge-invariant quantities, for example by rewriting the expressions in terms of the frequencies of the motion, or some other asymptotic observables. We have not worried about this here, as the goal was to provide a framework in which to compute waveform observables and perform a data analysis study. The specific parametrisation used to describe the waveform family does not affect those aims, although care will have to be taken when interpreting the observations that are made using these template waveforms.

The construction of ppE waveforms for modified gravity bumpy metrics is not yet complete, as we have here neglected explicit or direct corrections to the radiation-reaction force. As shown in~\cite{2009PhRvD..80l2003Y}, corrections to the conservative and dissipative sectors of any theory lead to waveform modifications that might be degenerate. Although this is definitely the case for comparable mass, quasi-circular inspirals, the generality of EMRI orbits might break these degeneracies.Thus, the work done here can be considered as a solid first-step toward generic ppE EMRI waveforms, where future studies should concentrate on the addition of dissipative effects.

Complete results for the corrections to the radiation-reaction terms in CS modified gravity were only  presented for circular orbits, as we had made the simplifying assumption that the change of the orbital inclination due to radiation reaction was a sub-dominant effect. This assumption made the computation of the corrections to $\dot{Q}$ more tractable, but it appears not to be valid in CS modified spacetimes. We were able to present results for circular orbits by imposing the physical requirement that circular orbits remain circular under radiation reaction. An extension of the $\dot{Q}$ calculation to the case of generic orbits should be pursued in the future. 

\begin{acknowledgments}
JG's work is supported by the Royal Society. NY acknowledges support from the National Aeronautics and Space Administration through Einstein Postdoctoral Fellowship Award Number PF0-110080 issued by the Chandra X-ray Observatory Center, which is operated by the Smithsonian Astrophysical Observatory for and on behalf of the National Aeronautics Space Administration under contract NAS8-03060. JG thanks the Aspen Centre for Physics for hospitality while this paper was being finished and we thank Carlos Sopuerta, Priscilla Canizares and Scott Hughes for useful discussions.
\end{acknowledgments}

\appendix
\section{Useful integrals}
\label{intapp}
Computation of the frequencies has relied in various places on the evaluation of integrals of the general form
\be
I_n = \int _0^\pi (1+e\cos\psi)^n {\rm d}\psi 
\ee
and we describe here how these may be computed.

For $n \geq 0$ we can expand the integrand as
\be
(1+e\cos\psi)^n = \sum_{k=0}^n \left(\begin{array}{c}n\\k\end{array}\right) e^k \cos^k\psi .
\ee
By writing $\cos\psi = ({\rm e}^{i\psi} + {\rm e}^{-i\psi})/2$, we find
\ba
\cos^k\psi &= \frac{1}{2^{k-1}} \sum_{l=0}^{\lfloor \frac{k-1}{2} \rfloor} \left[ \left(\begin{array}{c}k\\l\end{array}\right) \cos((k-2l) \psi) \right] \nonumber \\
&+ \left(\left\lfloor \frac{k}{2} \right\rfloor - \left\lfloor \frac{k-1}{2} \right\rfloor \right) \frac{1}{2^k}  \left(\begin{array}{c}k\\ k/2 \end{array}\right)
\ea
in which $\lfloor x\rfloor$ denotes the largest integer smaller than $x$. On integration over the interval $[0,\pi]$, all the terms in the summation give zero. The second term is only non-zero when $k$ is even and therefore only the even powers of $\cos\psi$ contribute to the result. We therefore find
\be
I_n  = \pi \sum_{k=0}^{\lfloor \frac{n}{2} \rfloor}\frac{n!}{(n-2k)! (k!)^2} \left(\frac{e}{2}\right)^{2k} 
\ee
in which $n! = n(n-1)...1$ is the usual factorial. If $n < 0$, the integral can be evaluated by first writing $\cos\psi =\cos^2(\psi/2) - \sin^2(\psi/2)$ and then using a substitution $t=\tan(\psi/2)$. This reduces the integral to
\be
I_n = \int_0^\infty \frac{2 (1+t^2)^{n-1}}{((1+e)+(1-e)t^2)^n} {\rm d}t .
\ee
This integrand can be decomposed as
\ba
\frac{(1+t^2)^{n-1}}{((1+e)+(1-e)t^2)^n} = \frac{\left(\frac{1+e}{1-e} + t^2 - \frac{2e}{1-e}\right)^{n-1}}{(1-e)^n \left(\frac{1+e}{1-e} + t^2\right)^n} \nonumber \\ = (1-e)^{-n} \sum_{k=0}^{n-1} \left(\begin{array}{c}n-1\\k\end{array}\right) \left(\frac{-2e}{1-e}\right)^k J_{k+1}\nonumber\\
\mbox{where } J_k = \int_0^\infty \frac{1}{\left(\frac{1+e}{1-e} + t^2\right)^k} {\rm d}t \hspace{3cm}\nonumber
\ea
Finally, writing $(1+e)/(1-e) = {\cal E}$ we can see through integration by parts
\ba
(3-2k) J_{k-1}  &= - 2 {\cal E} (k-1) J_k \nonumber \\
J_1&=\frac{\pi}{2\sqrt{\cal E}} \nonumber \\
\Rightarrow J_n &= \frac{(2n-3)!! \, \pi}{2^{n-1} {\cal E}^{n-1/2} (n-1)!} \qquad n \ge 2 \nonumber
\ea
where $n!! = n (n-2) (n-4) ....$.

The integrals we will need for the various special cases we discuss in this paper are
\ba
I_{-2} &= \frac{\pi}{(1-e^2)^{3/2}}, \qquad I_{-1} = \frac{\pi}{(1-e^2)^{1/2}}, \nonumber \\
I_0 &= I_1 = \pi, \qquad I_2=\pi \left(1+\frac{e^2}{2}\right), \nonumber \\
I_3 &= \pi \left(1+\frac{3e^2}{2}\right), \qquad I_4 = \pi \left(1+3e^2+\frac{3e^4}{8}\right)\nonumber \\
  \nonumber \\
 I_5 &= \pi \left(1+5e^2+\frac{15e^4}{8}\right) .
\ea

\section{Circular orbits Remain Circular}
\label{app:circ}
In this appendix, we study the proof of Kennefick and Ori~\cite{Apostolatos:1993nu,Kennefick:1995za} that circular orbits remain circular, but extended to the non-Kerr backgrounds analyzed in this paper.  Let us begin by recasting the radial equation of motion as an equation for the carter constant
\be
Q = H_{\GR}(r,E,L) + \delta H(r,E,L) - \Delta u_{r}^{2}\,,
\label{Q-def-app}
\ee
where as usual $\Delta = r^{2} - 2 M r + M^{2} a^{2}$, while $u_{r} = g_{rr} u^{r} = \rho^{2} \dot{r}/\Delta$. The potentials $H_{\GR}$ and $\delta H$ are defined via
\ba
H_{\GR} &= \Delta^{-1} \left[ E (r^{2} + M^{2} a^{2}) - M^{2} a L\right]^{2} 
\nonumber \\
&- M^{2} \left(L - a E\right)^{2} - r^{2}\,,
\nonumber \\
\delta H &= \frac{\delta R}{\Delta M^{2}}\,,
\ea
where the $\delta H$ in the CS limit becomes
\be
\delta H_{\CS} = \frac{5}{4} \frac{M E L}{\Delta} a \zeta \left(\frac{M}{r}\right)^{2}\,.
\ee

The action of an external force on a particle will force the evolution of the Carter constant. When this force is the radiation reaction force, then
\be
F_{\alpha} = \frac{D u_{\alpha}}{D \tau}\,,
\ee
where $D$ stands for covariant proper-time differentiation. The Carter constant then evolves as
\be
\dot{Q}= \frac{\partial Q}{\partial u_{\alpha}} F_{\alpha}\,.
\ee
Using the instantaneous circularity condition, $u_{r} = 0$, plus the chain rule, the above equation becomes
\be
\dot{Q} = Q_{,E} \dot{E}  + Q_{,L} \dot{L}. 
\ee
Using Eq.~\eqref{Q-def-app}, the evolution of $\dot{Q}$ for a circular orbit is simply
\be
\dot{Q} = H^{\GR}_{,E} \dot{E}  + H^{\GR}_{,L} \dot{L} + \frac{1}{\Delta M^{2}} \left(\delta R_{,E} \dot{E}  + \delta R_{,L} \dot{L}\right) \,, 
\label{app:final-dot-Q}
\ee
which in the CS limit becomes
\be
\dot{Q} = H^{\GR}_{,E} \dot{E}  + H^{\GR}_{,L} \dot{L} + \frac{5}{4} \frac{M^{2} E L}{\Delta} a \zeta \left(\frac{M}{r}\right)^{2} \left(\frac{\dot{E}}{E} + \frac{\dot{L}}{L} \right)\,. 
\ee

Let us now compute the rate of change of the Carter constant required to take a circular orbit into a new circular orbit. Consider then Eq.~\eqref{Q-def-app}, but solve for $\Delta u_{r}^{2}$. This new equation will then resemble that of a particle in an effective potential, given by $W(r,E,L,Q) \equiv H_{\GR} + \delta H - Q$. The stable circularity condition is equivalent to requiring that $W = 0$ and $W_{,r} = 0$. 

If the radiation reaction force acts on this instantaneously circular geodesic for a short time $\delta \tau$ only. After this, the orbit will change to a new geodesic with parameters ${\cal{I}}' \equiv (r',E',L',Q')$. Denoting the change by $\delta {\cal{I}} = {\cal{I}} - {\cal{I}}'$, the change in the potential is then
\be
\delta W = W_{,E} \delta E + W_{,L} \delta L + W_{,Q} \delta Q\,,
\ee
where we have used the fact that $W_{,r} = 0$ for circular orbits. 

For this equation to hold before and after the action of the radiation reaction force, we must have $\delta W = 0$, which allows us to solve for $\delta Q$. After dividing by $\delta \tau$ and taking the $\delta \tau \to 0$ limit, we find
\be
\dot{Q}_{\rm circ} =  - \frac{W_{,E}}{W_{,Q}} \dot{E} - \frac{W_{,L}}{W_{,Q}} \dot{L} \,. 
\ee
Using the definition $W$, and {\emph{assuming}} that $W_{,Q} = -1$, we then find
\be
\dot{Q}_{\rm circ} =  H^{\GR}_{,E}  \dot{E} + H^{\GR}_{,L} \dot{L} 
+ \frac{1}{\Delta M^{2}} \left( \delta R_{,E} \dot{E} + \delta R_{,L}\right) \dot{L}\,. 
\label{app:Qdot-circ}
\ee
The assumption that $W_{,Q} = -1$ relies on the condition that $\delta R$ be independent of $Q$. This condition is satisfied in CS gravity and in all ${\cal{B}}_{N}$ limits. It is not, however, satisfied if one modifies the ${\cal{B}}_{3}$ limit by letting $\gamma_{1,2} \neq 0$, or if one modifies the ${\cal{B}}_{4}$ limit by letting $\gamma_{1,3} \neq 2 \gamma_{1,2}$ and if one modifies the ${\cal{B}}_{5}$ limit by letting $a^{2}\gamma_{1,2} \neq 2 \gamma_{1,3} - \gamma_{1,4}$. 

The fact that $\dot{Q}_{\rm circ}$ in Eq.~\eqref{app:Qdot-circ} is identically equal to $\dot{Q}$ in Eq.~\eqref{app:final-dot-Q} means that even after the radiation-reaction force acts on an instantaneously circular orbit, the induced, instantaneous evolution of the orbit's Carter constant is such as to maintain instantaneous circularity. This result is identical to what is found for geodesics in a Kerr background. We can trace back the reason for this to the fact that the potential $\delta H$ (or equivalently $\delta R$) is independent of the Carter constant in all the cases studied here, including the CS one.  

\bibliography{master}
\end{document}